\documentclass[acmsmall]{acmart}
\usepackage{amsthm}
\usepackage[normalem]{ulem}
\usepackage[linesnumbered, noend]{algorithm2e}
\usepackage{amsmath}
\usepackage{subfig}
\usepackage{graphicx}
\usepackage{enumitem}
\usepackage{siunitx}
\usepackage{flushend}
\usepackage{worldflags}
\usepackage{appendix}

\AtBeginDocument{%
  }
\newtheorem{theorem}{Theorem}
\newcommand{\systemName}{\texttt{CarbonScaler}\xspace}
\newcommand{\algoName}{\texttt{Carbon Scaling Algorithm}\xspace}
\newcommand{\impName}{\texttt{Carbon AutoScaler}\xspace}
\newcommand{\simName}{\texttt{Carbon Advisor}\xspace}
\newcommand{\profilerName}{\texttt{Carbon Profiler}\xspace}
\newcommand{\agnosticPolicy}{\texttt{carbon-agnostic}\xspace}
\newcommand{\suspendPolicy}{\texttt{suspend-resume}\xspace}
\newcommand{\staticPolicy}{\texttt{static-scale}\xspace}

\setcopyright{acmlicensed}
\acmJournal{POMACS}
\acmYear{2023} \acmVolume{7} \acmNumber{3} \acmArticle{57} \acmMonth{12} \acmPrice{15.00}\acmDOI{10.1145/3626788}

\begin{document}
\title[CarbonScaler]{CarbonScaler: Leveraging Cloud Workload Elasticity for Optimizing Carbon-Efficiency}

\author{Walid A. Hanafy}
\orcid{0000-0001-5765-8194}
\affiliation{%
  \institution{University of Massachusetts Amherst}
  \city{Amherst}
  \state{MA}
  \country{USA}
  \postcode{01002}
}
\email{whanafy@cs.umass.edu}

\author{Qianlin Liang}
\orcid{0000-0003-4702-5689}
\affiliation{%
  \institution{University of Massachusetts Amherst}
  \city{Amherst}
  \state{MA}
  \country{USA}
  \postcode{01002}
}
\email{qliang@cs.umass.edu}

\author{Noman Bashir}
\orcid{0000-0001-9304-910X}
\affiliation{%
  \institution{University of Massachusetts Amherst}
  \city{Amherst}
  \state{MA}
  \country{USA}
  \postcode{01002}
}
\email{nbashir@umass.edu}

\author{David Irwin}
\orcid{0000-0003-1722-4927}
\affiliation{%
  \institution{University of Massachusetts Amherst}
  \city{Amherst}
  \state{MA}
  \country{USA}
  \postcode{01002}
}
\email{irwin@ecs.umass.edu}

\author{Prashant Shenoy}
\orcid{0000-0002-5435-1901}
\affiliation{%
  \institution{University of Massachusetts Amherst}
  \city{Amherst}
  \state{MA}
  \country{USA}
  \postcode{01002}
}
\email{shenoy@cs.umass.edu}

\renewcommand{\shortauthors}{Walid A. Hanafy et al.}

\begin{abstract}
Cloud platforms are increasing their emphasis on sustainability and reducing their operational carbon footprint. A common approach for reducing carbon emissions is to exploit the temporal flexibility inherent to many cloud workloads by executing them 
in periods with the greenest energy and suspending them at other times. Since such suspend-resume approaches can incur long delays in job completion times, we present a new approach that exploits the elasticity of batch workloads in the cloud to optimize their carbon emissions.  Our approach is based on the notion of ``carbon scaling,'' similar to cloud autoscaling, where a job dynamically varies its server allocation based on fluctuations in the carbon cost of the grid's energy. We develop a greedy algorithm for minimizing a job's carbon emissions via carbon scaling that is based on the well-known problem of marginal resource allocation. We implement a \systemName prototype in Kubernetes using its autoscaling capabilities and an analytic tool to guide the carbon-efficient deployment of batch applications in the cloud. We then evaluate CarbonScaler using real-world machine learning training and MPI jobs on a commercial cloud platform and show that it can yield i) 51\% carbon savings over carbon-agnostic execution; ii) 37\% over a state-of-the-art suspend-resume policy; and iii) 8\% over the best static scaling policy.

\end{abstract}



\maketitle

\section{Introduction}
\label{sec:intro}
Data centers worldwide consume over 200TWh of energy each year---comprising roughly 1\% of global electricity usage~\cite{Masanet2020RecalibratingGD}---and are poised to grow to 3-13\% of global electricity demand by 2030~\cite{Andrae2015OnGE,Jones2018HowTS}. The growth of hyper-scale cloud providers is fueling this rapid increase in energy use, resulting in a significant environmental impact by increasing carbon and greenhouse gas (GHG) emissions~\cite{guardian,ipcc-report}. For the past two decades, cloud providers have relentlessly focused on improving their data centers' energy-efficiency to reduce their operational costs---by driving down their power usage effectiveness (PUE) close to the optimal value of $1$.  As a result, optimizations, such as server consolidation, open-air cooling, and power infrastructure improvements, have yielded significant energy-efficiency gains. However, energy-efficiency improvements alone are insufficient to satisfy cloud data centers'  aggressive sustainability goals, since even energy-efficient data centers may generate significant carbon emissions from their energy use.   This has led to a new emphasis on carbon-efficient operations that directly target reducing data centers' overall carbon emissions~\cite{chien-cacm21}.

Carbon efficiency can be achieved through supply-side or demand-side methods. Supply-side methods include power purchase agreements (PPAs) from renewable generation sources, such as solar, wind, and hydro, which \emph{indirectly} offset a cloud data center's carbon emissions. Such optimizations yield net-zero operation~\cite{google-green-ppa,meta-climate-change, ms-buy-renewable} over a long period, such as a year, but offsets by themselves do not eliminate the instantaneous direct emissions at all times~\cite{hotair}. Consequently, supply-side optimizations must be combined with demand-side methods to reduce a cloud data center's instantaneous direct carbon emissions.  Demand-side optimizations exploit the fact that the carbon intensity of grid-supplied electricity varies both temporally and geographically. A common demand-side optimization is \emph{time shifting} delay-tolerant workloads to periods with the ``greenest'' electricity supply. Although not all cloud workloads are delay tolerant, many types of batch workloads exhibit significant temporal, performance, and even geographic flexibility.

One approach for leveraging the temporal flexibility above is to use \emph{suspend-resume} mechanisms~\cite{Wiesner2021LetsWA,Acun2022AHA,cloudcarbon, Radovanovic2021CarbonAwareCF}, where a scheduler suspends a job when electricity's carbon intensity rises (e.g., above some threshold) and resumes it when it drops (e.g., below the threshold). For example, Google recently adopted carbon-aware time-shifting in its Carbon-Intelligent Computing System~\cite{Radovanovic2021CarbonAwareCF}. While suspend-resume temporal shifting policies can reduce the carbon emissions of delay-tolerant workloads~\cite{Wiesner2021LetsWA}, they suffer from two drawbacks. First, the carbon intensity of grid-supplied electricity changes slowly, and there may be long periods (e.g., many hours) of high carbon periods where jobs remain suspended and make no progress. Such suspensions cause substantial delays in completion time, with 7-10$\times$ increases in completion times in some cases~\cite{ecovisor}.  Second, when batch jobs have limited temporal flexibility and thus cannot be significantly shifted, the effectiveness of these methods is significantly reduced.

To overcome these drawbacks, we present \systemName, a new approach that exploits the \emph{resource elasticity} of cloud workloads to dynamically vary the amount of resources allocated to applications in response to fluctuations in the carbon cost\footnote{We use the terms carbon intensity and carbon cost of electricity interchangeably.} of their energy supply. Our ``carbon scaling'' approach is analogous to cloud autoscaling, where the number of servers allocated to a cloud application varies dynamically over time~\cite{aws-autoscaling}. However, while cloud autoscalers generally respond to variations in applications' workload demand, often for request-based services, our ``carbon scaling'' approach responds to the carbon dynamics of electricity. In essence, carbon scaling scales up the servers allocated to an application when the carbon cost is low and gracefully scales them down when the cost increases. In contrast to the static allocation of suspend-resume approaches, carbon scaling enables faster progress during low carbon periods, which can potentially eliminate delays in job completion times while also reducing carbon emissions.

Designing cloud carbon scaler requires addressing two key design challenges: \emph{how much} to scale each application up or down and \emph{when}. Since different applications exhibit different scaling characteristics with respect to the number of allocated servers, a carbon scaler must take this scaling behavior into account when determining how much to scale up each application during low carbon periods. For example, an embarrassingly parallel job can opportunistically scale up significantly without increasing its overhead (thus increasing its carbon efficiency), while applications with scaling bottlenecks should scale up more judiciously. To maximize carbon savings, \systemName relies on its knowledge of the energy's future carbon intensity, application scalability profile, job length, and other execution constraints to compute a schedule to decide when to perform such scaling operations. However, carbon intensity forecasts, profile estimates, and the expected length are error-prone, requiring carbon scaling decisions to be robust to such errors.

In designing, implementing, and evaluating \systemName, we make the following contributions.

\begin{itemize}[leftmargin=*]
	\item We introduce the notion of carbon scaling for cloud applications, which maps to the well-known problem of marginal resource allocation for which greedy optimal solutions exist~\cite{greedy_optimal}. \systemName builds on these ideas to develop a greedy autoscaling algorithm that minimizes individual application's emissions by scaling the resources up and down in response to carbon cost variations. Further, CarbonScaler can substantially reduce or even eliminate the completion time delays seen in suspend-resume approaches.
    \item We implement a full prototype of \systemName in Kubernetes and use it to leverage a cloud application's elasticity and temporal flexibility to reduce its carbon footprint. We also implement our algorithm in 	\systemName's CarbonAdvisor tool, which enables analysis and simulated execution of cloud applications to evaluate their carbon savings before being deployed in the cloud.  CarbonAdvisor enables system designers to understand better how to minimize their carbon cost based on job characteristics, geographic region, and different run-time parameters.
    \item We evaluate \systemName against multiple baselines using numerous real-world batch applications, including machine learning training and MPI-based scientific jobs. Our results show that \systemName can yield up to i) 51\% carbon savings over a carbon-agnostic execution, ii) 37\% over the state-of-the-art suspend-resume policy, and iii) 8\% over the best static scaling policy.
\end{itemize}

\section{Background}
\label{sec:background}
This section provides background on sustainable data centers and carbon-aware scheduling.

\subsection{Sustainable Data Centers}

In addition to their long-standing emphasis on improving energy efficiency by reducing their PUE,\footnote{Power Usage Effectiveness (PUE) is the ratio of the total energy used by a data center to the energy used for computing.} cloud data centers have recently begun to focus on reducing their carbon footprint~\cite{Acun2022AHA, Radovanovic2021CarbonAwareCF}. 
This can be achieved by reducing operational carbon emissions measured in \unit{\gram CO2eq\per\kilo\watt\hour}, a.k.a. Scope 2 emissions~\cite{ghg}, resulting from electricity use, as well as by reducing embodied carbon---Scope 3 emissions---that arise during the manufacturing of data center hardware (e.g., servers).  Our work focuses on reducing Scope 2 emissions.  Cloud platforms have little direct (Scope 1) emissions, and optimizing embodied carbon of computing workloads is beyond the scope of this paper.

\begin{figure*}[t]
\centering
\begin{minipage}[t]{0.48\linewidth}
  \centering
	\includegraphics[width=\textwidth]{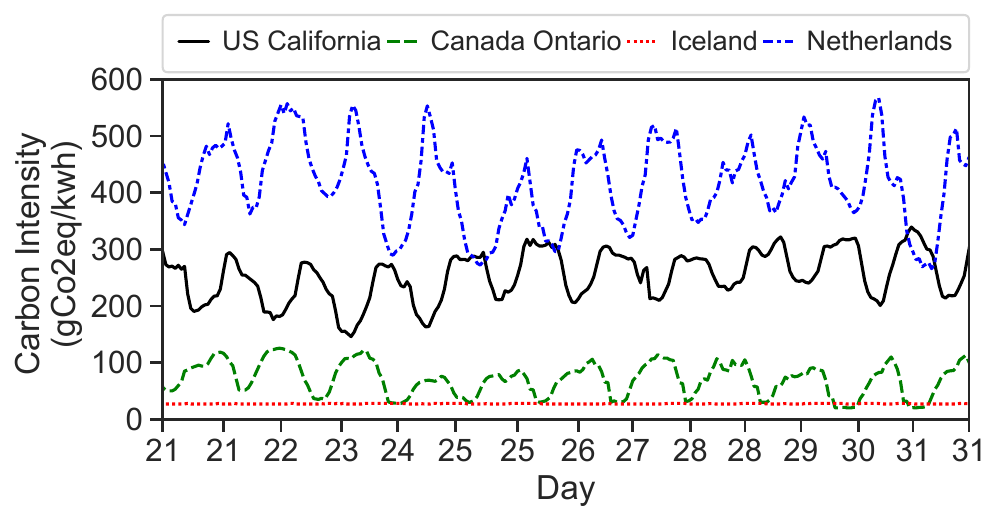}
 	\vspace{-0.85cm}
	\caption{\emph{Grid's carbon intensity shown over a 10 days period varies spatially and temporally.}}
	\label{fig:carbonTrace}
\end{minipage}
\hfill
\begin{minipage}[t]{0.44\linewidth}
  \centering
	\includegraphics[width=\textwidth]{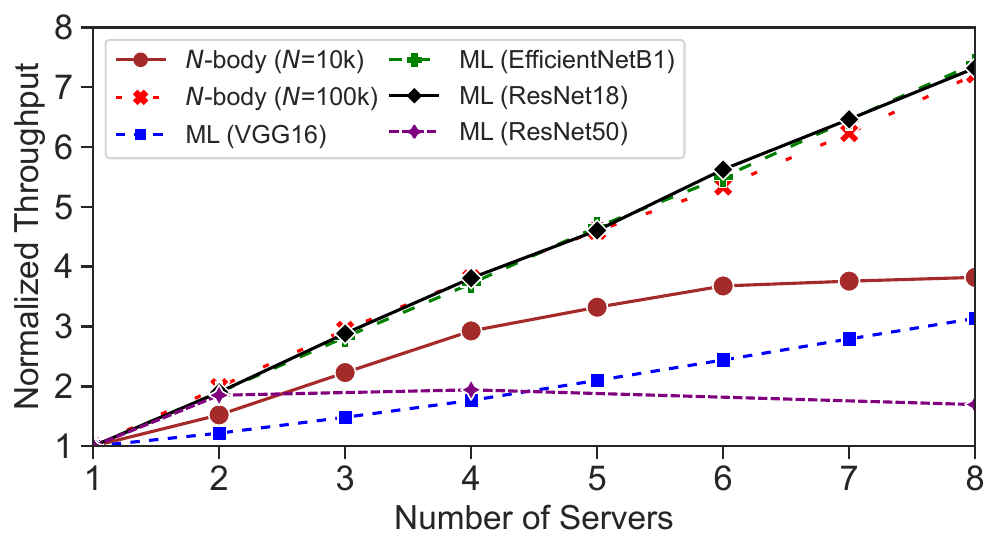}
	\vspace{-0.85cm}
	\caption{\emph{Scaling characteristics of common MPI jobs and machine learning training frameworks.}}
	\label{fig:workloadScale}
\end{minipage}
\vspace{-0.5cm}
\end{figure*}

\subsection{Carbon Intensity of Electricity}
\label{sec:carbon_intensity}

To reduce Scope 2 emissions, cloud data centers must track the carbon cost of their electricity supply and modulate their electricity consumption over time. The carbon cost of electricity depends on the source of generation. For example, a unit of energy generated by a coal plant will have a high carbon cost (i.e., emissions in terms of \unit{\gram CO2eq\per\kilo\watt\hour}), while energy generated from a hydroelectric plant will have no emissions. The electricity the grid supplies is produced by a mix of generation sources, and the resulting carbon cost is a weighted average of the corresponding sources.
Importantly, the generation mix varies from one region to another---based on the local power plants in each region---and also varies over time since the generation mix changes based on demand, the relative cost of generation, and intermittent generation from renewable sources.

Figure~\ref{fig:carbonTrace} depicts how the carbon cost differs by country/region and how it exhibits diurnal variations daily. In this case, Ontario tends to have a low but variable carbon cost because its energy mix consists of a large fraction of carbon-free nuclear and hydroelectric energy combined with some coal plants, which results in non-zero carbon intensity, and solar, which causes the diurnal fluctuations. California is similar but has a higher fraction of solar, which results in larger fluctuations, but also a higher fraction of coal plants, which elevates the average carbon intensity. The Netherlands also shows diurnal variation but with a higher average as it relies more on fossil-based electricity generation. By contrast, the carbon intensity of electricity in Iceland is nearly zero and flat due to its unique abundance of carbon-free geothermal energy.

\subsection{Carbon-aware Cloud Scheduling}
\label{sec:cloud-scheduling}
Many cloud workloads have both temporal flexibility and resource elasticity, which enables exploiting the temporal and spatial variations in energy's carbon intensity, as demonstrated in recent work~\cite{Wiesner2021LetsWA, cloudcarbon, Radovanovic2021CarbonAwareCF, sukprasert2023quantifying, warofeff}. To facilitate such efforts, commercial services, such as electricityMap~\cite{electricity-map} and WattTime~\cite{watttime}, have emerged that aggregate data from grids in different parts of the world and expose grid energy's current and forecasted carbon intensity to cloud providers and users in real-time. Researchers, in turn, are exploiting this data to design carbon-aware schedulers that dynamically shift workloads across time and space to reduce emissions. 

As mentioned above, temporal shifting involves moving delay-tolerant batch workloads to periods of low carbon intensity. In Figure~\ref{fig:carbonTrace}, for instance, rather than running a batch job continuously in a \emph{carbon-agnostic} manner, \emph{suspend-resume} approaches execute the job in the ``valleys'', where the carbon cost is low, and suspend the job during peak periods. This technique has been explored in recent work~\cite{Radovanovic2021CarbonAwareCF,Wiesner2021LetsWA,cloudcarbon,sukprasert2023quantifying}. Threshold-based suspend-resume scheduling policies suspend jobs whenever the carbon cost rises above a certain threshold, while deadline-based methods choose the $n$ lowest carbon cost periods between the arrival time and the deadline to execute the job.  Importantly, a key drawback of \emph{suspend-resume} methods, whether threshold-based or deadline-based, is that the carbon savings depend on the amount of time the user is willing to wait for their job to complete---a higher delay tolerance yields higher savings, but also a longer completion times.

Geographic or spatial shifting, in contrast, migrates jobs or workloads to regions with the greenest electricity grid~\cite{Moghaddam2014CarbonawareDC,Zheng2020MitigatingCA,Zhou2013CarbonAwareLB,cloudcarbon}.  However,  batch jobs often cannot exploit geographic shifting due to data privacy regulations, such as GDPR, that impose regional restrictions. Even when possible, spatially shifting jobs can incur high migration costs if it requires moving substantial state or data associated with the job. Since \systemName focuses on batch jobs, spatial shifting is outside the scope of this paper.  We discuss related work in spatial shifting in Section~\ref{sec:relatedwork}.

\subsection{Carbon-Aware Autoscaling}\label{sec:bg-scope}
Cloud workloads fall into two broad classes: interactive and batch. Since interactive workloads are latency-sensitive, they are not amenable to temporal shifting optimizations, and scaling is only beneficial in response to demand variations. Hence, our work focuses on distributed batch workloads, such as machine learning jobs, data analytics, and scientific computing simulations, which run on multiple machines. Given its benefits, elastic execution mechanisms are now built into many machine learning frameworks, such as Pytorch \cite{pytorch}, data processing frameworks, such as Spark \cite{elastic_spark}, as well as scheduling frameworks~\cite{kubeflow,elastic_hpc,sergeev2018horovod}.

Although autoscaling can be applied from a cluster or cloud service provider perspective, our work focuses on the cloud \emph{application's} perspective of carbon scaling, similar to cloud autoscaling. A typical autoscaler adjusts the number of servers based on the application demand, where higher demand can be measured in latency or average utilization of provisioned resources~\cite{aws-autoscaling,autoscaler}. However, \systemName adjusts the number of servers based on the carbon intensity of electricity. In both cases, the cloud application operates under an abstract view of the underlying servers and can allocate as many as needed.  We discuss carbon scaling from a cloud providers' perspective in Section~\ref{sec:discussion}.

Elastic scaling capabilities have enabled designing policies that scale resources based on electricity's carbon intensity~\cite{ecovisor}.  For example, a policy might scale up an application's resources when carbon is low and either halt or scale down when carbon is high. However, unlike traditional autoscalers, a distributed batch application often has communication bottlenecks that vary widely across applications and govern the scaling benefits. Figure~\ref{fig:workloadScale} depicts the scaling behavior of four deep learning training jobs, which use Horovod and PyTorch for elastic scaling and two MPI tasks that perform scientific computations. As shown, ResNet18 training and the larger $N$-body MPI computation show a linear increase in throughput as the number of servers increases, indicating linear scaling behavior. In contrast, the smaller $N$-body MPI computation exhibits diminishing growth in throughput with increased server allocation. Finally, VGG18 and ResNet50 training tasks exhibit a slower increase in throughput due to scaling bottlenecks. These differences in scaling behavior, as well as the variability in carbon intensity and execution constraints (e.g., start time and deadline), should be considered by a carbon scaling approach when optimizing for carbon savings.

\section{CarbonScaler Design}
\label{sec:carbon-scale-algo}
This section motivates \systemName in the context of prior work, formulates the carbon scaling problem, and then presents \systemName's design.

\begin{figure}[t]
    \centering
    \begin{tabular}{ccc}
    \includegraphics[width=0.29\textwidth]{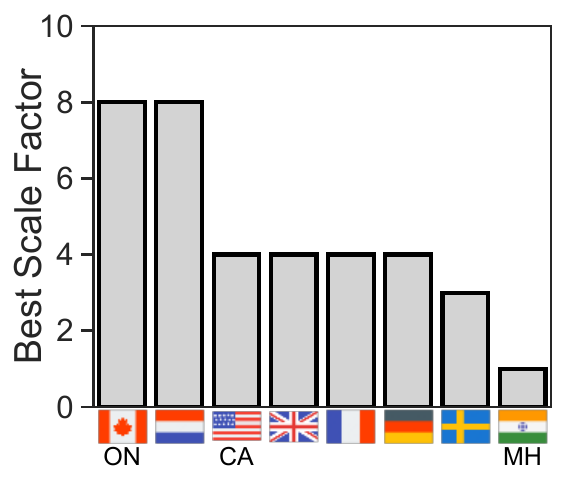} &
    \includegraphics[width=0.30\textwidth]{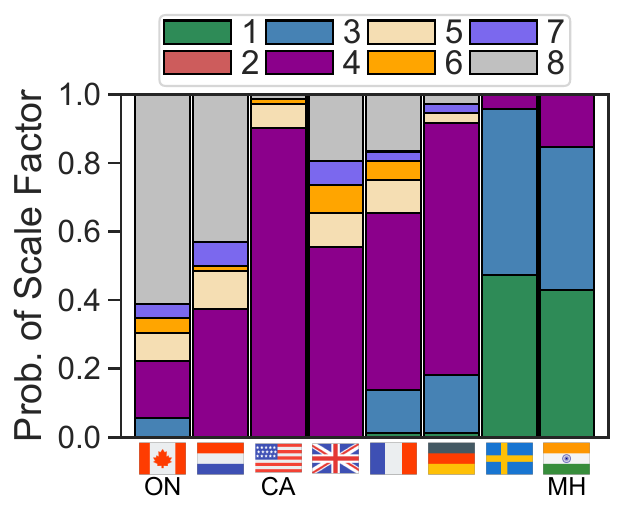} &
    \includegraphics[width=0.29\textwidth]{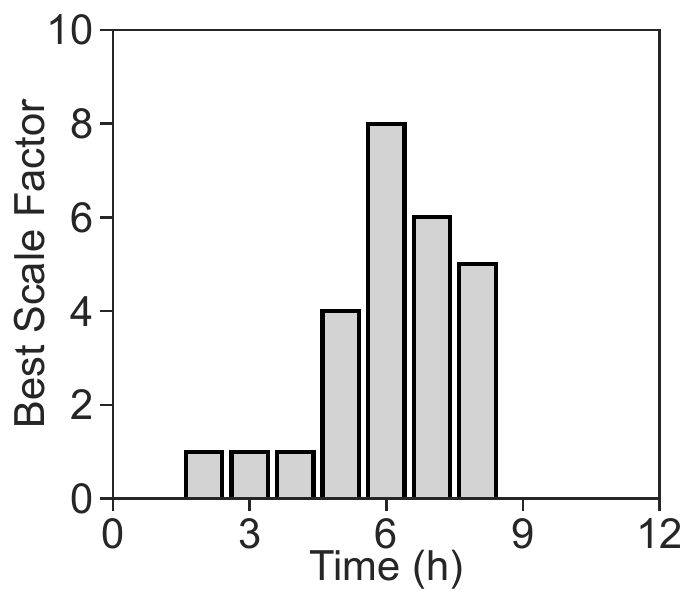} 
     \vspace{-0.1cm} \\
    (a) Effect of Region &  (b) Effect of Start Time & (c) Effect of Runtime \\
    \end{tabular}
    \vspace{-0.4cm}
    \caption{\emph{Best static scale factor varies across (a) geographical regions, (b) job start times, and (c) job execution.}}
    \vspace{-0.5cm}
    \label{fig:scale_factor_variations}
\end{figure}

\subsection{Motivation}\label{sec:motivation}

The \emph{suspend-resume} and temporal shifting policies proposed in prior work can reduce the carbon emissions of delay-tolerant workloads~\cite{Wiesner2021LetsWA}. However, they suffer from two drawbacks. First, the carbon intensity of grid-supplied electricity changes slowly, and thus, there may be arbitrarily long intervals (e.g., many hours) of high carbon periods where jobs remain suspended and make no progress. Such suspensions delay completion times, with a 7-10$\times$ increase in completion times in some cases~\cite{ecovisor}.  Second, when batch jobs have limited temporal flexibility and cannot be significantly shifted, the effectiveness of these methods is significantly reduced.

To overcome the drawbacks above, our paper presents \emph{CarbonScaler}, a new approach that exploits the \emph{resource elasticity} of cloud workloads to dynamically vary the amount of resources allocated to applications in response to fluctuations in the carbon cost of their energy supply. 
\systemName's key insight is that 
{\em the scale which yields the minimum carbon consumption, not only depends on the application characteristics but also the variations in carbon intensity across geographical regions, application start times within a given region, and the runtime of an application following a specific start time.}
Importantly, current approaches for selecting an application's scale factor do not apply directly to this context, necessitating a new approach. Specifically, analytic performance models of an application, such as those used in cloud auto-scaling approaches, only account for application performance characteristics and do not consider the impact of time-varying carbon intensity on the scale factor. Similarly, the state-of-the-art approach for leveraging workload elasticity demonstrates that this scale factor varies across applications~\cite{ecovisor} but does not provide an algorithm for choosing this scale factor or show how carbon intensity variations should be considered when doing so.

To demonstrate the impact of application characteristics and temporal variations in carbon intensity on the scale factor, we consider an {\em oracle} approach for choosing the best static scale factor for a 24hr job on a per-region, per start time, and per-timeslot for ML (ResNet 18).
Figure~\ref{fig:scale_factor_variations}(a) shows that the best static scale factor for a given application varies significantly, from 1$\times$ to 8$\times$, across geographical regions, as different regions exhibit different variations in carbon intensity. Figure~\ref{fig:scale_factor_variations}(b) presents the distribution of best static scale factors across all the possible start times for various regions for one of them. We observe that there is no single static scale that works for a given region due to the differences in their carbon intensity profiles.  In addition, the static scale must also be adapted depending on when an application executes. Finally, as shown in Figure~\ref{fig:scale_factor_variations}(c), the best static scale factor can even vary during application execution time, where the lowest carbon consumption is achieved by running the application with {\em five different} scaling factors.  Further, neither application performance models, which are inherently carbon-oblivious, nor state-of-the-art carbon-aware techniques, such as Ecovisor~\cite{ecovisor} or Wait Awhile\cite{Wiesner2021LetsWA}, can realize this oracle approach.

The dynamicity of choosing the best static scale factor motivates the design of \systemName, which adapts the operating scale factor for each application depending on where and when it executes. \systemName avoids computing the best static scale factor across application runs in an exhaustive brute-force manner and instead computes a carbon-aware schedule using a greedy approach. We next formulate the problem and present our dynamic scaling algorithm.

\subsection{Problem Formulation}
\label{sec:problem}
Similar to cloud autoscalers that scale each application \emph{independently}, a carbon scaler operates independently on each cloud application that wishes to optimize its carbon emissions.
When a new batch application arrives at time $t$, it specifies (i) the minimum number of servers, $m$, that it needs to run, where $m\geq1$, and (ii) the maximum number of servers $M$ that can be allocated to it, $M \geq m$.  The carbon scaler can then vary the servers allocated to the application between $m$ and $M$.   Suppose that $l$ is the estimated job length when executing on the baseline allocation of $m$ servers.\footnote{The job length $l$ can be estimated using profiling and modeling~\cite{optimus,performance_modeling} or using prior execution history. For example, \cite{weng2022mlaas} reports that 65\% of batch jobs see repeated execution at least five times within a two-month period.} By default, we assume that the desired job completion time is $T=t+l$, which means that jobs should complete ``on time'' with no delays.  Although $T$ must be at least $t+l$ for all jobs, some delay tolerant jobs have significant temporal flexibility and can {\em optionally} specify a longer completion time $T$ such that $T$$>$$t+l$. The value $T - (t+l)$ represents the slack available when executing the job. This slack captures the willingness of users to wait in order to  increase their carbon savings. The default case of $T=t+l$ assumes on-time completion and zero slack.

The completion time $T$ specifies the \emph{temporal flexibility} (delay tolerance) available to the job, while the maximum server allocation $M$ specifies the \emph{resource elasticity} of the job.  The parameters $T$ and $M$ can be specified differently to obtain a range of carbon scaling behaviors. For example, when $T=t+l$, the application has no temporal flexibility and cannot be subjected to suspend-resume mechanisms. In this case, the job can only exploit resource elasticity by scaling up to $M$ workers during low carbon periods and must be completed on time with no delays. In contrast, when $M=m$, the job has no resource flexibility, and the carbon scaler is limited to performing only suspend-resume optimizations with a static number of servers, $m$, while also ensuring it completes the job by the specified completion time $T$. Of course, when $T>t+l$ and $M>m$, a carbon scaler has the most flexibility and can exploit both resource elasticity and temporal shifting via suspend-resume.  Our goal is to design a carbon scaler that minimizes a job's carbon emissions subject to the available resource elasticity and temporal flexibility.

\subsection{Basic Design}
When a new batch job arrives, our system, which we refer to as \systemName, computes an \emph{initial} schedule for executing the job through completion. The execution schedule specifies how many servers to allocate to the batch job over time and when to dynamically change the allocation in response to variations in carbon cost. This initial schedule is based on a forecast of future carbon cost, as well as the expected progress of the job over time based on its resource allocation. 
As the job executes, \systemName adjusts its schedule periodically if it encounters forecast errors or deviations in the job's expected progress --- to ensure completion by the specified completion time $T$. Observed deviations can occur due to profiling errors, from network and locality interference~\cite{network_aware_scheduling}, or resource procurement denials. We discuss these issues further in \S\ref{sec:assumptions}.

\systemName assumes that carbon cost forecasts are available; commercial services~\cite{watttime,carboncast} provide such forecasts for up to four days with high accuracy in most locations.
Since the application specifies its temporal flexibility (in terms of completion time $T$) and its resource elasticity (in terms of the varying server allocation from $m$ to $M$), \systemName's schedule responds to fluctuations in forecasted carbon cost by scaling down or completely suspending the job when the carbon cost is high and opportunistically scaling up when the carbon cost is low.

Different clustered batch applications will have different scaling behaviors, as shown in Figure~\ref{fig:workloadScale}, which should be considered when scaling an application's server capacity between the specified range of $m$ to $M$. 
As noted in Figure~\ref{fig:workloadScale}, applications' throughput either increases sub-linearly or increases somewhat linearly initially and then shows diminishing returns with an additional increase in server capacity. This behavior is a direct consequence of Amdahl's law~\cite{amdahl-law}, which states that the speedup of a parallel application is limited by the amount of sequential code within it --- adding server capacity only speeds up the parallel component of the application. Software bottlenecks, such as synchronization overheads, also limit the ability to scale up.

\begin{figure}[t!]
    \centering
    \begin{tabular}{cc}
    \includegraphics[width=0.35\textwidth]{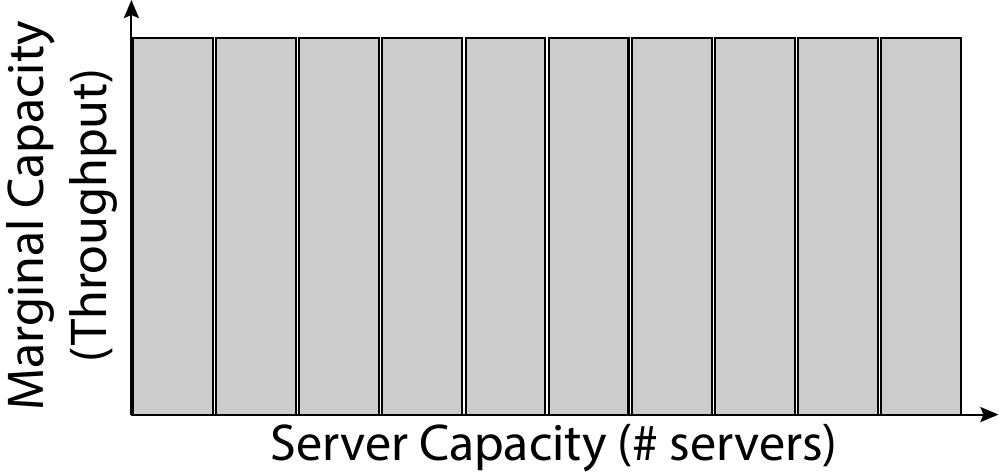} &
    \includegraphics[width=0.35\textwidth]{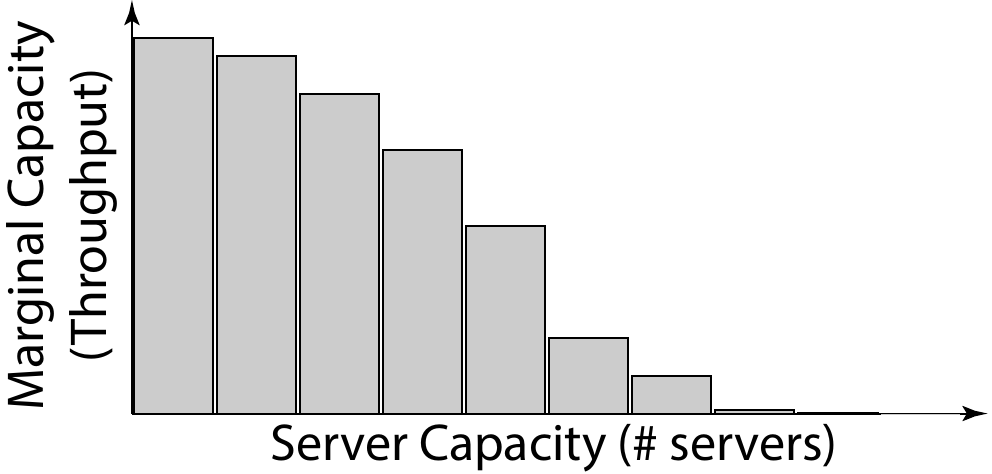}\vspace{-0.1cm} \\
    (a) Linear scaling & (b) Diminishing scaling
    \end{tabular}
    \vspace{-0.42cm}
    \caption{\emph{Example marginal capacity curves.}}
    \label{fig:scalabilityExample}
\end{figure}

\systemName considers this scaling behavior in terms of a \emph{marginal capacity curve}, shown in Figure~\ref{fig:scalabilityExample}, which captures the incremental increase in application capacity (i.e., throughput) for each unit increase in server capacity. 
The ideal case of  linear scaling translates to a \emph{flat} marginal capacity curve where each additional server results in a unit increase in (normalized) application capacity (see Figure~\ref{fig:scalabilityExample}(a)). Most applications will have a diminishing marginal capacity curve, where marginal capacity decreases monotonically with an increase in the server capacity (see Figure~\ref{fig:scalabilityExample}(b)).

The marginal capacity curve and the carbon intensity curve can then be used to scale the application up or down in a carbon-efficient manner. 
To do so, the marginal capacity curve is normalized by the forecasted carbon cost in each time step to compute the {\em marginal capacity per unit carbon} --- the marginal work done per unit carbon. 
\systemName then adds server capacity to the time slots that maximize the work done per unit of carbon. By doing so, \systemName allocates {\em more server resources when the carbon cost is low since more marginal work can be done at a lower carbon cost.} \systemName will incrementally add servers to various time slots until sufficient server capacity has been added to complete the job within the desired completion time $T$, thereby yielding a carbon-efficient execution schedule that optimizes the carbon emissions.

In practice, each application can have multiple marginal capacity curves, each representing a different phase of its execution. For example, a MapReduce job can have different scaling behaviors and marginal capacity curves for its map and reduce phases.  For ease of exposition, our discussion below assumes a single marginal capacity curve per application. However, our approach generalizes to multiple marginal capacity curves by considering the appropriate scaling curve in each time slot that corresponds to the current phase of the application's execution.

\subsection{Carbon Scaling Algorithm}\label{sec:algorithm}
\systemName relies on the knowledge of application scalability profile, carbon intensity forecast, and other job constraints to decide when to i) horizontally scale resources up or down or ii) suspend execution to ensure minimum carbon consumption. 
As noted earlier, when a new job arrives at time $t$, it specifies a \emph{desired} completion time (i.e., a ``deadline'') of time $T$. We also assume that the marginal capacity curve of the application is obtained by profiling the application offline (see Section~\ref{sec:profiler}) and is known at arrival time.  Finally, the algorithm takes the carbon cost forecast $c$, which we assume to be correct. We analyze the impact of inaccurate forecasts in Section~\ref{sec:assumptions}.

We assume that the interval $[t, T]$ is discretized into smaller fixed-length intervals (e.g., 15 minutes or an hour), and the number of servers allocated to the job can be changed at the start of each interval. 
Suppose that there are $n$ time intervals between $[t, T]$,  $n\geq 1$. 
Let $c_1, c_2, ..., c_n$ denote the forecasted carbon cost in each interval $i, i \in [t,T]$. 
Suppose that the marginal capacity curve is denoted by $MC_{m}, MC_{m+1}, ..., MC_{M}$, where $MC_j$ is the marginal capacity increase after allocating the $j$-th servers, $j \in [m, M]$. Since the estimated job length is $l$ when executing with minimum server capacity $m$, the total work the job needs to perform is $W = l \cdot MC_m$.  Our algorithm must compute a schedule where the aggregate server capacity allocated to the job over $[t,T]$ can perform this work before the completion time $T$, minimizing carbon emissions. 

\RestyleAlgo{ruled}
\newlength{\textfloatsepsave} 
\setlength{\textfloatsepsave}{\textfloatsep} 
\setlength{\textfloatsep}{0pt} 
\begin{algorithm}[t]
    \footnotesize
    \caption{\texttt{Carbon Scaling Algorithm()}}
    \label{alg:proposed_algorithm}
    \KwIn{Marginal capacity ($MC$), time slots $[t, T]$, carbon cost forecast ($c$), total work ($W$)}
    \KwOut{Execution Schedule $S$}
    $S \gets [0..0]$\;
    $L \gets []$ \;
    \For{$i \in [t, T]$}{
        \For{$j \in [m, M]$}{%
            $L$.append($i, j, MC_j/c_i$)\;
        }
    } \label{algline:marg-carb}
    $L \gets$ Sort($L$) ; \tcp{w.r.t. Norm. Marginal Cap.} \label{algline:greedy-start} 
    $w \gets 0$ \;
    \While{$w < W$ }{
        $i,j, * \gets L.pop()$; \tcp{next highest $MC_j/c_i$}
        $S[i] = j$; \tcp{increase allocation in slot $i$}
        $w.update(S)$ \;
    } \label{algline:greedy-end}
    \Return $S$
    \vspace{-0.1cm}
\end{algorithm}
\setlength{\textfloatsep}{\textfloatsepsave}
\setlength{\textfloatsep}{0pt}

The aforementioned carbon scaling problem is a marginal allocation problem of discrete resources, which is known to yield an optimal solution in many cases \cite{ greedy_optimal}. Our greedy \texttt{Carbon Scaling Algorithm}, detailed in Algorithm \ref{alg:proposed_algorithm}, builds on the algorithm and theoretical results in~\cite{greedy_optimal}. We provide the requirements and the optimality proof of our greedy \texttt{Carbon Scaling Algorithm} in appendix \ref{app:optimality}.
The Algorithm, first computes the {\em marginal capacity per unit carbon} in each time interval $i$ by normalizing the $MC$ curve with carbon cost $c_i$ in that interval (line~\ref{algline:marg-carb}).\footnote{If an application has multiple marginal capacity curves, we select the one for the execution phase in time slot $i$.} 
That is, the marginal capacity per unit carbon in time interval $i$ is ${MC_m}/c_i, MC_{m+1}/c_i, ..., MC_{M}/c_i$. The greedy algorithm then iteratively and incrementally allocates server capacity to various time slots in order of decreasing {\em marginal capacity per unit carbon} (lines~\ref{algline:greedy-start}-\ref{algline:greedy-end}). For each iteration, the algorithm chooses the interval $i$ from [1, n] such that allocating incremental server capacity to that time slot maximizes the work done per unit carbon (i.e., chooses the interval with the greatest $MC_j/c_i$ across all intervals). After allocating server capacity to that interval, it iteratively determines the next interval where allocating additional server capacity yields the next highest work done per unit of carbon. 

Note that our greedy algorithm may allocate additional capacity to the  same interval as the previous iteration of its marginal work done per unit carbon continues to be the highest across all intervals. Otherwise, a new time interval with the next highest marginal work done per unit is chosen for allocating server capacity.  Also, when a  time interval is initially chosen by the greedy algorithm for capacity allocation, it must be allocated the minimum requirement of $m$ servers, after which the allocation can be increased incrementally by one in subsequent steps. Similarly, if a time slot reaches the maximum allocation of $M$ servers, it is not considered further by the greedy algorithm. The process continues until sufficient capacity has been allocated across the $n$ time intervals to complete $W$ units of work.  This yields an initial schedule where each time interval has either a zero allocation (causing the job to be suspended in that period) or a non-zero allocation between $m$ and $M$, with the server allocation potentially changing at interval boundaries.

\begin{figure*}[t]
    \centering
    \begin{tabular}{cccc}
         \includegraphics[width=0.23\textwidth]{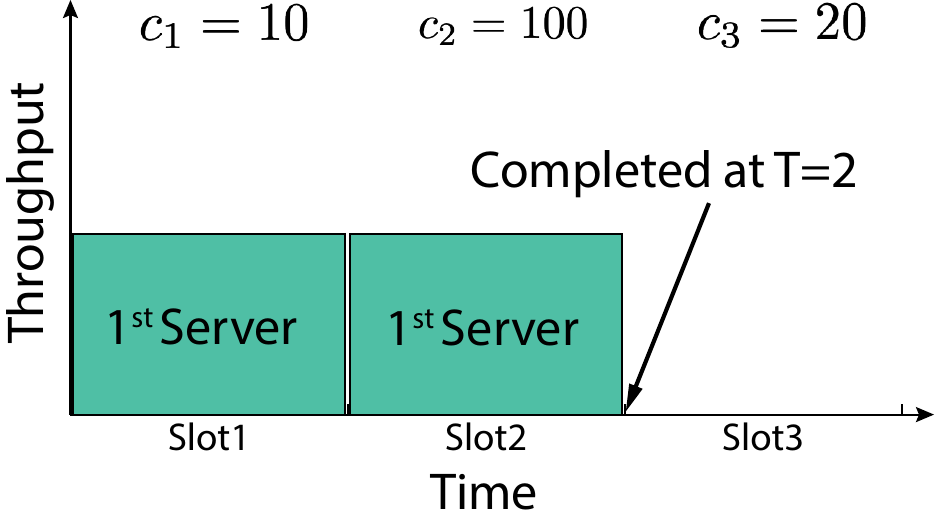} &
         \includegraphics[width=0.23\textwidth]{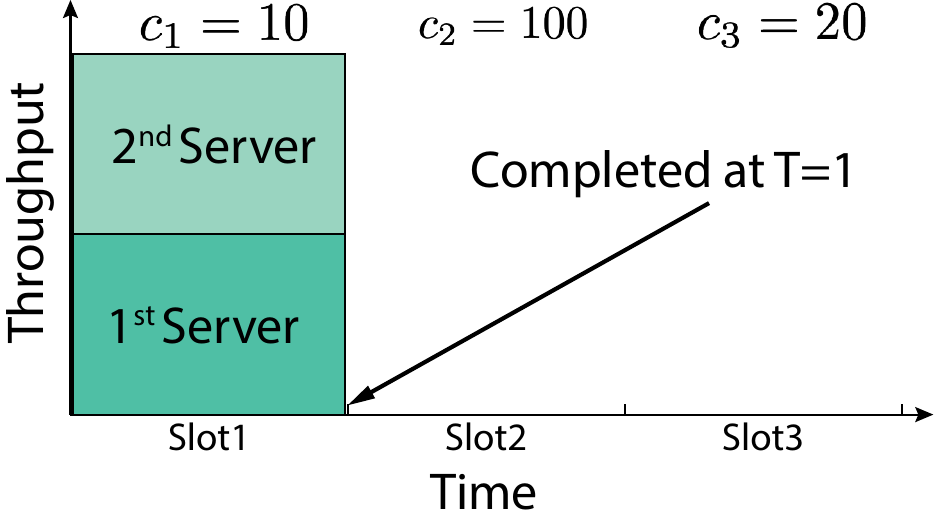} &
         \includegraphics[width=0.23\textwidth]{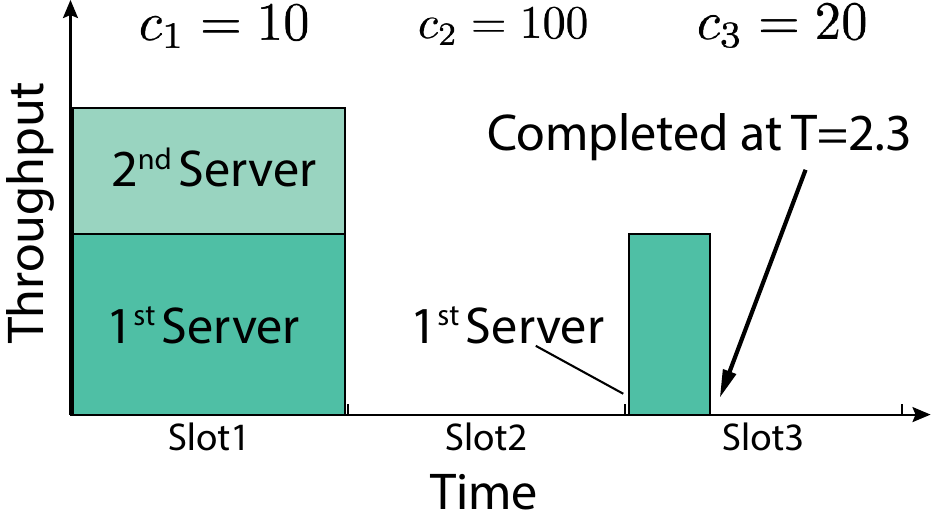} &
         \includegraphics[width=0.2\textwidth]{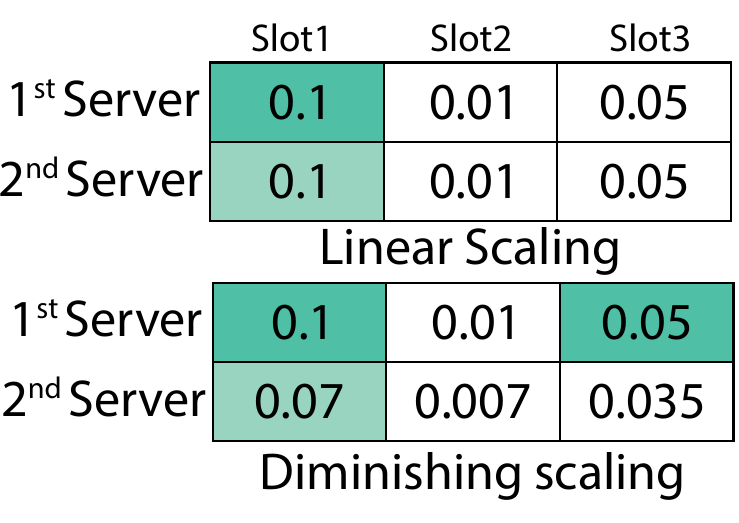}\vspace{-0.1cm}\\
         \footnotesize{(a) Carbon-agnostic}  & \footnotesize{(b) Linear scaling} & \footnotesize{(c) Diminishing scaling} & \footnotesize{(d) MC per unit carbon}
    \end{tabular}
    \vspace{-0.4cm}
    \caption{\emph{An illustrative example of our carbon scaling algorithm at work.}}
    \label{fig:algoExample}
\end{figure*}

\noindent{\bf Example.} To illustrate our carbon scaling algorithm, consider a job of length 2 that arrives at $t=0$ and needs to finish by $T=3$. Suppose that the job needs to execute on at least one server ($m=1$) and at most two servers ($M=2$).  Carbon-agnostic execution will run the job as soon as it arrives, and it will complete at time 2, as shown in Figure~\ref{fig:algoExample}(a). Suppose that the forecasted carbon cost in time slots 1, 2, and 3  is $c_1=10$ ("low"), $c_2=100$ ("high"), and $c_3=20$ ("medium"),  respectively.  First, assume that the job has ideal scaling behavior and a flat marginal capacity curve of $MC_1=1$ and $MC_2 =1$. The algorithm simply allocates two servers to the job in slot 1, since it has the lowest carbon cost and the highest marginal capacity per unit carbon. As shown in Figure~\ref{fig:algoExample}(b), such a job runs with two servers and terminates at the end of slot 1.  

Next, assume a job with a diminishing marginal capacity curve, given by $MC_1=1$ and $MC_2 =0.7$. Figure~\ref{fig:algoExample}(d) shows the marginal capacity per unit cost table ($MC_j/c_i$) for all three slots. The greedy algorithm allocates the first server to slot 1, since it has the highest marginal capacity per unit cost of 0.1.  In the next iteration, the greedy algorithm allocates a second server to slot 1 as it still has the highest marginal capacity per unit cost ($MC_2/c_1 = 0.07$).  Although two servers have been allocated, the total work done by these two servers is only 1.7 ($MC_1+MC_2$), which cannot complete the job of length 2 ($W=2$).  The algorithm then allocates another server to slot 3, which has the next highest marginal capacity per unit cost ($MC_1/c_3 = 0.05$). This yields a schedule where the job is given 2 servers in slot 1, zero in slot 2, and one server in slot 3.  The job only runs for one-third of slot 3 before it completes. The example also illustrates a tradeoff where \systemName reduces the emissions compared to carbon-agnostic execution (from 110 to 40 carbon units) but increases cloud costs by 15\%  due to the need for a third server. The  tradeoff between carbon saving and cost overheads is fundamental to carbon-aware computing, as demonstrated by prior work~\cite{warofeff}.

\noindent{\bf Periodic Schedule Recomputation.} Once the algorithm computes an initial schedule, \systemName can begin execution of the job by auto-scaling it up or down, or suspending it, in each time slot as per the schedule. \systemName continuously monitors the work done (``job progress'') and the emissions of the job over the course of its execution.  Recall that the initial schedule is computed based on a \emph{forecasted} carbon cost and an \emph{estimated} marginal capacity curve derived from profiling, both of which may have errors in their estimates.  Similar to weather forecasts, carbon forecasts can have errors, especially over the period of multiple days~\cite{dayahead_estimations, carboncast}.  Similarly, the marginal capacity curves may not be exact since production environments may differ somewhat from the profiling environment~\cite{network_aware_scheduling, optimus, performance_modeling}.   These errors can cause deviations in the expected work done or the expected carbon emissions  as estimated by the initial schedule. 

To be robust to carbon prediction or profile estimation errors, \systemName compares the expected work and carbon emissions to the estimates in the schedule at the end of each time interval.  If the deviations exceed a threshold, it recomputes the schedule for the remainder of the job's execution from the current time $t'$ to the completion time $T$. When doing so, \systemName can use an updated carbon forecast if available, since such forecasts are often updated every few hours, similar to weather forecasts. 
Thus, if the progress deviates from the plan (e.g., due to profiling errors), \systemName will recompute the schedule to ensure the highest carbon savings. Since some batch jobs can execute for days~\cite{Tirmazi2020BorgTN}, such schedule adjustments provide robustness to prediction errors and ensure timely job completion while minimizing carbon emissions. 


\noindent{\bf Run Time Complexity.} In Algorithm~\ref{alg:proposed_algorithm}, the time complexity of computing the marginal capacity per unit carbon (steps: 3-5) is $\mathcal{O}(n . M)$, list sorting is $\mathcal{O}(nM \log nM)$, and computing the schedule is $O(nM)$ (steps: 8-11).  The total complexity is $\mathcal{O}(nM + nM\log nM) \approx \mathcal{O}(nM\log nM)$.

\section{CarbonScaler Implementation}
\label{sec:carbon-scaler-system}

This section describes \systemName's implementation, which optimizes the carbon emissions of distributed batch cloud workloads. Our system comprises three main components: 
(1) \profilerName, which uses offline profiling to estimate marginal capacity ($MC$) curves and  energy usage of jobs, 
(2) \impName is our cloud-based carbon scaling system implemented in Kubernetes~\cite{kubernetes}, and 
(3) \simName, which simulates the execution of the jobs to estimate carbon reduction under different  deployment configurations.
\systemName is implemented in Go using $\sim$2.5$k$ SLOC. The code is available at \url{https://github.com/umassos/CarbonScaler}.

\begin{figure}[t]
	\centering
	\includegraphics[width=0.7\textwidth]{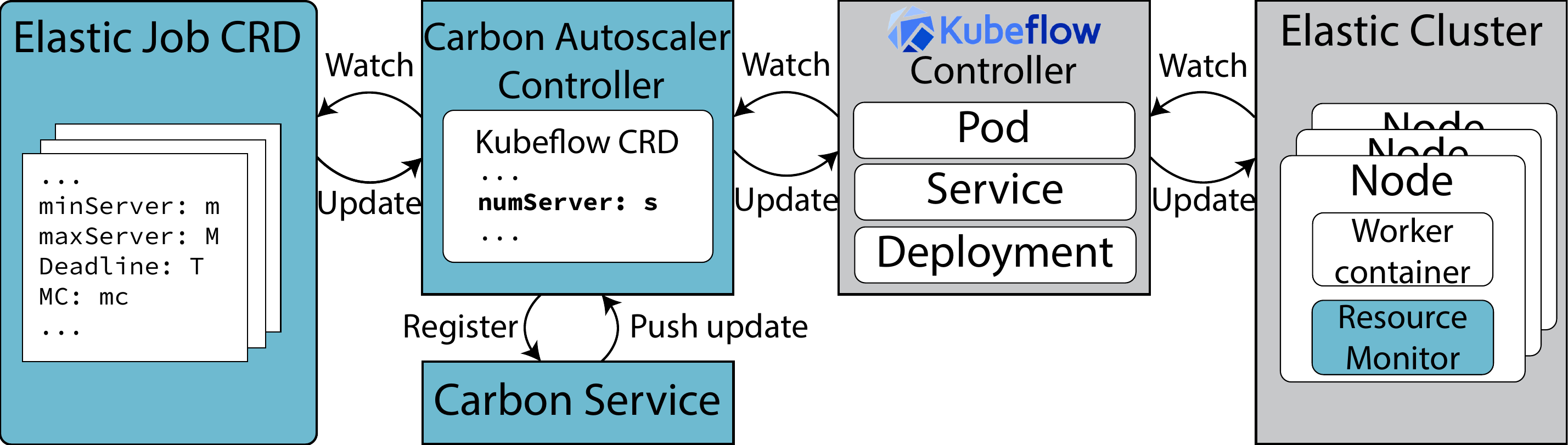}
	\vspace{-0.3cm}
	\caption{\emph{An overview of Carbon AutoScaler.}}
	\vspace{0.1cm}
	\label{fig:systemOverview}
\end{figure}

\subsection{Carbon Profiler}
\label{sec:profiler}
\systemName requires the marginal capacity curve of a job for carbon-aware scaling. \profilerName performs a one-time offline profiling of a new job to derive its marginal capacity curve. To do so, it runs the job with server allocations ranging from the job-specified minimum number of servers, $m$, to the maximum number of servers, $M$, and records the work done at each allocation.
To minimize the profiling overhead, \profilerName runs the job for a small, configurable amount of time $\alpha$ (up to a few minutes) and varies the resource allocation with a granularity $\beta$, which depends on $M$. If $\beta > 1$, \profilerName interpolates the recorded measurements to obtain a complete marginal capacity curve.  
Finally, the marginal capacity curves are valid for a computing environment identical to the profiling environment. The scaling behavior and the expected savings may change if the environment is significantly different, necessitating environment-specific profiling or an online update of the capacity curves.  
\systemName also allows substituting \profilerName with alternative workload profiling approaches from prior work~\cite{paleo,Oyama2016,Pei2019,justus2018,cai2017neuralpower,optimus, performance_modeling}. 


\subsection{Carbon AutoScaler}
Figure~\ref{fig:systemOverview} shows an overview of \impName that uses Kubeflow~\cite{kubeflow} to implement our \algoName from \S\ref{sec:algorithm}. 
The incoming elastic batch applications use Kubernetes' Custom Resource Definition (CRD), written in \texttt{.yaml} format. \impName follows Kubernetes standards in defining its user-facing interface. In this case, the user extends the normal job specification by adding extra \impName-specific maps that provide scaling and scheduling information,  including minimum $m$ and maximum $M$ number of servers, completion time $T$, and an estimated job length $l$. The user also specifies methods for obtaining the marginal capacity curve, where the current default is profiling. The user then submits the jobs to \impName using standard using Kubernetes APIs such as \texttt{kubectl}.

We implement \impName as a controller that sits on top of the Kubeflow training operator and leverages its core resource management functionality for clustered batch jobs, such as ML training and MPI. 
\impName first runs the \algoName to compute the initial schedule for each job. To do so, \impName tracks carbon intensity using a dedicated service that provides the instantaneous and forecasted carbon intensity. Then, \impName informs the Kubeflow training operator to execute the schedule by modifying the Kubeflow job specification to scale the resources allocated to the job, such as the number of replicas.  \impName is also in charge of maintaining the job status of the Kubeflow operator.   
\impName implements resource-level and application-level monitoring. 
\impName implements additional Kubernetes services to monitor resource usage, energy usage, and carbon usage over time. We track CPU usage using Kubernetes Metrics Server~\cite{metrics-server}, CPU energy usage using Running Average Power Limiting (RAPL)~\cite{david2010rapl} interfaces and PowerAPI~\cite{bourdon2013powerapi}, and GPU energy usage using NVIDIA Data Center GPU Manager (DCGM)~\cite{nvidia-dcgm}. The resource and power monitoring can include other resources such as storage and network. 
\impName monitors application-level metrics such as progress and throughput via application-level interfaces. 

Finally, \impName registers a reconcile callback function, which is called when the carbon intensity changes and when applications report their progress. This enables \impName to detect divergence in progress, throughput, or carbon intensity. \systemName then recomputes the schedule as explained in \S\ref{sec:algorithm}.

\subsection{Carbon Advisor}\label{sec:advisor}
\simName enables pre-deployment analysis of the carbon scaling algorithm in an environment that simulates the operation of \impName. \simName takes, as input, a carbon trace, job start time, deadline, job length, and \systemName-specific parameters, such as range of server allocations $[m, M]$ and marginal capacity curve. 
The fidelity of \simName depends on the accuracy of the marginal capacity profile for the application. In Section~\ref{sec:evaluation}, we demonstrate the high fidelity of \simName in estimating the carbon savings from different carbon-aware scaling policies. The \simName simulates the running of the job and reports savings for carbon-aware scaling policies. Additionally, the \simName enables simulating various kinds of errors to ensure the robustness of the predictions, as described in Section~\ref{sec:assumptions}.
The simple plug-and-play nature of the tool allows application developers to perform what-if scenarios and explore a wide range of parameters before actual deployment. For example, users can explore the benefits of extending their waiting time and its impact on carbon savings. \simName also enables key high-level analysis by default, such as computing the distribution of carbon savings across different start times of the year.  Finally, to facilitate initial exploration, we plan to provide carbon traces and marginal capacity curves used in the paper alongside the tool.

\section{Experimental Evaluation}
\label{sec:evaluation}

This section  evaluates the performance of \systemName using our prototype implementation, described in Section~\ref{sec:carbon-scaler-system}. We augment the prototype evaluation results with additional large-scale analysis that leverages \simName. 

\vspace{-0.25cm}
\subsection{Experimental Setup}
\label{sec:setup}
\vspace{-0.05cm}
\noindent\textbf{Workload.} 
Table~\ref{tab:workloads} describes the elastic workloads we use for evaluating \systemName and their specifications. The workloads span both CPU- and GPU-intensive applications such as the $N$-body problem~\cite{nbodysimulation} implemented using MPI~\cite{mpi} and machine learning models, including ResNet~\cite{resnet}, EfficientNet~\cite{efficient-net}, and VGG~\cite{efficient-net} implemented using Pytorch~\cite{pytorch}. The table shows the base configurations and power measurements for jobs that need 24hrs to finish. The chosen workloads have a wide-range of scaling characteristics (shown in Figure~\ref{fig:workloadScale}), configurations, and energy requirements.

\noindent\textbf{Infrastructure.}
We deployed \systemName in two different settings to demonstrate its adaptability to the underlying infrastructure. For CPU-intensive workloads, we used a local computing cluster consisting of 8 servers, each equipped with a 16-core Xeon CPU E5-2620, connected through a 10G network. For GPU-intensive workloads, we deployed \systemName on Amazon Web Services (AWS) using 8 \texttt{p2.xlarge} instances, each equipped with NVIDIA K80 GPU. 

\begin{table}[t]

\resizebox{0.7\columnwidth}{!}{%
\footnotesize
\begin{tabular}{||c|c|c|c|c||}
\hline \hline
\textbf{Name} & \textbf{Implementation} & \textbf{Epochs} & \textbf{BatchSize} & \textbf{Power (W)} \\ \hline
$N$-Body Simulation (10,000)  & MPI     & 138000 & NA  & CPU~(60)  \\ \hline
$N$-Body Simulation (100,000) & MPI     & 1500   & NA  & CPU~(60)  \\ \hline
Resnet18 (Tiny ImageNet)   & Pytorch & 173    & 256 & CPU+GPU~(210)   \\ \hline
EfficientNetB1 (ImageNet)        & Pytorch & 45     & 96  & CPU+GPU~(210) \\ \hline
VGG16 (ImageNet)           & Pytorch & 31     & 96  & CPU+GPU~(210) \\ \hline \hline
\end{tabular}%
}
\caption{\emph{Details of elastic workloads in evaluation. Epochs represent the number of epochs needed for a 24hr job.}}
\label{tab:workloads}
\end{table}

\begin{figure}[t]
    \centering
    \includegraphics[trim={14.5cm 7cm 14.5cm 7cm},clip, width=0.6\linewidth]{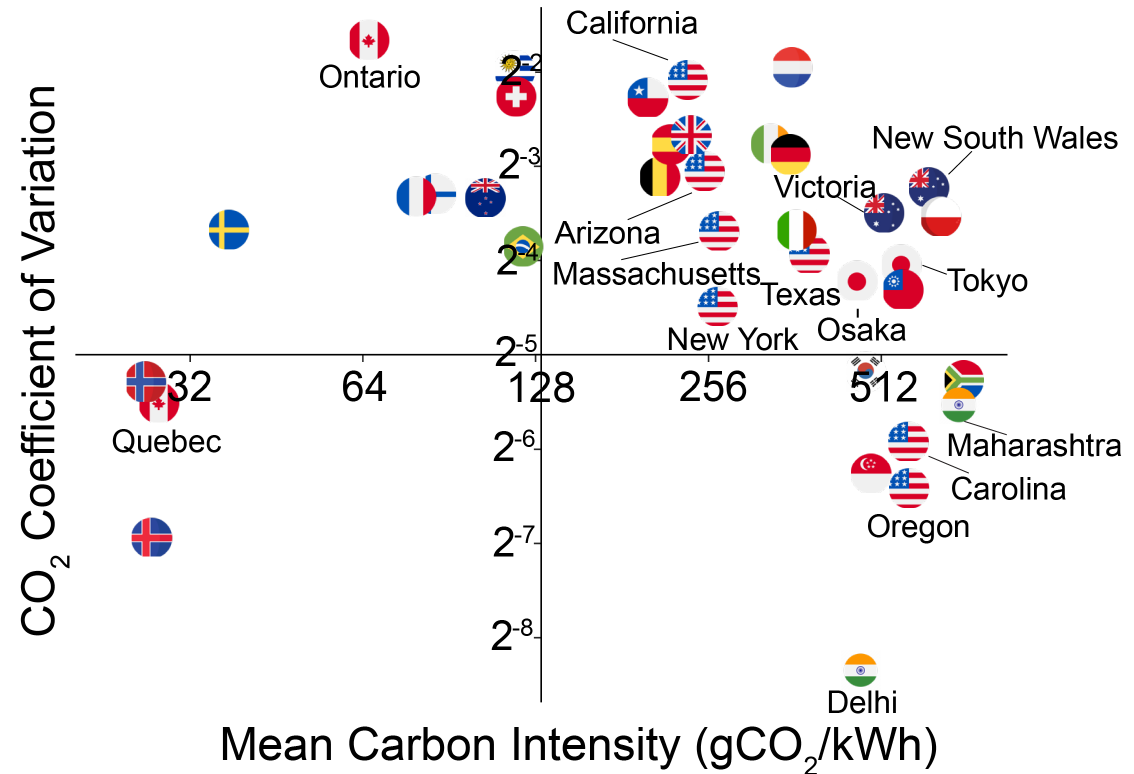}
    \vspace{-0.3cm}
    \caption{\emph{Most cloud regions globally have a high carbon cost, but also show significant daily variations, providing an opportunity for CarbonScaler to optimize carbon emissions.}}
    \label{fig:intensity_variability}
\end{figure}

\noindent\textbf{Carbon Traces.} 
We collected carbon traces for different geographical locations using electricityMap~\cite{electricity-map}, an online service that provides real-time and archival carbon intensity information. We use average carbon intensity values, measured in grams of carbon dioxide equivalent per kilowatt-hour (\unit{\gram CO2eq\per\kilo\watt\hour}), provided at hourly granularity. The collected carbon traces span from January 2020 to December 2022, we specify the duration for each trace where it is used. 

To choose representative regions for our evaluation, we analyzed the average carbon intensity and the coefficient of variation (computed as standard deviation over mean) for different AWS regions. 
Figure~\ref{fig:intensity_variability} shows the results for 37 regions. 
Most regions have high carbon intensity but also show high daily variations, while some have low carbon intensity with similarly high daily variations. Since \suspendPolicy and \systemName rely on these high variations to reduce emissions, the figure indicates that both techniques will be effective in the majority of low-carbon as well as high-carbon cloud regions. A few cloud regions have stable carbon costs (i.e., low variations), including low carbon regions such as Iceland and Sweden, and high carbon regions such as India and Singapore. 
The effectiveness of \suspendPolicy and \systemName is diminished in such cloud regions as changing the execution time and scale does not alter the carbon intensity. Still, such regions are a small minority of the total cloud regions in a global cloud platform such as AWS.
Based on this analysis, we choose Netherlands (\worldflag[width=2mm]{NL}) as a representative high carbon region and Ontario, Canada (\worldflag[width=2mm]{CA}) as an example of a low carbon region for our subsequent experiments. Nonetheless, we evaluate the potential savings across regions in Section~\ref{sec:trace-character}.

\noindent{\bf Baselines Policies.} 
We evaluate the performance of \systemName against three baseline policies: \agnosticPolicy, \suspendPolicy, and \staticPolicy.  The \agnosticPolicy is a simple policy that runs a job without considering carbon emissions and represents the status quo.  The \suspendPolicy policy is inspired by prior work~\cite{Wiesner2021LetsWA,cloudcarbon}. 
As mentioned in \S\ref{sec:cloud-scheduling}, \suspendPolicy can be implemented in two ways: threshold-based, which uses a carbon threshold to suspend-resume a job in a deadline-unaware manner, and deadline-based, which chooses the $k$ lowest carbon periods before the specified deadline for execution. In this case, \suspendPolicy defaults to \agnosticPolicy policy when the completion time equals the job length ($T=l$), i.e., no slack,  since execution cannot be deferred. This policy acts as a baseline for temporal shifting scenarios where we assume a job has a completion time higher than the job length ($T>l$). Finally, \staticPolicy is another policy inspired by prior work~\cite{ecovisor}, where an application picks the lowest carbon intensity points and runs with a certain \emph{static} scale factor to utilize the carbon intensity variations better. This is our default baseline for scenarios where we evaluate \systemName for its ability to leverage workload elasticity and scaling. Unless stated otherwise, we report the mean across 15 runs for our cloud experiments and 100 runs for \simName's simulated executions. 

\noindent
\textbf{Carbon Advisor Fidelity.} To demonstrate the fidelity of the simulator, we compare the carbon savings estimates from \simName with the results from various real experiments in the evaluation. \simName estimates have an average error of less than 5\%, demonstrating the high accuracy of our simulation results based on \simName.


\begin{figure}[t]
    \centering
    \includegraphics[width=0.54\textwidth]{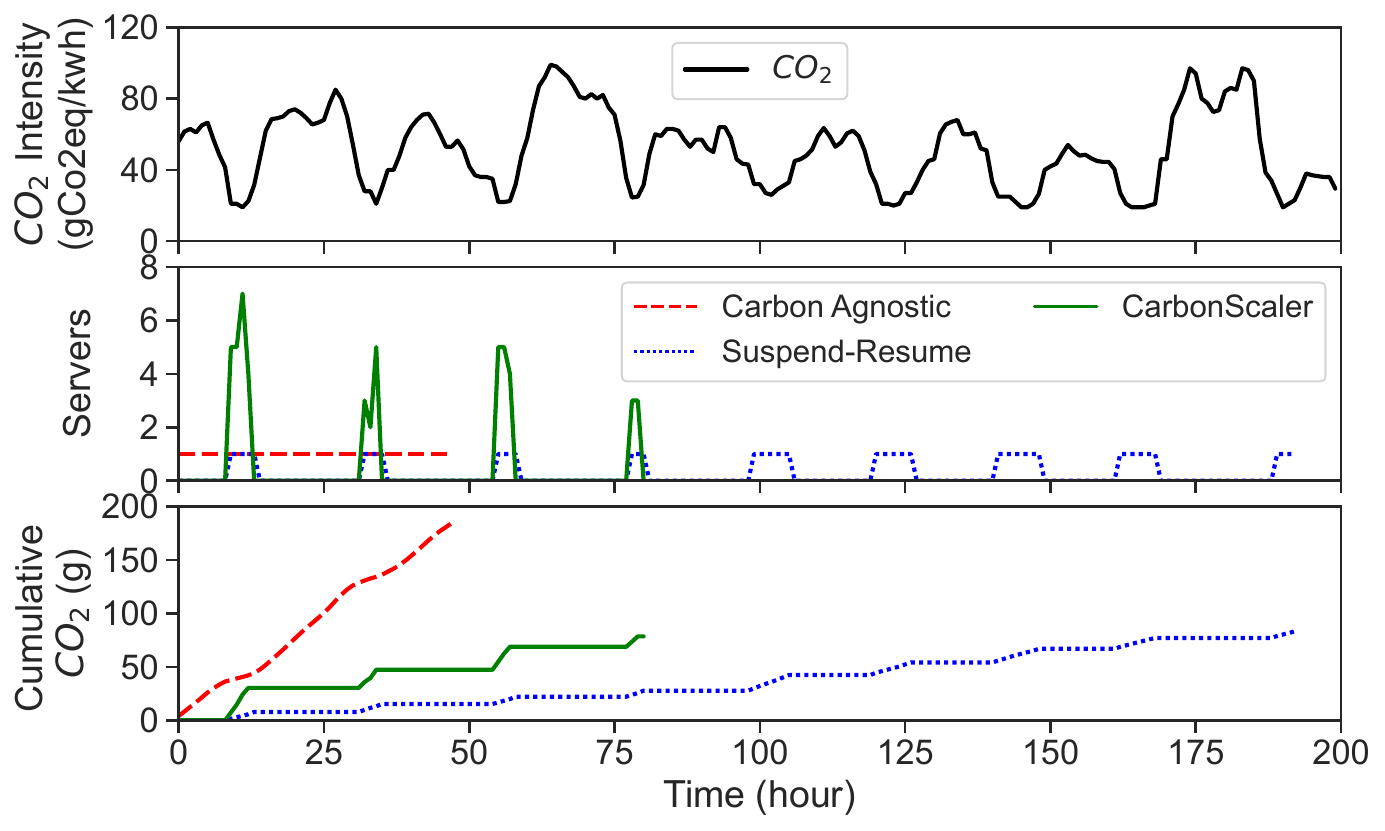}
    \vspace{-0.45cm}
    \caption{\emph{CarbonScaler in action for a 48hrs long $N$-body MPI job ($N$=100k), where $T=2\times l$.}}
    \vspace{-0.0cm}
    \label{fig:illustrative_example}
\end{figure}

\vspace{-0.2cm}
\subsection{CarbonScaler in Action}
\vspace{-0.1cm}
To show \systemName in action, we ran a 48hr $N$-body MPI job on our CPU cluster and compared its execution to the threshold-based \suspendPolicy (deadline-unaware) and \agnosticPolicy policies.  As shown in Figure~\ref{fig:illustrative_example},
the \agnosticPolicy policy starts the job as soon as it arrives and finishes in 48hrs at the cost of 184g of $CO_2$ emissions. The \suspendPolicy policy \emph{suspends} the job during high carbon intensity periods and waits for the carbon intensity to fall below a threshold (25$^{th}$ percentile in this case) to \emph{resume} the job. 
By leveraging temporal flexibility, \suspendPolicy saved 45\% carbon compared to the \agnosticPolicy policy but increased the job completion time by 4$\times$. 
Finally, we set the desired completion time $T$ to be 96hrs, i.e., $T=2\times l$, and execute our proposed \systemName policy.  \systemName scales the number of servers depending on the application's scaling properties and the carbon cost at a given time.
As a result, \systemName achieves 42\% carbon saving over \agnosticPolicy policy.  \systemName achieves comparable savings with \suspendPolicy while also reducing the job completion time to 2$\times$ of \agnosticPolicy policy.




\vspace{-0.3cm}
\subsection{Impact of Workload Elasticity}\label{sec:eval_elasticity}
\vspace{-0.1cm}
The two key aspects that impact carbon savings from \systemName are temporal flexibility and workload elasticity.  While prior work necessitates temporal flexibility for carbon savings, \systemName can achieve significant savings by leveraging workload elasticity even when no temporal flexibility is available. The extent of savings depends on the scalability properties of the workload: a highly scalable job (with flat or close to flat marginal capacity) can achieve higher savings, as illustrated for the simple workload in Figure~\ref{fig:algoExample}. To demonstrate the elasticity effect, we limit the job completion time to the job length, i.e., $T=l$, which means no temporal flexibility is available. We run 24hrs long jobs for various applications in Table~\ref{tab:workloads} using \agnosticPolicy policy, \staticPolicy (2$\times$), and \systemName. 

\begin{figure}[t]
    \centering
    \begin{tabular}{ccc}
    \includegraphics[width=0.48\textwidth]{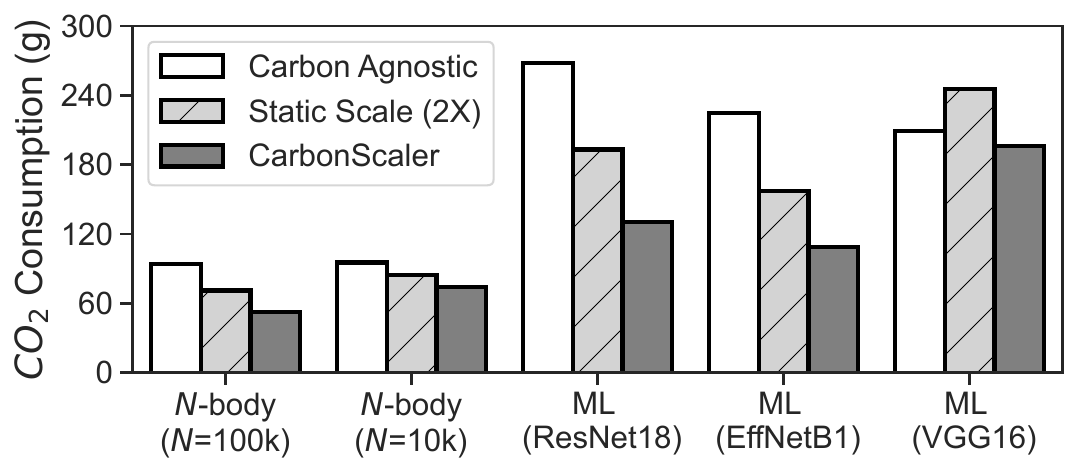} &
    \includegraphics[width=0.46\textwidth]{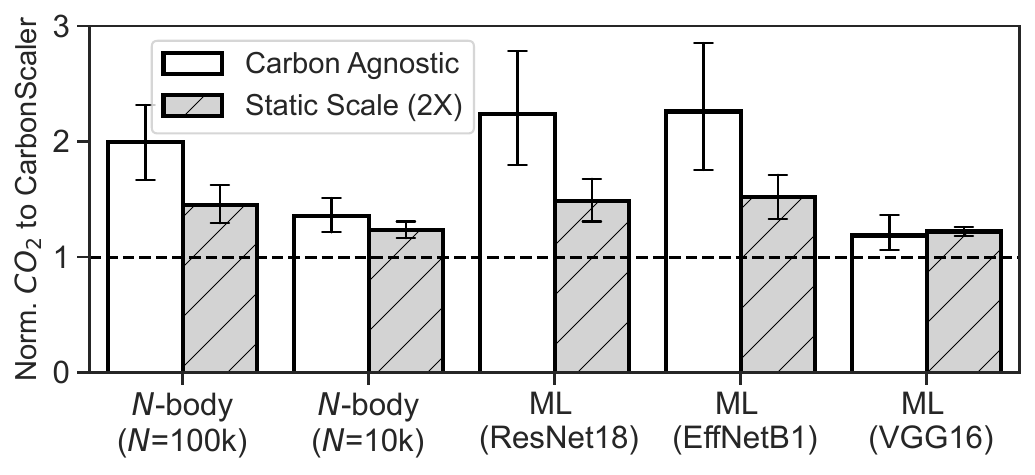} \vspace{-0.1cm} \\
    (a) Carbon footprint & (b) Performance w.r.t \systemName \\
    \end{tabular}
    \vspace{-0.45cm}
    \caption{\emph{Carbon footprint and performance of different workloads scheduled under carbon-agnostic, static-scale (2$\times$), and CarbonScaler, in Ontario, Canada, where $T=l$ (i.e., no slack and on-time completion).}}
    \label{fig:scalabilty-effect}
\end{figure}

\begin{figure}[t]
    \centering
    \begin{tabular}{ccc}
    \includegraphics[width=0.32\textwidth]{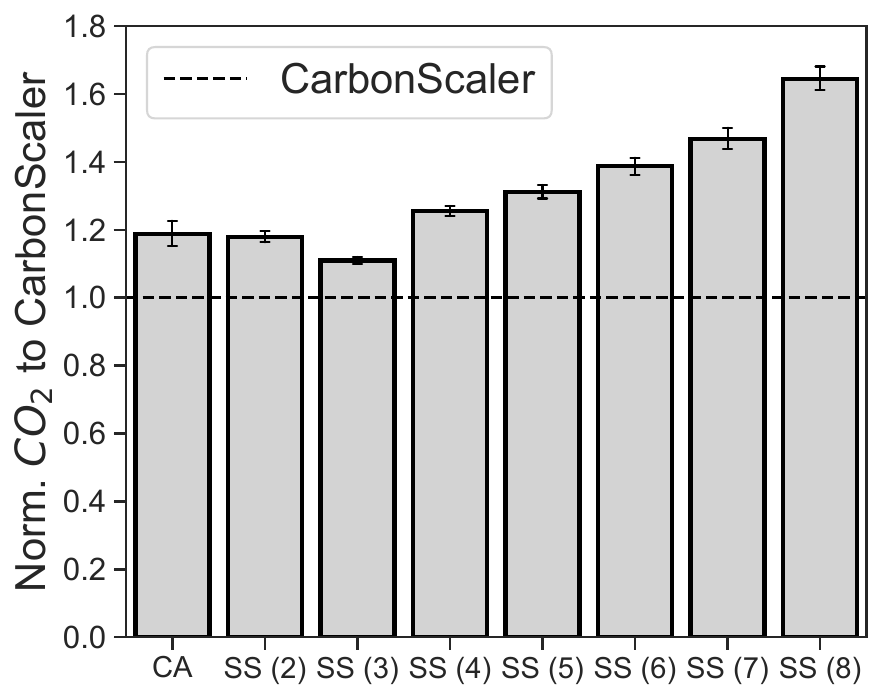} &
    \includegraphics[width=0.32\textwidth]{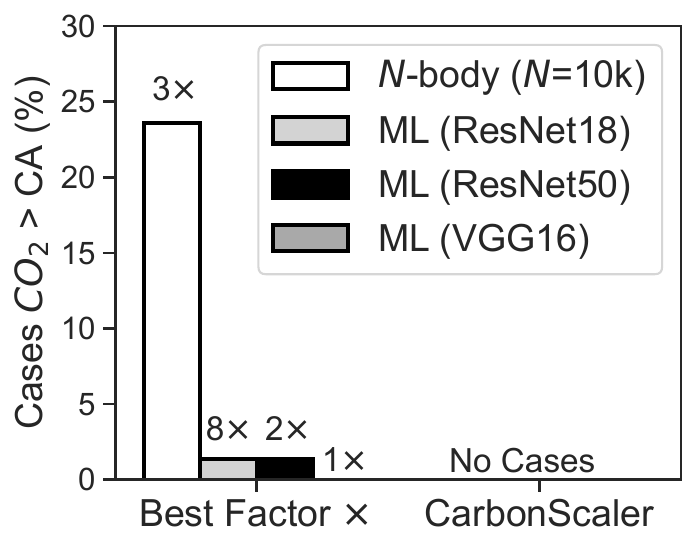} &
    \includegraphics[width=0.3\textwidth]{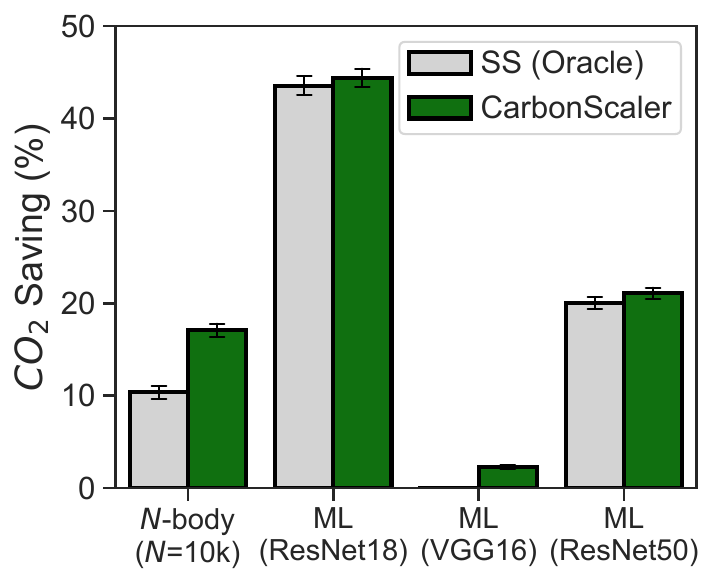}
    \vspace{-0.1cm} \\
    (a) & (b) &  (c)
    \\
    \end{tabular}
    \vspace{-0.4cm}
    \caption{\emph{\systemName vs. the best static scale (SS) factor in Ontario, Canada. Carbon emissions of various static scale factors compared to \systemName (a), percentage of start times when a policy consumes more carbon than \agnosticPolicy (b), and static scale oracle against \systemName for multiple applications (c).}}
    \label{fig:static_scaling_ca_on}
\end{figure}

\begin{figure}[t]
    \centering
    \begin{tabular}{ccc}
    \includegraphics[width=0.31\textwidth]{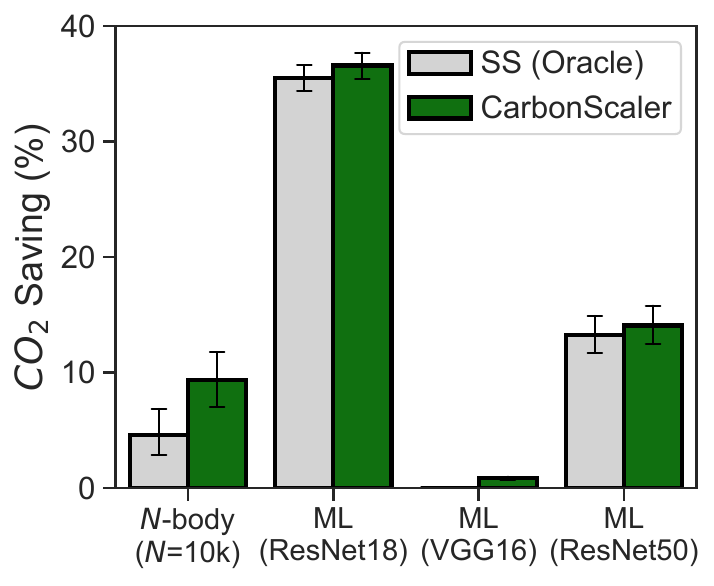} &
    \includegraphics[width=0.31\textwidth]{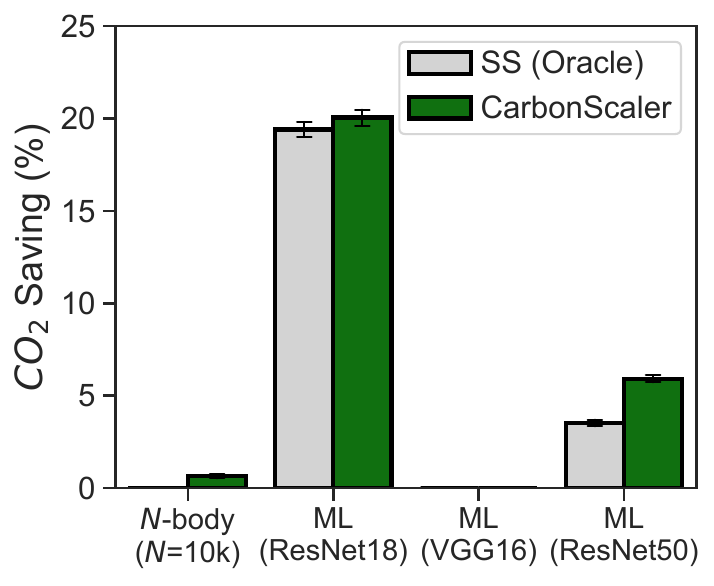} &
    \includegraphics[width=0.32\textwidth]{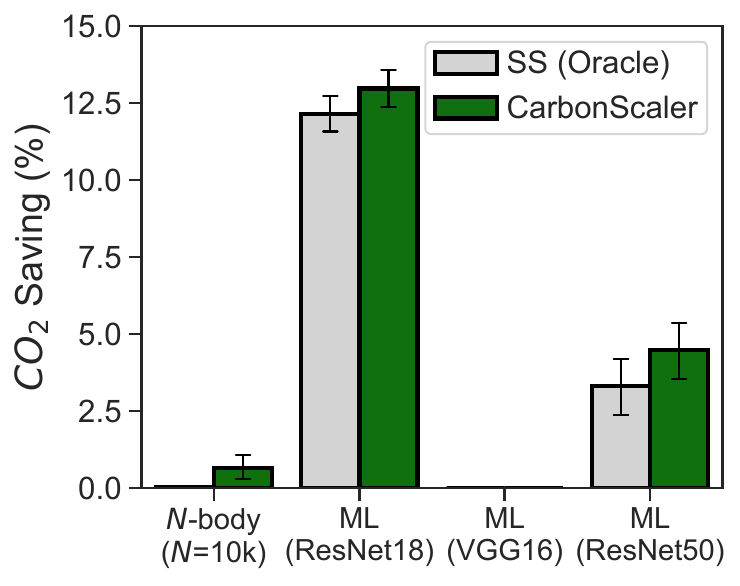}
    \vspace{-0.1cm} \\
    (a) Netherlands & (b) California, US &  (c) Texas, US
    \\
    \end{tabular}
    \vspace{-0.4cm}
    \caption{\emph{Comparing \systemName with static scale oracle in multiple regions.}}
    \vspace{0.1cm}
    \label{fig:cs_ss_regions}
\end{figure}

Figure~\ref{fig:scalabilty-effect} shows the performance of the three policies for different workloads. Figure~\ref{fig:scalabilty-effect}(a) compares the absolute carbon footprint of the three policies and shows that the \systemName highest savings are for highly scalable workloads.
For example, for $N$-body ($N$=100k) and ML (ResNet18), \systemName saves up to 140 and 63 (\unit{\gram CO2eq}) compared to \agnosticPolicy and \staticPolicy (2$\times$), respectively. 
To demonstrate the superiority of \systemName, independent of the task and start-time dependent carbon consumption, we compare the normalized carbon savings of different policies to \systemName.
Figure~\ref{fig:scalabilty-effect}(b) compares the performance of all policies to \systemName, where the whiskers represent the 95$^{th}$ percentile confidence interval and the horizontal line represents the performance of \systemName. The figure shows that, aside from the saving, workloads, and start times, \systemName demonstrates the ability to outperform all other policies. In particular, \systemName uses 33\% and 20\% less carbon than \agnosticPolicy and \staticPolicy (2$\times$), respectively. 
The figure also shows that, since the \staticPolicy does not consider the job's scalability properties, it can instead \emph{increase} the carbon consumption for some workloads by as much as 20\% by scaling the job beyond a single \emph{optimal} scale factor. 
On the other hand, \systemName is cognizant of scaling behavior and picks a different scale at each time slot that has the highest work done per unit carbon cost, yielding minimum carbon consumption.

To further demonstrate \systemName benefits over the best static scale factor, we use \simName to compare \systemName against oracle-based static scale factors. 
Figure \ref{fig:static_scaling_ca_on}(a) shows the performance of all scale factors and \systemName for $N$-body ($N$=10k). The static scaling consumes 17-65\% more carbon than \systemName. While the \staticPolicy policy can reduce carbon emissions compared to \agnosticPolicy for scale factors 2 and 3, it can consume more carbon at higher scale factors due to the non-linear scalability of the workloads. 
The potential increase in carbon consumption is not only true for an arbitrary non-optimal scale factor; even the best scale factor for each start time can consume more carbon than \agnosticPolicy. Figure~\ref{fig:static_scaling_ca_on}(b) shows the probability that the best scale factor (on top of each bar) yields a higher consumption than the \agnosticPolicy operation. As shown, certain instances always exist where this best scale factor performs worse than \agnosticPolicy. Perhaps the only exception is ML (VGG16), a non-scalable application, where the best scale factor is the \agnosticPolicy (1$\times$).

As opposed to \systemName, the best static scale factor may not be optimal for all the time slots during the execution of a job, resulting in higher carbon emissions. In Figure \ref{fig:static_scaling_ca_on}(c), we show the additional savings from adapting the scale factor during the execution of a job for multiple applications. 
As demonstrated, \systemName outperforms the static scale oracle by 1.2\% to 8\%, depending on the job's scalability characteristics.
Figure \ref{fig:cs_ss_regions} extends the evaluation of \ref{fig:static_scaling_ca_on}(c) and shows how \systemName outperforms the oracle \staticPolicy in different regions, even when carbon savings are limited.
However, it is worth noting that static state oracle is the artifact of our implementation.  Neither application performance models, which are inherently carbon-oblivious, nor state-of-the-art carbon-aware techniques, such as Ecovisor~\cite{ecovisor} or Wait Awhile\cite{Wiesner2021LetsWA}, can realize this optimal oracle approach.

\noindent\emph{\textbf{Key Takeaway.} CarbonScaler better leverages the workload elasticity by choosing dynamic scale factors depending on the job scalability characteristics and carbon intensity for each start time for the job and each time slot during a job's execution.}

\begin{figure}[t]
    \centering
    \begin{tabular}{cc}
    \includegraphics[width=0.49\textwidth]{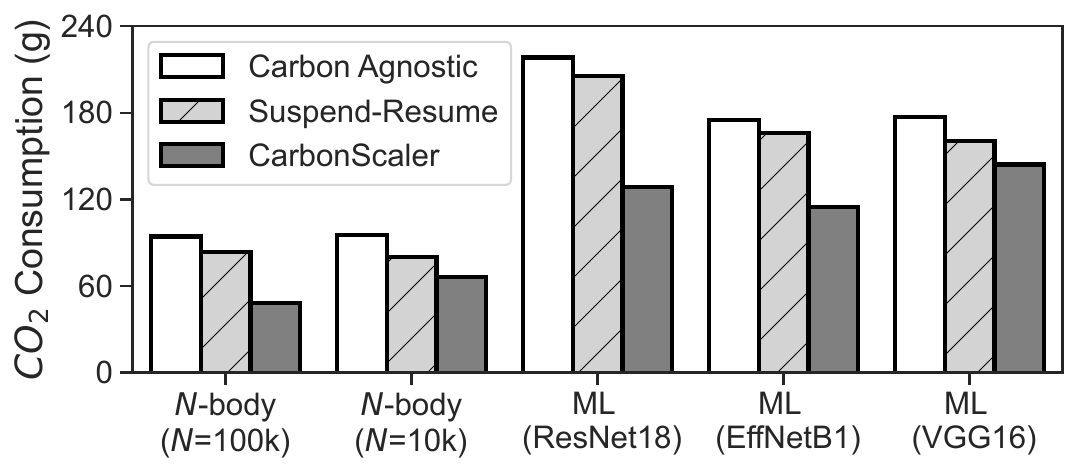} &
    \includegraphics[width=0.47\textwidth]{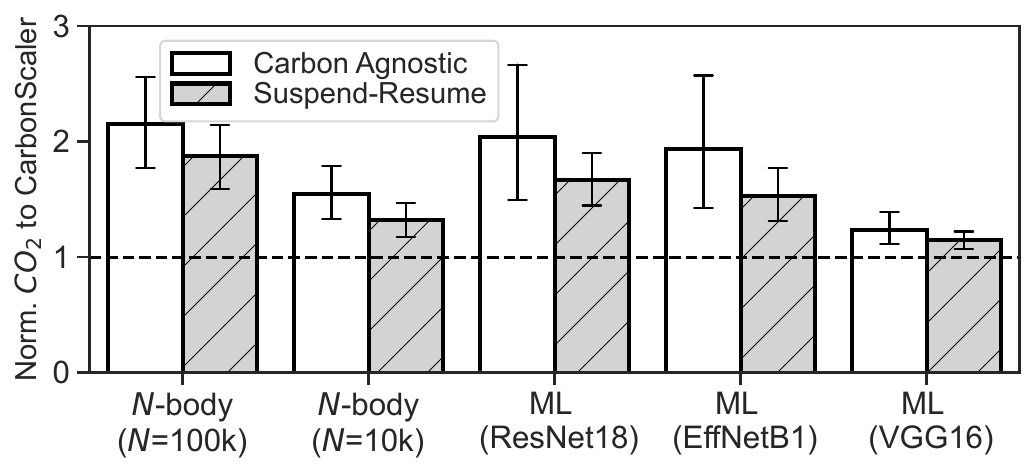} \vspace{-0.1cm}
    \\
     \multicolumn{2}{c}{(a) Ontario, Canada} \vspace{-0.00cm}\\
    \includegraphics[width=0.49\textwidth]{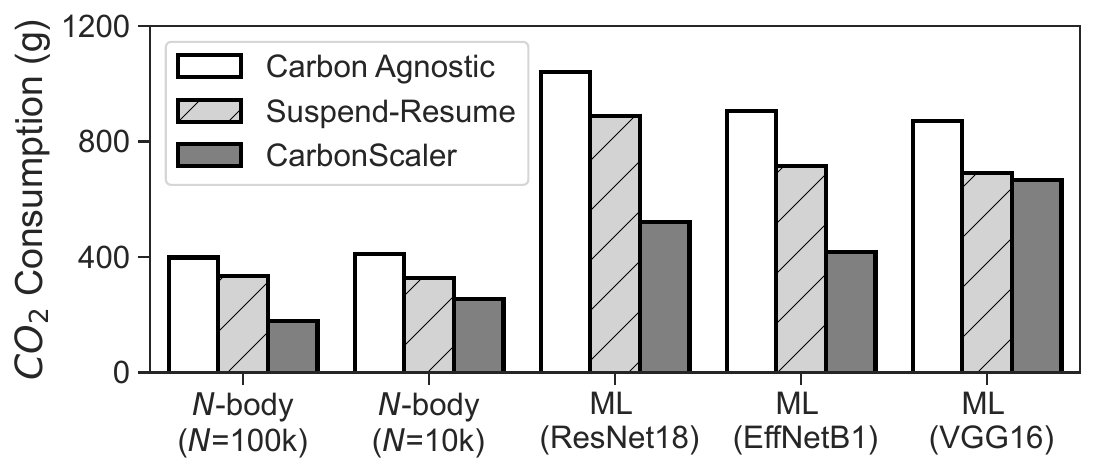}
     & \includegraphics[width=0.47\textwidth]{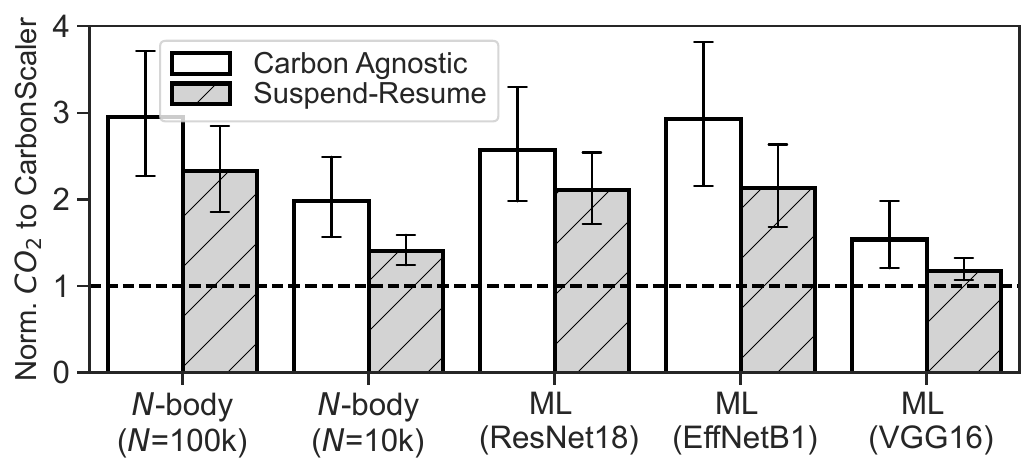}
     \vspace{-0.1cm}
    \\
     \multicolumn{2}{c}{(b) Netherlands} \vspace{-0.2cm} \\
    \end{tabular}
    \vspace{-0.15cm}
    \caption{\emph{Carbon footprint and normalized performance of different workloads and policies, where $T=1.5\times l$.}}
    \vspace{0.1cm}
    \label{fig:time_shifting_apps_locations}
\end{figure}

\vspace{-0.3cm}
\subsection{Impact of Temporal Flexibility}
\vspace{-0.05cm}
In addition to workload elasticity, temporal flexibility can be an important source of carbon savings for delay-tolerant jobs. 
We evaluate the impact of temporal flexibility by running workloads from Table~\ref{tab:workloads} using \agnosticPolicy policy, \suspendPolicy policy, and \systemName with extended completion times where $T>l$.
To ensure that the \suspendPolicy respects the job-specified completion time, we use the deadline-aware version of the \suspendPolicy policy~\cite{Wiesner2021LetsWA}.
Figure~\ref{fig:time_shifting_apps_locations} shows the carbon consumption (left) and performance of different policies (right) when \emph{running} the workloads with 24 hrs length $l$, and 36 hrs as completion time $T$, $T=1.5\times l$, across two locations.
\systemName is better at exploiting the temporal flexibility and outperforms the \suspendPolicy policy for all workloads. As shown, \systemName is able to save 36\% and 22\% compared to \agnosticPolicy and \suspendPolicy in Ontario, Canada, and 51\% and 37\% in the Netherlands for the highly scalable ML (ResNet18). On the other hand, for less scalable workloads such as ML(VGG16), most of the carbon savings of \systemName stem from time-shifting, yielding comparable savings to \suspendPolicy. The right column of the figure also demonstrates the superiority of \systemName aside from the carbon savings, which is start-time and task dependent.

\begin{figure*}[t]
\centering
\begin{minipage}[t]{0.48\linewidth}
  \centering
   \includegraphics[width=\textwidth]{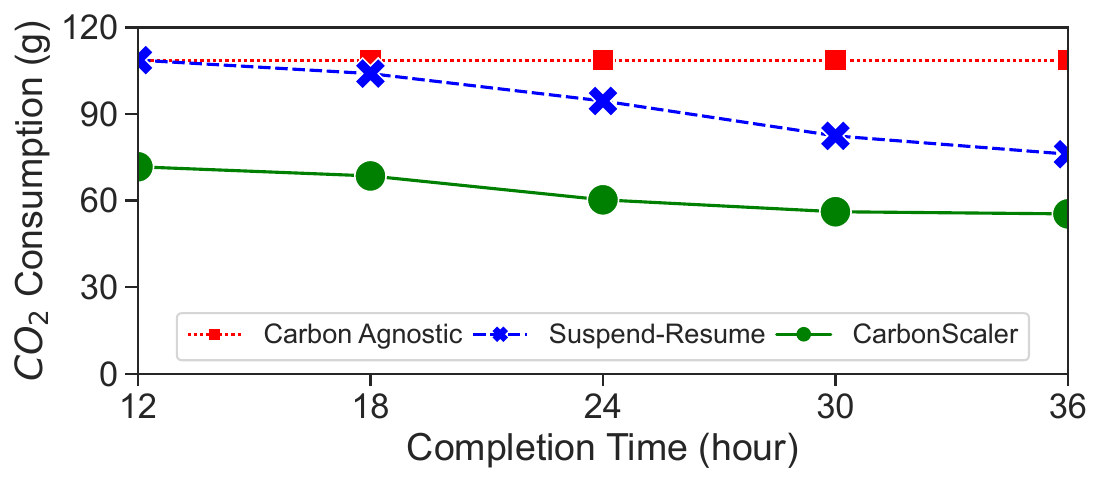}
    \vspace{-0.8cm}
    \caption{\emph{Effect of completion time on the carbon footprint of a 12hrs long ResNet18 job in Ontario, Canada.}}
    \label{fig:job_deadline}
\end{minipage}
\hfill
\begin{minipage}[t]{0.468\linewidth}
   \includegraphics[width=\textwidth]{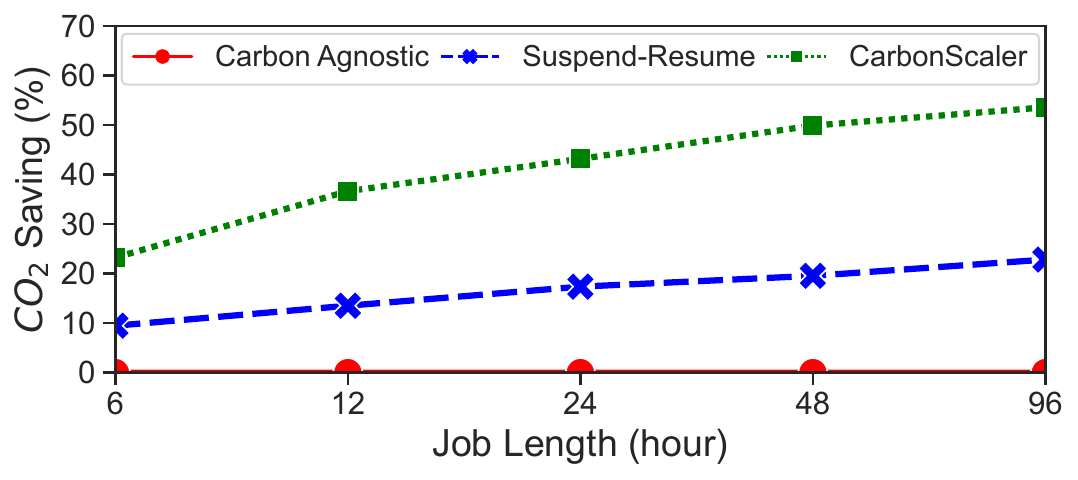}
    \vspace{-0.8cm}
    \caption{\emph{Effect of job length on CO$_2$ savings for an $N$-Body ($N$=100k) job in Ontario, Canada, $T=1.5\times l$.}}
    \label{fig:job_length_n_body}
\end{minipage}
\end{figure*}

\noindent{\bf Effect of Completion Time.} 
Prior results have demonstrated that temporal flexibility can yield significant savings. 
Figure~\ref{fig:job_deadline} evaluates the gain in carbon savings with  increasing temporal flexibility (higher desired completion time $T$). 
We run a 12hrs ML training job (ResNet18) and configure it to complete in 12hrs ($T=l$) up to 36hrs ($T=3\times l$). 
For higher  completion times, more low carbon slots become available, which allows \systemName and \suspendPolicy to reduce the carbon consumption by 30-45\% and 0-32\%, respectively. 
\systemName achieves higher savings by using a higher scale factor during the lowest carbon slots and only picks a higher carbon slot if it gives a better marginal work done per unit carbon. 
For very high completion times, the savings of \systemName over \suspendPolicy diminish, since it begins to prefer job suspensions over high scale factors to avoid the impact of non-linear scaling behavior.

\noindent{\bf Effect of Job Length.}
The length of a job is another key factor in determining carbon savings.  
As the job length increases, more low-carbon slots become available as the grid's carbon intensity generally has a diurnal pattern.
To evaluate the impact of job length, we varied the job length from 6 hours to 96 hours and used our \simName to analyze the estimated carbon savings.
Figure~\ref{fig:job_length_n_body} shows the carbon savings of different policies, against a \agnosticPolicy baseline, for the $N$-body($N$=100k) MPI workload when $T = 1.5 \times l$.
\systemName outperforms \suspendPolicy and \agnosticPolicy over various job lengths. The carbon savings increase with job length 
since there are more low-carbon time slots to choose from, providing opportunities for greater savings.  Overall, \systemName achieves 30\% more savings than \suspendPolicy for long batch jobs.

\begin{figure}[t]
    \centering
    \begin{tabular}{cc}
    \includegraphics[width=0.485\textwidth]{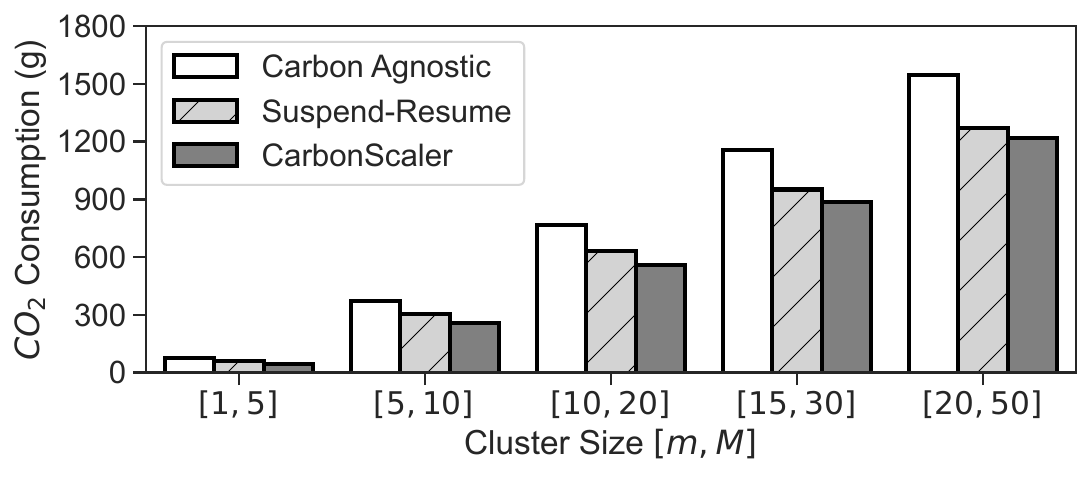} \vspace{-0.1cm} &
    \includegraphics[width=0.465\textwidth]{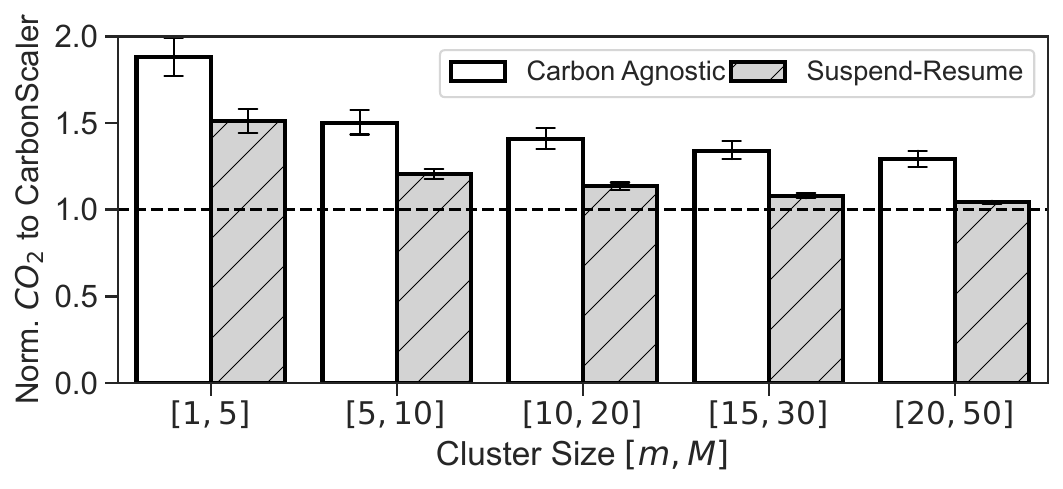}\vspace{-0.1cm} \\
    (a) Carbon Consumption (g)  &
     (b) Performance w.r.t. \systemName (\%) \\
    \end{tabular}
    \vspace{-0.3cm}
    \caption{\emph{Carbon consumption and normalized performance of 24-hour $N$-body ($N$=100k) MPI job with different cluster sizes in Ontario, Canada, where $T=1.5\times l$.}}
    \label{fig:job_largescale}
\end{figure}

\vspace{0.1cm}
\noindent{\bf Effect of Cluster Size.}
Our experiments thus far have used a lower bound of 1 server ($m=1$) and an upper bound of 8 servers ($M=8$) for workloads due to cluster size and cloud cost constraints.  
However, larger batch jobs execute on larger clusters, with larger $m$ and $M$.
For example, certain HPC and ML training applications run on tens or even hundreds of servers in the cloud~\cite{hpcML, bert} and can only be executed on a large number of servers $m\gg 1$. 
To evaluate the efficacy of \systemName for large clusters, we extrapolated the marginal capacity curve for the current $N$-body($N=100k)$ job. Then, we use \simName to estimate how carbon savings change when running progressively bigger jobs on increasing cluster sizes while keeping the job length unchanged at 24hrs.

Figure~\ref{fig:job_largescale}(a) compares the carbon consumption across cluster sizes. The figure shows that, although the savings percentages diminish with larger cluster sizes as they are less dynamic, the absolute carbon savings increase.
Figure~\ref{fig:job_largescale}(b) depicts the relation between policies aside from the size-dependent carbon consumption. As shown, \systemName can obtain 30--42\% additional savings than \agnosticPolicy, and \suspendPolicy achieves the same savings of 17\% over \agnosticPolicy policy across all cluster sizes.  The \suspendPolicy achieves this static saving since it suspends the job in the same high carbon periods regardless of the cluster size. Lastly, the figure shows that the savings difference between \systemName and \suspendPolicy reduces as the cluster size increases since the marginal capacity curve shows diminishing gains for larger cluster sizes.

\vspace{0.05cm}
\noindent \emph{\textbf{Key Takeaway.} CarbonScaler exploits  temporal flexibility to outperform suspend-resume policy across regions with different carbon costs and over different job lengths, completion times, and cluster sizes.} 

\begin{figure}[t]
    \centering
    \begin{tabular}{ccc}
    \includegraphics[width=0.31\textwidth]{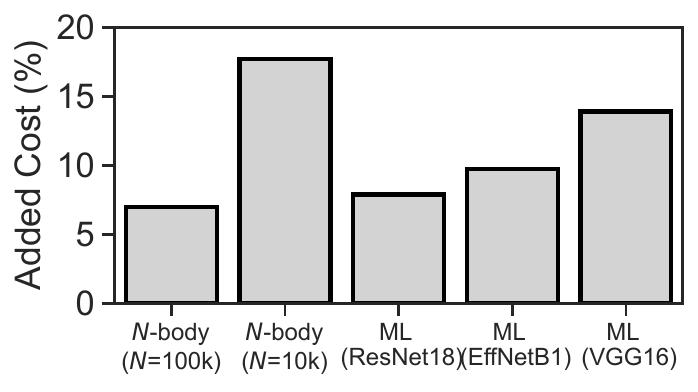}  &
    \includegraphics[width=0.31\textwidth]{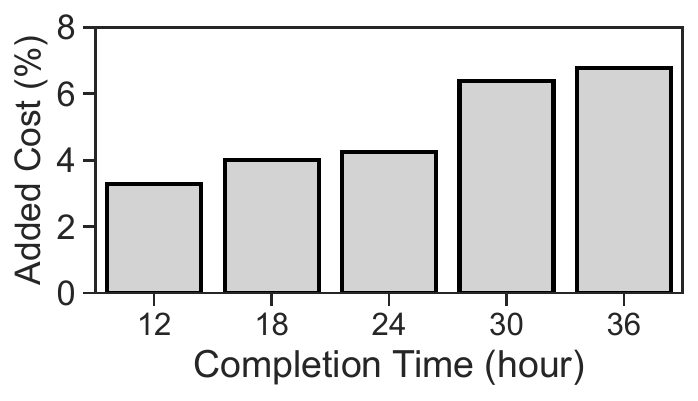} &
    \includegraphics[width=0.31\textwidth]{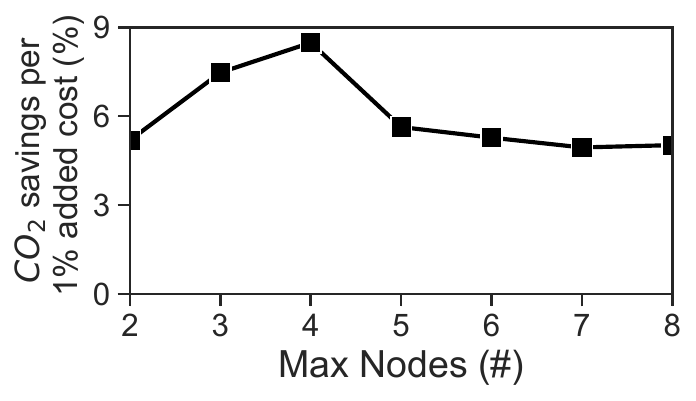}
    \vspace{-0.05cm}\\
    (a) Workloads (Fig.~\ref{fig:scalabilty-effect})  & (b) Deadlines (Fig.~\ref{fig:job_deadline}) & (c) Carbon-Cost tradeoff  \\
    \end{tabular}
    \vspace{-0.4cm}
    \caption{\emph{Monetary cost overhead of CarbonScaler over carbon-agnostic execution for different scenarios.
    }}
    \label{fig:added_cost}
\end{figure}

\vspace{-0.2cm}
\subsection{Monetary Cost Overhead}
\vspace{-0.05cm}
\label{sec:monetary-overhead}
As discussed in Section~\ref{sec:algorithm}, for the workloads with diminishing marginal capacity curves, \systemName can potentially incur extra cloud costs quantified as the additional cloud compute-hours needed compared to the \agnosticPolicy policy. 
In Figure~\ref{fig:added_cost}, we present the effect of workload scalability, extended completion time, and degree of flexibility on the added cost of \systemName.
Figure~\ref{fig:added_cost}(a) shows that the highly scalable workloads such as $N$-body ($N=100$) and ML (ResNet18) that yield the highest savings under \systemName cost only 5-10\% higher than a \agnosticPolicy policy. The less scalable workloads incur higher costs for the same carbon savings. 
It is worth noting that the \staticPolicy would also incur similar overheads as the cost depends on the scaling properties of the workload \cite{warofeff}. 
For ML (ResNet18) workload, Figure~\ref{fig:job_deadline} demonstrates that as the job completion time increases, the added cost increases up to 7\% and then plateaus with a further increase in completion time. 
This is because, at higher job completion times, there are more low-carbon slots available where \systemName can scale higher. Importantly, across both scenarios, the added cost never increases beyond 18\%.  
Finally, in figure~\ref{fig:added_cost} (c), we leverage \simName to highlight the tradeoff between carbon savings and cost overheads across different degrees of flexibility for ML (ResNet18). The figure shows that there exists a flexibility degree that yields the highest carbon savings of almost 9\% per each \% of added cost. 

\noindent \emph{\textbf{Key Takeaway.} 
The cloud cost overhead of CarbonScaler is small but depends on the scalability properties of the workloads (the higher the scalability, the lower the cost overhead). Furthermore, there may be a sweet spot across various dimensions that yields the highest savings per unit of added cost.
}

\vspace{-0.2cm}
\subsection{Impact of Carbon Cost Dynamics}
\label{sec:trace-character}
\vspace{-0.05cm}
Since achievable carbon savings depend on the temporal characteristics of the carbon costs within a cloud region, which significantly vary across regions, we next evaluate the impact of regions and carbon intensity variability on carbon savings.

\begin{figure*}[t]
\centering
\begin{minipage}[t]{0.48\linewidth}
  \vspace{0pt}
    \includegraphics[width= 1.05\textwidth]{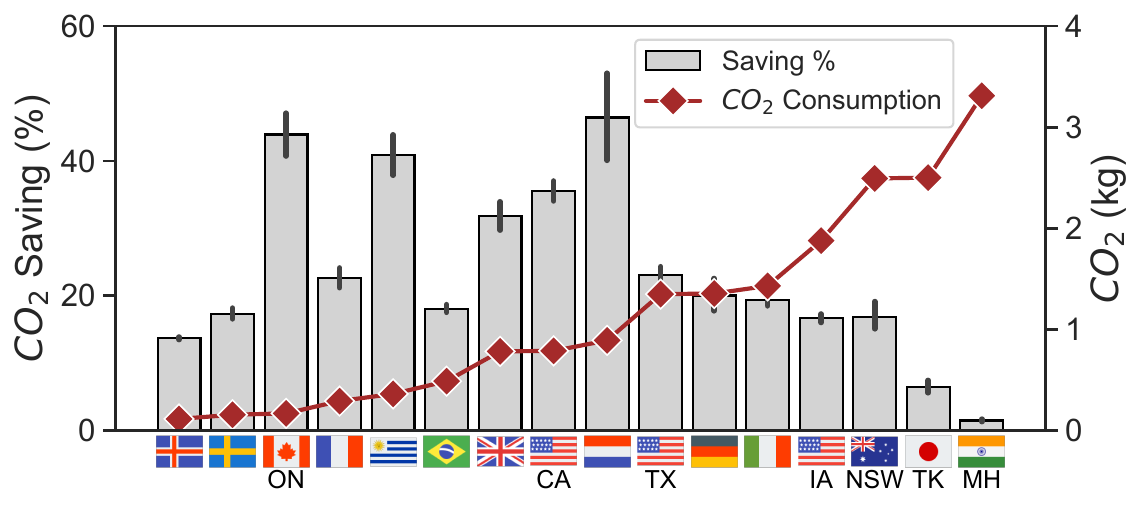}
    \vspace{-0.7cm}
    \caption{\emph{Carbon consumption (kg) and savings (\%), for an ML (ResNet18) job, where $T=l$, across geographical regions (carbon intensity increases from left to right).}}
    \vspace{0.1cm}
    \label{fig:trace_resnet18}
\end{minipage}
\hfill
\begin{minipage}[t]{0.48\linewidth}
\vspace{0pt}\raggedright
  \begin{tabular}{cc}
    \includegraphics[width=0.475\textwidth]{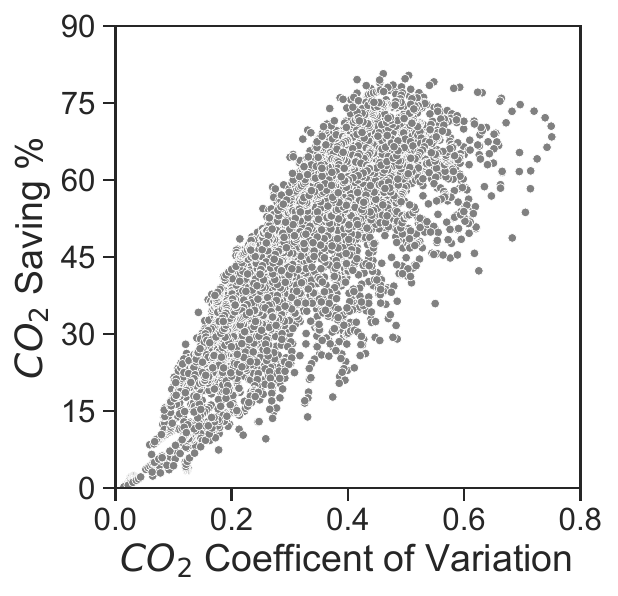}& \hspace{-0.5cm}
    \includegraphics[width=0.475\textwidth]{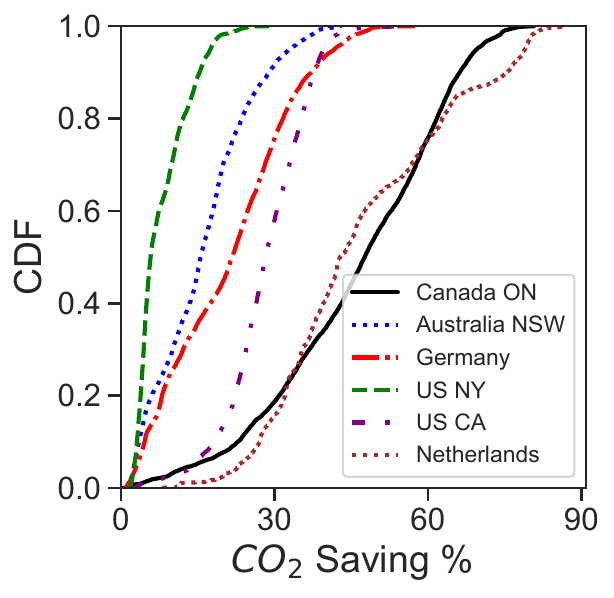}
    \vspace{-0.1cm} \\    
    (a) & (b) \\
    \end{tabular}
    \vspace{-0.4cm}
    \caption{\emph{Effect of variation (a) and location (b) on \emph{realized} savings for an ML (ResNet18) job ($T=l=24$hrs).}}
    \label{fig:achivable_saving}
\end{minipage}
\end{figure*}

\noindent{\bf Carbon Savings Across Cloud Regions.}
To assess the effect of regions on carbon savings, we use \simName to compute carbon savings achieved by a 24hrs long ML (ResNet18) job, with $T=l$, across 16 different AWS cloud regions.
Figure~\ref{fig:trace_resnet18} provides several insights about the average relative and absolute carbon savings compared to the \agnosticPolicy policy.
First, the figure shows that the carbon emissions of the same job can vary by an {\em order of magnitude} depending on which cloud region is used to execute it. 
Second,  \systemName is able to achieve significant carbon savings (in \%) for most regions, with a median and average savings of 16\% and 19\%, respectively. 
So long as the carbon costs exhibit diurnal variations, \systemName can reduce the job's  emissions over the \agnosticPolicy policy regardless of whether it runs in a low or high carbon region. 
Finally, Figure~\ref{fig:trace_resnet18} shows that India's (\worldflag[width=2mm]{IN}) region is an exception: while it has high absolute carbon costs, its low hourly variations prevent \systemName from generating much savings.

\vspace{0.05cm}
\noindent{\bf Effect of Variability.} 
As noted earlier, regions with variable carbon cost tend to generate higher carbon savings. 
This is because the high variations in such regions provide more low carbon periods to exploit for carbon reductions. 
We use the coefficient of variation, standard deviation divided by mean, as a metric to quantify the variability of the region. 
Figure~\ref{fig:achivable_saving}(a) shows the carbon savings, for each starting point of the year, for a 24hrs ML (ResNet18) job with no excess time for Ontario, Canada using \simName. 
The carbon savings are highly correlated with the coefficient of variation, with a Pearson coefficient of 0.82. 
However, even a highly variable location like Ontario has a small fraction of hours when savings are less than 20\%, a fraction that will vary depending on the region. 
Figure~\ref{fig:achivable_saving}(b) presents the distribution of carbon savings and compare regions with different average coefficient of variation. 
Note that the curves on the right are better as they lead to high carbon savings most of the time. 
The regions represented by the curves are also strictly ordered by their coefficient of variation, which means that a coefficient of variation can be used to rank regions, when mean carbon cost is comparable, for their carbon saving potential.

\vspace{0.05cm}
\noindent \emph{\textbf{Key Takeaway.} CarbonScaler achieves carbon savings for most cloud regions regardless of their absolute carbon cost. In addition, higher diurnal variations in carbon cost translate to greater savings.}

\subsection{Robustness to Errors}
\label{sec:assumptions}
In prior experiments, we assumed that the carbon forecasts are perfect and applications are profiled on an environment similar to what they eventually run on, yielding highly accurate marginal capacity curves. 
However, in practice, these assumptions may not always be true, and we evaluate the impact of deviation from the ground-truth for these two factors.

\begin{figure}[t]
  \centering
  \includegraphics[width=0.6\textwidth]{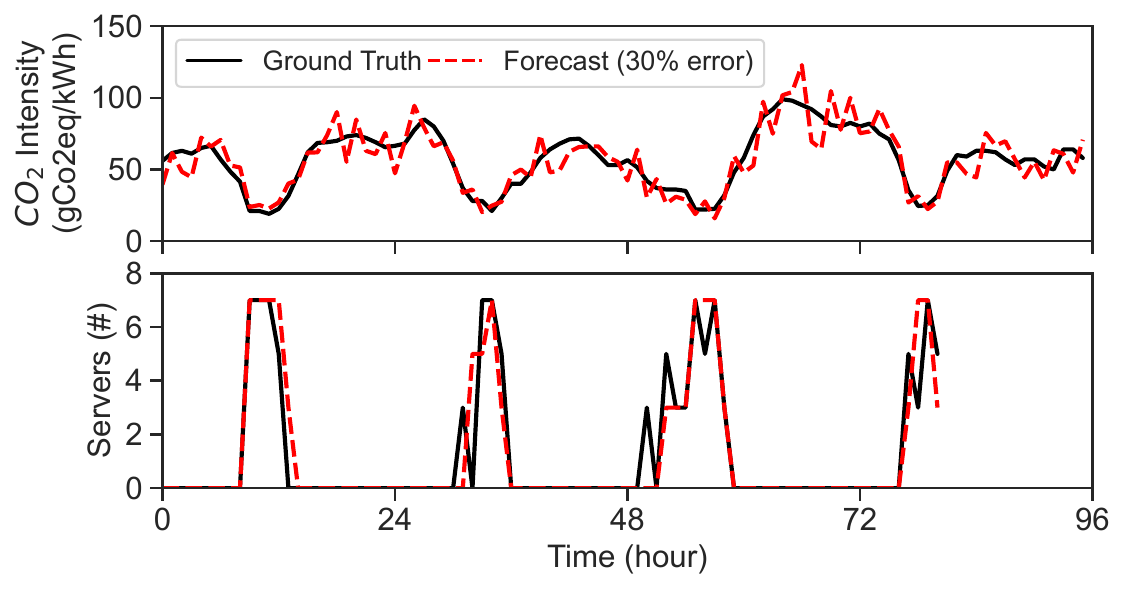}
    \vspace{-0.5cm}
    \caption{\emph{Illustrative example of error in carbon forecasts for an $N$-Body ($N=100k$) workload.}}
    \label{fig:error_nbody100_30}
\end{figure}

\vspace{0.05cm}
\noindent{\bf Effect of Carbon Forecast Error.}
Carbon forecasts are easily available through online tools and services such as~\cite{watttime, dayahead_estimations, carboncast}, with a reported mean accuracy of 6.4\%. 
More importantly, the fidelity of \systemName does not depend on the actual magnitude of the carbon forecast and instead relies on correctly identifying the hills (high carbon slots) and valleys (low carbon slots) in the carbon trace, which can be predicted with high accuracy. 
To illustrate this effect, we generate carbon traces with forecast errors of up to 30\% by adding a uniformly random error in the range of -X\% to X\% for an error of X\%. 
Figure~\ref{fig:error_nbody100_30}(top) shows an example ground-truth and forecasted (X = 30\% error) carbon cost time-series. 
While an erroneous forecast deviates from ground-truth at certain points, it still retains the hills and valleys, leading to harmonious schedules in both cases.

\begin{figure*}[t]
\centering
\begin{minipage}[t]{0.32\linewidth}
  \centering
  \includegraphics[width=\textwidth]{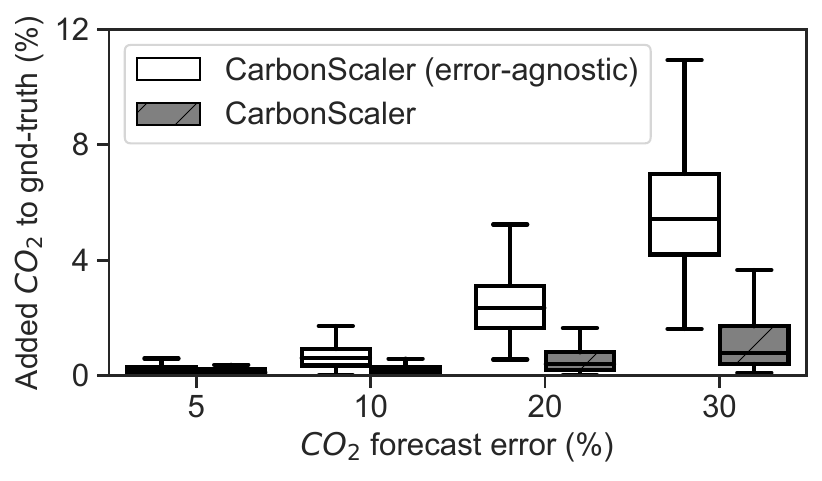}
  \vspace{-0.8cm}
    \caption{\emph{Effect of carbon forecast errors for an $N$-body ($N$=100k) job.}}
    \label{fig:error_nbody100}
\end{minipage}
\hfill
\begin{minipage}[t]{0.32\linewidth}
  \centering
   \includegraphics[width=\textwidth]{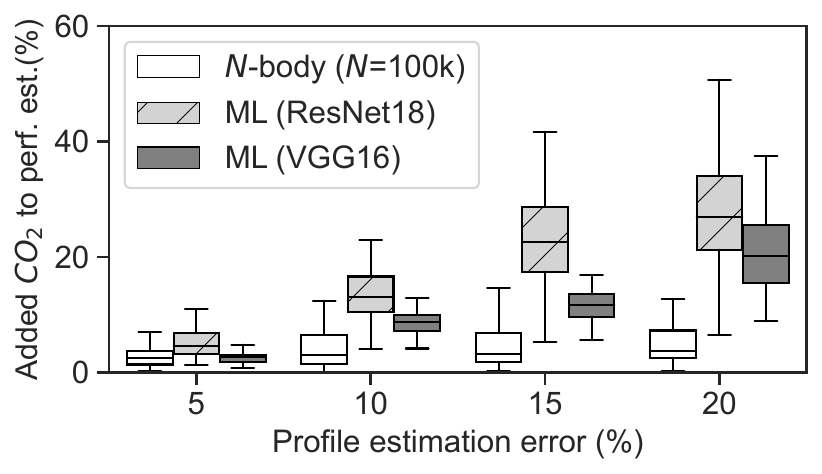}
   \vspace{-0.8cm}
    \caption{\emph{Effect of errors in profiled marginal capacity curves.}}
    
    \label{fig:profile_error}
\end{minipage}
\hfill
\begin{minipage}[t]{0.32\linewidth}
    \centering
    \includegraphics[width=\textwidth]{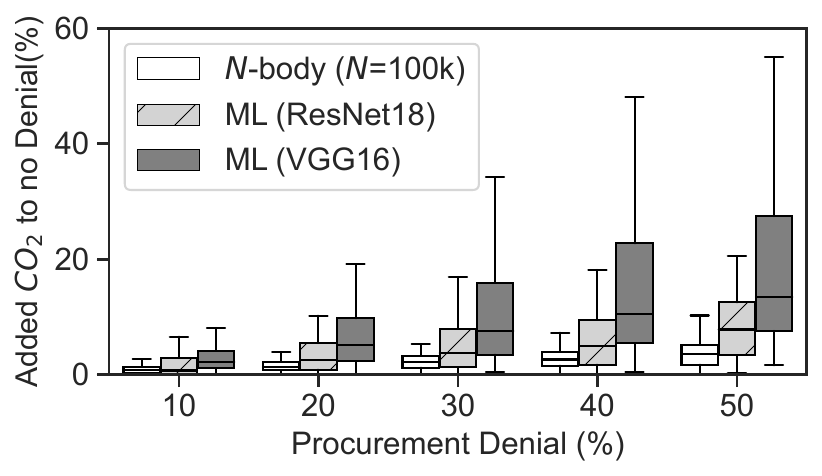}
    \vspace{-0.8cm}
    \caption{\emph{Carbon overhead of the server procurement denial.}}
    \label{fig:denial_effect}
\end{minipage}
\vspace{0.1cm}
\end{figure*}

To further quantify the effect of forecast errors, 
we compare the performance of \systemName with perfect carbon forecast to an error-agnostic variant of \systemName that is oblivious to forecast errors, and \systemName that recompute the schedule when the realized forecast error exceeds 5\%.
Figure~\ref{fig:error_nbody100} shows the carbon overhead over the perfect forecast scenario. 
The results highlight the resiliency of \systemName to forecast errors, as a 30\% forecast error resulted in merely 4\% added carbon at 95$^{th}$ percentile.

\vspace{0.05cm}
\noindent{\bf Effect of Profiling Errors.}
The marginal capacity curves generated by the \profilerName can become erroneous if the environment characteristics, such as network bottlenecks \cite{network_aware_scheduling}, change during the execution. 
This can impact the carbon savings of a given job if scaling behavior changes due to deviation from actual marginal capacity curves.
To evaluate the effect of erroneous profiles, we added uniformly random errors to the marginal capacity curves and measured the carbon consumption using \simName.  
Figure~\ref{fig:profile_error} shows the carbon overhead over \systemName with accurate marginal capacity profiles. 
The results show that the magnitude of error depends on the application power consumption and scalability behavior, e.g., the $N$-body job is less affected by errors as it has low power consumption and scales somewhat linearly. 
Additionally, we only show the results for the initial phase of execution, where errors persist. 
\systemName's error-handling mechanism of updating marginal capacity curves, when they deviate, corrects the errors, and net overhead over the entire execution of the workload would be considerably small.

\vspace{0.05cm}
\noindent{\bf Impact of Server Procurement Denial.}
Since \systemName dynamically scales each job independently, similar to cloud autoscalers, many jobs may request cloud servers during low carbon periods, creating a high demand for servers during such periods. Thus, jobs may end up competing with one another for additional servers, which can cause the cloud platform to deny some requests for new instances, to avoid failures. For example, it is not uncommon to see denials for popular GPU instances during work hours, even in the absence of carbon scaling.
To evaluate the effect of such denials, we run a 24hr job with 48hr completion time, $(T=2\times l)$, with different probabilities of random procurement denials. In such cases \systemName keeps retrying its request and then recomputes the schedule to mitigate the impact of denials on job completion.
Figure~\ref{fig:denial_effect} illustrates that the carbon overhead, compared to a no-denial scenario, increases as the denial percentage increases. 
The overhead's magnitude depends on a job's scalability behavior. 
For example, a highly scalable $N$-body job incurs 5\% overhead, while a non-scalable ML job (VGG16) incurs up to 15\% overhead compared to the best schedule.

\vspace{0.05cm}
\noindent \emph{\textbf{Key Takeaway.} CarbonScaler only depends on carbon cost trends, and simple recomputations achieve savings comparable to the perfect estimation. The potential overheads of profiling error can be overcome by updating marginal capacity curves as they start to deviate. Finally, resource availability can impact the achievable savings, but the magnitude depends on the scalability properties of the workloads.} 

\subsection{System Overheads}
\systemName incurs two types of systemic overheads in its execution. 
First, \systemName incurs switching overhead, which is the overhead of scaling or suspending, as the number of resources changes over time. 
The scaling overhead is a function of the application state size (e.g., the number of parameters in ML models). 
Although \systemName did not account for this overhead in its scheduling decisions, in our experiments, the scaling overhead was between 20-40 seconds. We note that \suspendPolicy incurs similar overheads as the state is scale-independent. 
The second source of overhead is the time needed by \profilerName to obtain marginal capacity curves. 
As mentioned in \S\ref{sec:profiler}, profiling time can be configured using profile duration $\alpha$ at each allocation level, and granularity $\beta$ of allocations profiled. 
We use $\alpha=1$ minute, and $\beta=1$, i.e., we profile across all possible allocation levels. 
Thus, the one-time profiling took 40 minutes, where each workload in Figure~\ref{fig:scalabilty-effect} took only 8 minutes. 

\vspace{0.08cm}
\noindent \emph{\textbf{Key Takeaway.} CarbonScaler's  systemic overheads are small, configurable, and generally occur once.} 

\section{Discussion}
\label{sec:discussion}
\systemName takes an application-centric approach to reduce the carbon footprint of cloud workloads. While addressing potential second-order effects is outside the scope of \systemName, we discuss the implications for cloud operators when customers operate in a carbon-aware manner.


\vspace{0.08cm}
\noindent{\bf Capacity Constraints.} 
Cloud operators have different optimization goals and constraints than their tenants. The conflicts are handled through the \emph{pay-as-you-go} pricing model, which hides the underlying constraints, objectives, and potential second-order effects from customers. Additionally, datacenters are designed for peak demand to handle workloads that exhibit diurnal patterns, where they increase at certain times of the day and are correlated between customers. As a result, they typically have low utilization, usually between 40-60\%~\cite{osti_1372902, Barroso2009TheDA, noman_eurosys21}, providing enough headroom to handle peaks from carbon-aware demand shifting. 
Carbon savings are achieved by aligning the demand with the carbon intensity. However, as more and more customers try to increase their carbon efficiency, the compute and power demand will increase at certain periods beyond the datacenter capacity. This will require cloud operators to handle such spikes by adopting a dynamic pricing model, enforce fair sharing limits, or by denying resource acquisition requests if needed. The modeling of such dynamic pricing and carbon-aware fair shares and how \systemName will respond is outside the scope of this paper. For acquisition denials, we demonstrate that \systemName is robust to such denials (see Figure~\ref{fig:denial_effect}).

\vspace{0.08cm}
\noindent{\bf Datacenter Energy Optimizations.}
\systemName is an application-centric approach to reducing the carbon footprint of executing workloads in the cloud, which can be used by organizations that are setting ambitious goals for reducing the carbon footprint of their operations~\cite{netflix-net-zero, twitter-net-zero, vmware-net-zero}. While cloud customers' behavior impacts the cloud datacenter operation, the pay-as-you-go model hides that from the customer. Internally, cloud operators can deploy several optimizations, such as forecasting demand and putting servers into a deep sleep or turning them off completely, offering resources at a discounted price, and procuring location-specific renewable energy. Many cloud operators are already experimenting with such optimizations. Examples include variable capacity computing at Google~\cite{Radovanovic2021CarbonAwareCF, variable_capacity_challanges}, spot VMs offered by AWS~\cite{aws-spot}, and 24/7 renewable energy procurements by all the major cloud providers~\cite{google-green-ppa, epa_green_energy_contracts}. Considering the impact of such operator-side optimizations is outside the scope of a customer-oriented approach like \systemName.

\noindent{\bf Holistic Emissions Reduction.}
A datacenter's carbon emissions arise from manufacturing hardware like servers (embodied emissions) and operating these resources (operational emissions). 
While both emission types are important, they require distinct optimizations~\cite{hotcarbon-embodied}. 
For instance, cutting embodied emissions involves extending device lifespan and choosing low operational carbon suppliers~\cite{li2023sustainable, Acun2022AHA}. 
How cloud operators and customers leverage such techniques for optimizing embodied carbon is outside the scope of this paper.  Instead, in this paper, we focus on reducing operational carbon emissions by modulating how and when we execute our workloads.


\section{Related Work}
\label{sec:relatedwork}
\noindent{\bf Batch Scheduler.}
HPC schedulers have focused on achieving high utilization and performance efficiency. Traditional batch schedulers such as Slurm~\cite{2003slurm} and Torque~\cite{2006torque} focus on fixed-sized clusters and employed multiple policies to optimize turnaround~\cite{ CUNHA201735}, utilization~\cite{ REUTHER201876}, and energy\cite{garg2011environment}. Recent schedulers such as  Borg, Kubernetes, and Mesos~\cite{kubernetes, 2011mesos} have utilized the elasticity of cloud resources while considering the monetary cost. In both cases, sustainability concerns have influenced operational and scheduling decisions, leading to optimization objectives such as reducing carbon consumption. In the rest of this section, we discuss recent research on carbon-aware scheduling.

\vspace{0.08cm}
\noindent{\bf Energy Accounting.}
Reporting carbon consumption depends on a cluster's ability to account for an individual tenant's energy consumption.
 \systemName currently focuses on CPU and GPU resources as they are 1) highly correlated with total energy consumption \cite{vmaccounting}, 2) software tools such as (RAPL)~\cite{david2010rapl} and \texttt{nvidia-smi} ~\cite{nvidia-dcgm}, are available on modern processors and GPUs. However, our accounting methods can be generalized to other server resources as shown in ~\cite{bourdon2013powerapi, vmaccounting, process_power_accounting, smartwatts}. Such accounting techniques are vital for holistic carbon optimization since cloud service providers such as Microsoft \cite{ms-carbon-accouting}, and AWS \cite{aws-carbon-accouting} are starting to offer basic carbon management capabilities. 

\vspace{0.08cm}
\noindent{\bf Temporal Shifting.} Temporal shifting by
delaying execution of batch jobs from high carbon slots to lower carbon slots has been explored in ~\cite{Wiesner2021LetsWA, cloudcarbon, Radovanovic2021CarbonAwareCF}.  The Let's wait-a-while \cite{Wiesner2021LetsWA} approach uses temporal shifting to reduce the carbon footprint of batch workloads using threshold and deadline-based methods and by exploiting overnight or weekend hours to extract savings. In \cite{cloudcarbon}, the authors highlight the implications of scheduling AI workloads in different settings and suggest temporal shifting to minimize the carbon footprint. Finally, in \cite{Radovanovic2021CarbonAwareCF}, the authors employ a virtual limit on resources when carbon  cost is high to force the scheduler to shift workloads to lower carbon periods.  As noted in \S \ref{sec:intro}, a limitation of temporal shifting approaches is that they delay job completion times and may also require users to specify deadlines for jobs. In contrast, \systemName employs resource elasticity to scale and complete jobs in a timely manner and can additionally exploit temporal flexibility whenever available.

\vspace{0.08cm}
\noindent{\bf Spatial Shifting.} 
Prior work has studied spatial shifting to select the region with the lowest carbon footprint to execute newly arriving jobs. The authors of \cite{Acun2022AHA, cloudcarbon, Zheng2020MitigatingCA, sukprasert2023quantifying} explore the spatial selection to achieve lower carbon cost. The authors of \cite{Acun2022AHA} explore data center and power upgrade plans to allow more carbon-efficient execution, while \cite{cloudcarbon, sukprasert2023quantifying} explore cloud data center regions and potential carbon savings. Lastly, \cite{Zheng2020MitigatingCA} exploits migration to avoid energy curtailment. While we study the benefits of using different geographic regions to run carbon scaling jobs in \S \ref{sec:trace-character}, a full analysis of combining spatial shifting with carbon scaling is outside the scope of this paper.

\section{Conclusion}
Many compute-intensive cloud workloads, such as ML training and scientific computations, have inherent resource elasticity and temporal flexibility that can be leveraged to optimize carbon emission reductions. 
To exploit this opportunity, we propose \systemName that judiciously scales up or down an application, based on its scalability behavior and carbon cost, to minimize its carbon emissions.
We implement \systemName as a cloud-based autoscaler implemented using Kubernetes and a simulation-based advisory tool to facilitate pre-deployment analysis. 
We demonstrate the efficacy of \systemName in reducing carbon emissions for various workloads, job configurations, and cloud regions. 
We demonstrated that using real-world machine learning training and MPI jobs on a commercial cloud platform, \systemName can yield i) 51\% carbon savings over carbon-agnostic execution, ii) 37\% over a suspend-resume policy, and iii) 8\% over the best static scaling policy.
In the future, we plan to extend \systemName into a cluster-wide scheduler to address the challenges of resource heterogeneity, resource pressure, priorities, and power management.

\label{sec:conclusion}

\begin{acks}
We thank the anonymous reviewers and our shepherd Anshul Gandhi, for their valuable comments, which improved the quality of this paper. This research is supported by NSF grants 2211302, 2211888, 2213636, 2105494, US Army contract W911NF-17-2-0196, VMware, and Amazon Web Services.
\end{acks}

\bibliographystyle{ACM-Reference-Format}
\bibliography{references}


\begin{thebibliography}{77}


\ifx \showCODEN    \undefined \def \showCODEN     #1{\unskip}     \fi
\ifx \showDOI      \undefined \def \showDOI       #1{#1}\fi
\ifx \showISBNx    \undefined \def \showISBNx     #1{\unskip}     \fi
\ifx \showISBNxiii \undefined \def \showISBNxiii  #1{\unskip}     \fi
\ifx \showISSN     \undefined \def \showISSN      #1{\unskip}     \fi
\ifx \showLCCN     \undefined \def \showLCCN      #1{\unskip}     \fi
\ifx \shownote     \undefined \def \shownote      #1{#1}          \fi
\ifx \showarticletitle \undefined \def \showarticletitle #1{#1}   \fi
\ifx \showURL      \undefined \def \showURL       {\relax}        \fi
\providecommand\bibfield[2]{#2}
\providecommand\bibinfo[2]{#2}
\providecommand\natexlab[1]{#1}
\providecommand\showeprint[2][]{arXiv:#2}

\bibitem[Aarseth(1985)]%
        {nbodysimulation}
\bibfield{author}{\bibinfo{person}{J.~Sverre Aarseth}.}
  \bibinfo{year}{1985}\natexlab{}.
\newblock \showarticletitle{12 - Direct Methods for N-Body Simulations}.
\newblock In \bibinfo{booktitle}{\emph{Multiple Time Scales}}.
  \bibinfo{publisher}{Academic Press}, \bibinfo{pages}{377--418}.
\newblock
\showISBNx{978-0-12-123420-1}
\urldef\tempurl%
\url{https://doi.org/10.1016/B978-0-12-123420-1.50017-3}
\showDOI{\tempurl}


\bibitem[Acun et~al\mbox{.}(2023)]%
        {Acun2022AHA}
\bibfield{author}{\bibinfo{person}{Bilge Acun}, \bibinfo{person}{Benjamin Lee},
  \bibinfo{person}{Fiodar Kazhamiaka}, \bibinfo{person}{Kiwan Maeng},
  \bibinfo{person}{Udit Gupta}, \bibinfo{person}{Manoj Chakkaravarthy},
  \bibinfo{person}{David Brooks}, {and} \bibinfo{person}{Carole-Jean Wu}.}
  \bibinfo{year}{2023}\natexlab{}.
\newblock \showarticletitle{Carbon Explorer: A Holistic Framework for Designing
  Carbon Aware Datacenters}. In \bibinfo{booktitle}{\emph{Proceedings of the
  28th ACM International Conference on Architectural Support for Programming
  Languages and Operating Systems, Volume 2}} (Vancouver, BC, Canada)
  \emph{(\bibinfo{series}{ASPLOS 2023})}. \bibinfo{publisher}{Association for
  Computing Machinery}, \bibinfo{address}{New York, NY, USA},
  \bibinfo{pages}{118–132}.
\newblock
\showISBNx{9781450399166}
\urldef\tempurl%
\url{https://doi.org/10.1145/3575693.3575754}
\showDOI{\tempurl}


\bibitem[Amdahl(1967)]%
        {amdahl-law}
\bibfield{author}{\bibinfo{person}{Gene~M Amdahl}.}
  \bibinfo{year}{1967}\natexlab{}.
\newblock \showarticletitle{Validity of the {S}ingle {P}rocessor {A}pproach to
  {A}chieving {L}arge {S}cale {C}omputing {C}apabilities}. In
  \bibinfo{booktitle}{\emph{Proceedings of the Spring Joint Computer
  Conference}}.
\newblock


\bibitem[Andrae and Edler(2015)]%
        {Andrae2015OnGE}
\bibfield{author}{\bibinfo{person}{Anders S.~G. Andrae} {and}
  \bibinfo{person}{Tomas Edler}.} \bibinfo{year}{2015}\natexlab{}.
\newblock \showarticletitle{On Global Electricity Usage of Communication
  Technology: Trends to 2030}.
\newblock \bibinfo{journal}{\emph{Challenges}} \bibinfo{volume}{6},
  \bibinfo{number}{1} (\bibinfo{year}{2015}), \bibinfo{pages}{117--157}.
\newblock
\showISSN{2078-1547}
\urldef\tempurl%
\url{https://doi.org/10.3390/challe6010117}
\showDOI{\tempurl}


\bibitem[Armbrust et~al\mbox{.}(2015)]%
        {elastic_spark}
\bibfield{author}{\bibinfo{person}{Michael Armbrust},
  \bibinfo{person}{Tathagata Das}, \bibinfo{person}{Aaron Davidson},
  \bibinfo{person}{Ali Ghodsi}, \bibinfo{person}{Andrew Or},
  \bibinfo{person}{Josh Rosen}, \bibinfo{person}{Ion Stoica},
  \bibinfo{person}{Patrick Wendell}, \bibinfo{person}{Reynold Xin}, {and}
  \bibinfo{person}{Matei Zaharia}.} \bibinfo{year}{2015}\natexlab{}.
\newblock \showarticletitle{{Scaling Spark in the Real World: Performance and
  Usability}}.
\newblock \bibinfo{journal}{\emph{Proc. VLDB Endow.}} \bibinfo{volume}{8},
  \bibinfo{number}{12} (\bibinfo{date}{aug} \bibinfo{year}{2015}),
  \bibinfo{pages}{1840–1843}.
\newblock
\showISSN{2150-8097}
\urldef\tempurl%
\url{https://doi.org/10.14778/2824032.2824080}
\showDOI{\tempurl}


\bibitem[AWS(2022)]%
        {aws-autoscaling}
\bibfield{author}{\bibinfo{person}{AWS}.} \bibinfo{year}{2022}\natexlab{}.
\newblock \bibinfo{title}{AWS Auto Scaling}.
\newblock \bibinfo{howpublished}{\url{https://aws.amazon.com/autoscaling/}}.
\newblock


\bibitem[Barroso and H{\"o}lzle(2009)]%
        {Barroso2009TheDA}
\bibfield{author}{\bibinfo{person}{Luiz~Andr{\'e} Barroso} {and}
  \bibinfo{person}{Urs H{\"o}lzle}.} \bibinfo{year}{2009}\natexlab{}.
\newblock \bibinfo{booktitle}{\emph{The Datacenter as a Computer: An
  Introduction to the Design of Warehouse-Scale Machines}}.
\newblock \bibinfo{publisher}{Springer Nature}, \bibinfo{address}{Europe}. 189
  pages.
\newblock


\bibitem[Bashir et~al\mbox{.}(2021)]%
        {noman_eurosys21}
\bibfield{author}{\bibinfo{person}{Noman Bashir}, \bibinfo{person}{Nan Deng},
  \bibinfo{person}{Krzysztof Rzadca}, \bibinfo{person}{David Irwin},
  \bibinfo{person}{Sree Kodak}, {and} \bibinfo{person}{Rohit Jnagal}.}
  \bibinfo{year}{2021}\natexlab{}.
\newblock \showarticletitle{Take it to the {L}imit: {P}eak {P}rediction-driven
  {R}esource {O}vercommitment in {D}atacenters}. In
  \bibinfo{booktitle}{\emph{Proceedings of the Sixteenth European Conference on
  Computer Systems}} (Online Event, United Kingdom)
  \emph{(\bibinfo{series}{EuroSys '21})}. \bibinfo{publisher}{Association for
  Computing Machinery}, \bibinfo{address}{New York, NY, USA},
  \bibinfo{pages}{556–573}.
\newblock
\showISBNx{9781450383349}
\urldef\tempurl%
\url{https://doi.org/10.1145/3447786.3456259}
\showDOI{\tempurl}


\bibitem[Bashir et~al\mbox{.}(2023)]%
        {hotcarbon-embodied}
\bibfield{author}{\bibinfo{person}{Noman Bashir}, \bibinfo{person}{David
  Irwin}, {and} \bibinfo{person}{Prashant Shenoy}.}
  \bibinfo{year}{2023}\natexlab{}.
\newblock \showarticletitle{On the {P}romise and {P}itfalls of {O}ptimizing
  {E}mbodied {C}arbon}. In \bibinfo{booktitle}{\emph{Proceedings of the 2nd
  Workshop on Sustainable Computer Systems (HotCarbon)}}.
  \bibinfo{publisher}{ACM}, \bibinfo{address}{New York, NY, USA},
  \bibinfo{numpages}{6}~pages.
\newblock


\bibitem[Bashir et~al\mbox{.}(2022)]%
        {hotair}
\bibfield{author}{\bibinfo{person}{Noman Bashir}, \bibinfo{person}{David
  Irwin}, \bibinfo{person}{Prashant Shenoy}, {and} \bibinfo{person}{Abel
  Souza}.} \bibinfo{year}{2022}\natexlab{}.
\newblock \showarticletitle{Sustainable Computing -- Without the Hot Air}. In
  \bibinfo{booktitle}{\emph{HotCarbon: Workshop on Sustainable Computer Systems
  Design and Implementation}}. \bibinfo{publisher}{ACM}, \bibinfo{address}{New
  York, NY, USA}, \bibinfo{numpages}{7}~pages.
\newblock


\bibitem[Bourdon et~al\mbox{.}(2013)]%
        {bourdon2013powerapi}
\bibfield{author}{\bibinfo{person}{Aur{\'e}lien Bourdon}, \bibinfo{person}{Adel
  Noureddine}, \bibinfo{person}{Romain Rouvoy}, {and} \bibinfo{person}{Lionel
  Seinturier}.} \bibinfo{year}{2013}\natexlab{}.
\newblock \showarticletitle{Powerapi: {A} {S}oftware {L}ibrary to {M}onitor the
  {E}nergy {C}onsumed at the {P}rocess-level}.
\newblock \bibinfo{journal}{\emph{ERCIM News}} (\bibinfo{year}{2013}).
\newblock


\bibitem[Boyle and Junod(2023)]%
        {twitter-net-zero}
\bibfield{author}{\bibinfo{person}{Seán Boyle} {and} \bibinfo{person}{Casey
  Junod}.} \bibinfo{year}{2023}\natexlab{}.
\newblock \bibinfo{title}{Accelerating our climate commitments on {E}arth
  {D}ay}.
\newblock
  \bibinfo{howpublished}{\url{https://blog.twitter.com/en_us/topics/company/2022/accelerating-our-climate-commitments-on-earth-day}}.
\newblock


\bibitem[Cai et~al\mbox{.}(2017)]%
        {cai2017neuralpower}
\bibfield{author}{\bibinfo{person}{Ermao Cai}, \bibinfo{person}{Da-Cheng Juan},
  \bibinfo{person}{Dimitrios Stamoulis}, {and} \bibinfo{person}{Diana
  Marculescu}.} \bibinfo{year}{2017}\natexlab{}.
\newblock \showarticletitle{Neuralpower: {P}redict and {D}eploy
  {E}nergy-efficient {C}onvolutional {N}eural {N}etworks}. In
  \bibinfo{booktitle}{\emph{Asian Conference on Machine Learning}}.
\newblock


\bibitem[Chien(2021)]%
        {chien-cacm21}
\bibfield{author}{\bibinfo{person}{A. Chien}.} \bibinfo{year}{2021}\natexlab{}.
\newblock \showarticletitle{Driving the {C}loud to {T}rue {Z}ero {C}arbon}.
\newblock \bibinfo{journal}{\emph{Communication of the ACM}}
  \bibinfo{volume}{64}, \bibinfo{number}{2} (\bibinfo{date}{February}
  \bibinfo{year}{2021}).
\newblock


\bibitem[Colmant et~al\mbox{.}(2015)]%
        {process_power_accounting}
\bibfield{author}{\bibinfo{person}{Maxime Colmant}, \bibinfo{person}{Mascha
  Kurpicz}, \bibinfo{person}{Pascal Felber}, \bibinfo{person}{Lo\"{\i}c
  Huertas}, \bibinfo{person}{Romain Rouvoy}, {and} \bibinfo{person}{Anita
  Sobe}.} \bibinfo{year}{2015}\natexlab{}.
\newblock \showarticletitle{Process-Level Power Estimation in VM-Based
  Systems}. In \bibinfo{booktitle}{\emph{Proceedings of the Tenth European
  Conference on Computer Systems}} (Bordeaux, France)
  \emph{(\bibinfo{series}{EuroSys '15})}. \bibinfo{publisher}{Association for
  Computing Machinery}, \bibinfo{address}{New York, NY, USA}, Article
  \bibinfo{articleno}{14}, \bibinfo{numpages}{14}~pages.
\newblock
\showISBNx{9781450332385}
\urldef\tempurl%
\url{https://doi.org/10.1145/2741948.2741971}
\showDOI{\tempurl}


\bibitem[Cunha et~al\mbox{.}(2017)]%
        {CUNHA201735}
\bibfield{author}{\bibinfo{person}{Renato~L.F. Cunha},
  \bibinfo{person}{Eduardo~R. Rodrigues}, \bibinfo{person}{Leonardo~P. Tizzei},
  {and} \bibinfo{person}{Marco~A.S. Netto}.} \bibinfo{year}{2017}\natexlab{}.
\newblock \showarticletitle{{Job placement advisor based on turnaround
  predictions for HPC hybrid clouds}}.
\newblock \bibinfo{journal}{\emph{Future Generation Computer Systems}}
  \bibinfo{volume}{67} (\bibinfo{year}{2017}), \bibinfo{pages}{35--46}.
\newblock
\showISSN{0167-739X}
\urldef\tempurl%
\url{https://doi.org/10.1016/j.future.2016.08.010}
\showDOI{\tempurl}


\bibitem[David et~al\mbox{.}(2010)]%
        {david2010rapl}
\bibfield{author}{\bibinfo{person}{Howard David}, \bibinfo{person}{Eugene
  Gorbatov}, \bibinfo{person}{Ulf~R Hanebutte}, \bibinfo{person}{Rahul Khanna},
  {and} \bibinfo{person}{Christian Le}.} \bibinfo{year}{2010}\natexlab{}.
\newblock \showarticletitle{R{A}{P}{L}: {M}emory {P}ower {E}stimation and
  {C}apping}. In \bibinfo{booktitle}{\emph{ACM/IEEE International Symposium on
  Low-Power Electronics and Design (ISLPED)}}.
\newblock


\bibitem[Devlin et~al\mbox{.}(2019)]%
        {bert}
\bibfield{author}{\bibinfo{person}{Jacob Devlin}, \bibinfo{person}{Ming-Wei
  Chang}, \bibinfo{person}{Kenton Lee}, {and} \bibinfo{person}{Kristina
  Toutanova}.} \bibinfo{year}{2019}\natexlab{}.
\newblock \showarticletitle{{BERT}: Pre-training of Deep Bidirectional
  Transformers for Language Understanding}. In
  \bibinfo{booktitle}{\emph{Proceedings of the 2019 Conference of the North
  {A}merican Chapter of the Association for Computational Linguistics: Human
  Language Technologies, Volume 1 (Long and Short Papers)}}.
  \bibinfo{publisher}{Association for Computational Linguistics},
  \bibinfo{address}{Minneapolis, Minnesota}, \bibinfo{pages}{4171--4186}.
\newblock
\urldef\tempurl%
\url{https://doi.org/10.18653/v1/N19-1423}
\showDOI{\tempurl}


\bibitem[Dodge et~al\mbox{.}(2022)]%
        {cloudcarbon}
\bibfield{author}{\bibinfo{person}{Jesse Dodge}, \bibinfo{person}{Taylor
  Prewitt}, \bibinfo{person}{Remi Tachet~des Combes}, \bibinfo{person}{Erika
  Odmark}, \bibinfo{person}{Roy Schwartz}, \bibinfo{person}{Emma Strubell},
  \bibinfo{person}{Alexandra~Sasha Luccioni}, \bibinfo{person}{Noah~A. Smith},
  \bibinfo{person}{Nicole DeCario}, {and} \bibinfo{person}{Will Buchanan}.}
  \bibinfo{year}{2022}\natexlab{}.
\newblock \showarticletitle{Measuring the Carbon Intensity of AI in Cloud
  Instances}. In \bibinfo{booktitle}{\emph{2022 ACM Conference on Fairness,
  Accountability, and Transparency}} \emph{(\bibinfo{series}{FAccT '22})}.
\newblock
\showISBNx{9781450393522}


\bibitem[EC2(2022)]%
        {aws-spot}
EC2 \bibinfo{year}{2022}\natexlab{}.
\newblock \bibinfo{title}{Amazon {E}{C}2 {S}pot {I}nstances}.
\newblock \bibinfo{howpublished}{\url{https://aws.amazon.com/ec2/spot/}}.
\newblock


\bibitem[EPA(2023)]%
        {epa_green_energy_contracts}
\bibfield{author}{\bibinfo{person}{EPA}.} \bibinfo{year}{2023}\natexlab{}.
\newblock \bibinfo{title}{Green Power Partnership Long-term Contracts}.
\newblock
\newblock
\urldef\tempurl%
\url{https://www.epa.gov/greenpower/green-power-partnership-long-term-contracts}
\showURL{%
\tempurl}


\bibitem[Federgruen and Groenevelt(1986)]%
        {greedy_optimal}
\bibfield{author}{\bibinfo{person}{Awi Federgruen} {and} \bibinfo{person}{Henri
  Groenevelt}.} \bibinfo{year}{1986}\natexlab{}.
\newblock \showarticletitle{{The Greedy Procedure for Resource Allocation
  Problems: Necessary and Sufficient Conditions for Optimality}}.
\newblock \bibinfo{journal}{\emph{Oper. Res.}} \bibinfo{volume}{34},
  \bibinfo{number}{6} (\bibinfo{date}{dec} \bibinfo{year}{1986}),
  \bibinfo{pages}{909–918}.
\newblock
\showISSN{0030-364X}


\bibitem[Fieni et~al\mbox{.}(2020)]%
        {smartwatts}
\bibfield{author}{\bibinfo{person}{Guillaume Fieni}, \bibinfo{person}{Romain
  Rouvoy}, {and} \bibinfo{person}{Lionel Seinturier}.}
  \bibinfo{year}{2020}\natexlab{}.
\newblock \showarticletitle{SmartWatts: Self-Calibrating Software-Defined Power
  Meter for Containers}. In \bibinfo{booktitle}{\emph{2020 20th IEEE/ACM
  International Symposium on Cluster, Cloud and Internet Computing (CCGRID)}}.
  \bibinfo{pages}{479--488}.
\newblock
\urldef\tempurl%
\url{https://doi.org/10.1109/CCGrid49817.2020.00-45}
\showDOI{\tempurl}


\bibitem[Forum(1994)]%
        {mpi}
\bibfield{author}{\bibinfo{person}{Message~P Forum}.}
  \bibinfo{year}{1994}\natexlab{}.
\newblock \bibinfo{booktitle}{\emph{MPI: A Message-Passing Interface
  Standard}}.
\newblock \bibinfo{type}{{T}echnical {R}eport}. \bibinfo{address}{USA}.
\newblock


\bibitem[Fox et~al\mbox{.}(2017)]%
        {elastic_hpc}
\bibfield{author}{\bibinfo{person}{William Fox}, \bibinfo{person}{Devarshi
  Ghoshal}, \bibinfo{person}{Abel Souza}, \bibinfo{person}{Gonzalo~P. Rodrigo},
  {and} \bibinfo{person}{Lavanya Ramakrishnan}.}
  \bibinfo{year}{2017}\natexlab{}.
\newblock \showarticletitle{{E-HPC: A Library for Elastic Resource Management
  in HPC Environments}}. In \bibinfo{booktitle}{\emph{Proceedings of the 12th
  Workshop on Workflows in Support of Large-Scale Science}} (Denver, Colorado)
  \emph{(\bibinfo{series}{WORKS '17})}. \bibinfo{publisher}{Association for
  Computing Machinery}, \bibinfo{address}{New York, NY, USA}, Article
  \bibinfo{articleno}{1}, \bibinfo{numpages}{11}~pages.
\newblock
\showISBNx{9781450351294}
\urldef\tempurl%
\url{https://doi.org/10.1145/3150994.3150996}
\showDOI{\tempurl}


\bibitem[Gandhi et~al\mbox{.}(2012)]%
        {autoscaler}
\bibfield{author}{\bibinfo{person}{Anshul Gandhi}, \bibinfo{person}{Mor
  Harchol-Balter}, \bibinfo{person}{Ram Raghunathan}, {and}
  \bibinfo{person}{Michael~A. Kozuch}.} \bibinfo{year}{2012}\natexlab{}.
\newblock \showarticletitle{{AutoScale: Dynamic, Robust Capacity Management for
  Multi-Tier Data Centers}}.
\newblock \bibinfo{journal}{\emph{ACM Trans. Comput. Syst.}}
  \bibinfo{volume}{30}, \bibinfo{number}{4}, Article \bibinfo{articleno}{14}
  (\bibinfo{date}{nov} \bibinfo{year}{2012}), \bibinfo{numpages}{26}~pages.
\newblock
\showISSN{0734-2071}
\urldef\tempurl%
\url{https://doi.org/10.1145/2382553.2382556}
\showDOI{\tempurl}


\bibitem[Garg et~al\mbox{.}(2011)]%
        {garg2011environment}
\bibfield{author}{\bibinfo{person}{Saurabh~Kumar Garg},
  \bibinfo{person}{Chee~Shin Yeo}, \bibinfo{person}{Arun Anandasivam}, {and}
  \bibinfo{person}{Rajkumar Buyya}.} \bibinfo{year}{2011}\natexlab{}.
\newblock \showarticletitle{{Environment-conscious scheduling of HPC
  applications on distributed cloud-oriented data centers}}.
\newblock \bibinfo{journal}{\emph{J. Parallel and Distrib. Comput.}}
  \bibinfo{volume}{71}, \bibinfo{number}{6} (\bibinfo{year}{2011}),
  \bibinfo{pages}{732--749}.
\newblock


\bibitem[Google(2022)]%
        {google-green-ppa}
\bibfield{author}{\bibinfo{person}{Google}.} \bibinfo{year}{2022}\natexlab{}.
\newblock \bibinfo{title}{Google’s Green PPAs: What, How, and Why}.
\newblock
  \bibinfo{howpublished}{\url{https://static.googleusercontent.com/media/www.google.com/en//green/pdfs/renewable-energy.pdf}}.
\newblock


\bibitem[Hanafy et~al\mbox{.}(2023)]%
        {warofeff}
\bibfield{author}{\bibinfo{person}{Walid~A Hanafy}, \bibinfo{person}{Roozbeh
  Bostandoost}, \bibinfo{person}{Noman Bashir}, \bibinfo{person}{David Irwin},
  \bibinfo{person}{Mohammad Hajiesmaili}, {and} \bibinfo{person}{Prashant
  Shenoy}.} \bibinfo{year}{2023}\natexlab{}.
\newblock \showarticletitle{{The War of the Efficiencies: Understanding the
  Tension between Carbon and Energy Optimization}}. In
  \bibinfo{booktitle}{\emph{Proc. 2nd ACM Workshop on Hot Topics in Sustainable
  Computing Systems (HotCarbon'23)}}.
\newblock


\bibitem[Harvey(2021)]%
        {guardian}
\bibfield{author}{\bibinfo{person}{Fiona Harvey}.}
  \bibinfo{year}{2021}\natexlab{}.
\newblock \bibinfo{title}{The {G}uardian, {M}ajor {C}limate {C}hanges
  {I}nevitable and {I}rreversible – {I}{P}{C}{C}'s {S}tarkest {W}arning
  {Y}et}.
\newblock
  \bibinfo{howpublished}{\url{https://www.theguardian.com/science/2021/aug/09/humans-have-caused-unprecedented-and-/irreversible-change-to-climate-scientists-warn}}.
\newblock


\bibitem[He et~al\mbox{.}(2016)]%
        {resnet}
\bibfield{author}{\bibinfo{person}{Kaiming He}, \bibinfo{person}{Xiangyu
  Zhang}, \bibinfo{person}{Shaoqing Ren}, {and} \bibinfo{person}{Jian Sun}.}
  \bibinfo{year}{2016}\natexlab{}.
\newblock \showarticletitle{Deep {R}esidual {L}earning for {I}mage
  {R}ecognition}. In \bibinfo{booktitle}{\emph{Proceedings of the IEEE
  conference on computer vision and pattern recognition (CVPR)}}.
\newblock


\bibitem[Hindman et~al\mbox{.}(2011)]%
        {2011mesos}
\bibfield{author}{\bibinfo{person}{Benjamin Hindman}, \bibinfo{person}{Andy
  Konwinski}, \bibinfo{person}{Matei Zaharia}, \bibinfo{person}{Ali Ghodsi},
  \bibinfo{person}{Anthony~D. Joseph}, \bibinfo{person}{Randy Katz},
  \bibinfo{person}{Scott Shenker}, {and} \bibinfo{person}{Ion Stoica}.}
  \bibinfo{year}{2011}\natexlab{}.
\newblock \showarticletitle{Mesos: A {P}latform for {F}ine-{G}rained {R}esource
  {S}haring in the {D}ata {C}enter}. In \bibinfo{booktitle}{\emph{USENIX
  Symposium on Networked Systems Design and Implementation (NSDI)}}.
  \bibinfo{publisher}{USENIX Association}, \bibinfo{address}{Boston, MA},
  \bibinfo{pages}{14}.
\newblock
\urldef\tempurl%
\url{https://www.usenix.org/conference/nsdi11/mesos-platform-fine-grained-resource-sharing-data-center}
\showURL{%
\tempurl}


\bibitem[Inc.(2023)]%
        {vmware-net-zero}
\bibfield{author}{\bibinfo{person}{VMware Inc.}}
  \bibinfo{year}{2023}\natexlab{}.
\newblock \bibinfo{title}{Journey to {N}et {Z}ero}.
\newblock
  \bibinfo{howpublished}{\url{https://www.vmware.com/company/net-zero.html}}.
\newblock


\bibitem[Institute(2022)]%
        {ghg}
\bibfield{author}{\bibinfo{person}{World~Resource Institute}.}
  \bibinfo{year}{2022}\natexlab{}.
\newblock \bibinfo{booktitle}{\emph{GreenHouseGas Protocol}}.
\newblock
\urldef\tempurl%
\url{https://ghgprotocol.org/}
\showURL{%
\tempurl}


\bibitem[Jacobs et~al\mbox{.}(2017)]%
        {hpcML}
\bibfield{author}{\bibinfo{person}{Sam~Ad\'{e} Jacobs}, \bibinfo{person}{Nikoli
  Dryden}, \bibinfo{person}{Roger Pearce}, {and} \bibinfo{person}{Brian
  Van~Essen}.} \bibinfo{year}{2017}\natexlab{}.
\newblock \showarticletitle{Towards Scalable Parallel Training of Deep Neural
  Networks}. In \bibinfo{booktitle}{\emph{Proceedings of the Machine Learning
  on HPC Environments}} \emph{(\bibinfo{series}{MLHPC'17})}.
\newblock


\bibitem[Jalaparti et~al\mbox{.}(2015)]%
        {network_aware_scheduling}
\bibfield{author}{\bibinfo{person}{Virajith Jalaparti}, \bibinfo{person}{Peter
  Bodik}, \bibinfo{person}{Ishai Menache}, \bibinfo{person}{Sriram Rao},
  \bibinfo{person}{Konstantin Makarychev}, {and} \bibinfo{person}{Matthew
  Caesar}.} \bibinfo{year}{2015}\natexlab{}.
\newblock \showarticletitle{Network-Aware Scheduling for Data-Parallel Jobs:
  Plan When You Can}. In \bibinfo{booktitle}{\emph{Proceedings of the 2015 ACM
  Conference on Special Interest Group on Data Communication}} (London, United
  Kingdom) \emph{(\bibinfo{series}{SIGCOMM '15})}.
  \bibinfo{publisher}{Association for Computing Machinery},
  \bibinfo{address}{New York, NY, USA}, \bibinfo{pages}{407–420}.
\newblock
\showISBNx{9781450335423}
\urldef\tempurl%
\url{https://doi.org/10.1145/2785956.2787488}
\showDOI{\tempurl}


\bibitem[Jones(2018)]%
        {Jones2018HowTS}
\bibfield{author}{\bibinfo{person}{Nicola Jones}.}
  \bibinfo{year}{2018}\natexlab{}.
\newblock \showarticletitle{How to {S}top {D}ata {C}entres from {G}obbling {U}p
  the {W}orld’s {E}lectricity}.
\newblock \bibinfo{journal}{\emph{Nature}} (\bibinfo{year}{2018}).
\newblock


\bibitem[Justus et~al\mbox{.}(2018)]%
        {justus2018}
\bibfield{author}{\bibinfo{person}{Daniel Justus}, \bibinfo{person}{John
  Brennan}, \bibinfo{person}{Stephen Bonner}, {and}
  \bibinfo{person}{Andrew~Stephen McGough}.} \bibinfo{year}{2018}\natexlab{}.
\newblock \showarticletitle{Predicting the Computational Cost of Deep Learning
  Models}. In \bibinfo{booktitle}{\emph{2018 IEEE International Conference on
  Big Data (Big Data)}}.
\newblock


\bibitem[Kansal et~al\mbox{.}(2010)]%
        {vmaccounting}
\bibfield{author}{\bibinfo{person}{Aman Kansal}, \bibinfo{person}{Feng Zhao},
  \bibinfo{person}{Jie Liu}, \bibinfo{person}{Nupur Kothari}, {and}
  \bibinfo{person}{Arka~A. Bhattacharya}.} \bibinfo{year}{2010}\natexlab{}.
\newblock \showarticletitle{Virtual Machine Power Metering and Provisioning}.
  In \bibinfo{booktitle}{\emph{Proceedings of the 1st ACM Symposium on Cloud
  Computing}} (Indianapolis, Indiana, USA) \emph{(\bibinfo{series}{SoCC '10})}.
  \bibinfo{publisher}{Association for Computing Machinery},
  \bibinfo{address}{New York, NY, USA}, \bibinfo{pages}{39–50}.
\newblock
\showISBNx{9781450300360}
\urldef\tempurl%
\url{https://doi.org/10.1145/1807128.1807136}
\showDOI{\tempurl}


\bibitem[Kubeflow(2022)]%
        {kubeflow}
\bibfield{author}{\bibinfo{person}{Kubeflow}.} \bibinfo{year}{2022}\natexlab{}.
\newblock \bibinfo{title}{Kubeflow: The Machine Learning Toolkit for
  Kubernetes}.
\newblock \bibinfo{howpublished}{\url{https://www.kubeflow.org/}}.
\newblock
\newblock
\shownote{Accessed: 2022-10-03}.


\bibitem[Kubernetes(2022)]%
        {kubernetes}
\bibfield{author}{\bibinfo{person}{Kubernetes}.}
  \bibinfo{year}{2022}\natexlab{}.
\newblock \bibinfo{title}{Kubernetes: {P}roduction-grade {C}ontainer
  {O}rchestration}.
\newblock \bibinfo{howpublished}{\url{https://kubernetes.io/}}.
\newblock
\newblock
\shownote{Accessed: 2022-10-03}.


\bibitem[Li et~al\mbox{.}(2023)]%
        {li2023sustainable}
\bibfield{author}{\bibinfo{person}{Baolin Li}, \bibinfo{person}{Siddharth
  Samsi}, \bibinfo{person}{Vijay Gadepally}, {and} \bibinfo{person}{Devesh
  Tiwari}.} \bibinfo{year}{2023}\natexlab{}.
\newblock \bibinfo{title}{{Sustainable HPC: Modeling, Characterization, and
  Implications of Carbon Footprint in Modern HPC Systems}}.
\newblock
\newblock
\showeprint[arxiv]{2306.13177}~[cs.DC]


\bibitem[Maji et~al\mbox{.}(2022a)]%
        {carboncast}
\bibfield{author}{\bibinfo{person}{Diptyaroop Maji}, \bibinfo{person}{Prashant
  Shenoy}, {and} \bibinfo{person}{Ramesh~K. Sitaraman}.}
  \bibinfo{year}{2022}\natexlab{a}.
\newblock \showarticletitle{{CarbonCast: Multi-Day Forecasting of Grid Carbon
  Intensity}}. In \bibinfo{booktitle}{\emph{Proceedings of the 9th ACM
  International Conference on Systems for Energy-Efficient Buildings, Cities,
  and Transportation}} (Boston, Massachusetts) \emph{(\bibinfo{series}{BuildSys
  '22})}. \bibinfo{publisher}{Association for Computing Machinery},
  \bibinfo{address}{New York, NY, USA}, \bibinfo{pages}{198–207}.
\newblock
\showISBNx{9781450398909}
\urldef\tempurl%
\url{https://doi.org/10.1145/3563357.3564079}
\showDOI{\tempurl}


\bibitem[Maji et~al\mbox{.}(2022b)]%
        {dayahead_estimations}
\bibfield{author}{\bibinfo{person}{Diptyaroop Maji}, \bibinfo{person}{Ramesh~K.
  Sitaraman}, {and} \bibinfo{person}{Prashant Shenoy}.}
  \bibinfo{year}{2022}\natexlab{b}.
\newblock \showarticletitle{{DACF: Day-Ahead Carbon Intensity Forecasting of
  Power Grids Using Machine Learning}}. In
  \bibinfo{booktitle}{\emph{Proceedings of the Thirteenth ACM International
  Conference on Future Energy Systems}} \emph{(\bibinfo{series}{e-Energy'22})}.
\newblock


\bibitem[Maps(2022)]%
        {electricity-map}
\bibfield{author}{\bibinfo{person}{Electricity Maps}.}
  \bibinfo{year}{2022}\natexlab{}.
\newblock \bibinfo{title}{Electricity {M}ap}.
\newblock \bibinfo{howpublished}{\url{https://www.electricitymap.org/map}}.
\newblock


\bibitem[Masanet et~al\mbox{.}(2020)]%
        {Masanet2020RecalibratingGD}
\bibfield{author}{\bibinfo{person}{Eric~R. Masanet}, \bibinfo{person}{Arman
  Shehabi}, \bibinfo{person}{Nuoa Lei}, \bibinfo{person}{Sarah~J. Smith}, {and}
  \bibinfo{person}{Jonathan~G. Koomey}.} \bibinfo{year}{2020}\natexlab{}.
\newblock \showarticletitle{Recalibrating {G}lobal {D}ata {C}enter {E}nergy-use
  {E}stimates}.
\newblock \bibinfo{journal}{\emph{Science}} (\bibinfo{year}{2020}).
\newblock


\bibitem[Masson-Delmotte et~al\mbox{.}(2021)]%
        {ipcc-report}
\bibfield{author}{\bibinfo{person}{Val{\'e}rie Masson-Delmotte},
  \bibinfo{person}{Panmao Zhai}, \bibinfo{person}{Anna Pirani},
  \bibinfo{person}{Sarah~L Connors}, \bibinfo{person}{Clotilde P{\'e}an},
  \bibinfo{person}{Sophie Berger}, \bibinfo{person}{Nada Caud},
  \bibinfo{person}{Yang Chen}, \bibinfo{person}{Leah Goldfarb},
  \bibinfo{person}{Melissa~I Gomis}, {et~al\mbox{.}}}
  \bibinfo{year}{2021}\natexlab{}.
\newblock \bibinfo{booktitle}{\emph{Summary for {P}olicymakers. In: {C}limate
  {C}hange 2021: The {P}hysical {S}cience {B}asis. {C}ontribution of {W}orking
  {G}roup {I} to the {S}ixth {A}ssessment {R}eport of the {I}ntergovernmental
  {P}anel on {C}limate {C}hange}}.
\newblock \bibinfo{type}{{T}echnical {R}eport}. \bibinfo{institution}{United
  Nation Intergovernmental Panel on Climate Change (IPCC)}.
\newblock


\bibitem[META(2022)]%
        {meta-climate-change}
\bibfield{author}{\bibinfo{person}{META}.} \bibinfo{year}{2022}\natexlab{}.
\newblock \bibinfo{title}{How We're Helping Fight Climate Change}.
\newblock
  \bibinfo{howpublished}{\url{https://about.fb.com/news/2021/06/2020-sustainability-report-how-were-helping-fight-climate-change/}}.
\newblock


\bibitem[Microsoft(2022a)]%
        {aws-carbon-accouting}
\bibfield{author}{\bibinfo{person}{Microsoft}.}
  \bibinfo{year}{2022}\natexlab{a}.
\newblock \bibinfo{title}{AWS Customer Carbon Footprint Tool}.
\newblock
  \bibinfo{howpublished}{\url{https://aws.amazon.com/blogs/aws/new-customer-carbon-footprint-tool/}}.
\newblock


\bibitem[Microsoft(2022b)]%
        {ms-carbon-accouting}
\bibfield{author}{\bibinfo{person}{Microsoft}.}
  \bibinfo{year}{2022}\natexlab{b}.
\newblock \bibinfo{title}{Microsoft Carbon accouting tool}.
\newblock
  \bibinfo{howpublished}{\url{https://www.microsoft.com/en-us/sustainability/emissions-impact-dashboard}}.
\newblock


\bibitem[Microsoft(2022c)]%
        {ms-buy-renewable}
\bibfield{author}{\bibinfo{person}{Microsoft}.}
  \bibinfo{year}{2022}\natexlab{c}.
\newblock \bibinfo{title}{Microsoft is {C}hanging the {W}ay {I}t {B}uys
  {R}enewable {E}nergy}.
\newblock
  \bibinfo{howpublished}{\url{https://www.theverge.com/2021/7/14/22574431/microsoft-renewable-energy-purchases}}.
\newblock


\bibitem[Moghaddam et~al\mbox{.}(2014)]%
        {Moghaddam2014CarbonawareDC}
\bibfield{author}{\bibinfo{person}{Fereydoun~Farrahi Moghaddam},
  \bibinfo{person}{Reza~Farrahi Moghaddam}, {and} \bibinfo{person}{Mohamed
  Cheriet}.} \bibinfo{year}{2014}\natexlab{}.
\newblock \showarticletitle{Carbon-aware {D}istributed {C}loud: {M}ulti-level
  {G}rouping {G}enetic {A}lgorithm}.
\newblock \bibinfo{journal}{\emph{Cluster Computing}} (\bibinfo{year}{2014}).
\newblock


\bibitem[NVIDIA(2022)]%
        {nvidia-dcgm}
\bibfield{author}{\bibinfo{person}{NVIDIA}.} \bibinfo{year}{2022}\natexlab{}.
\newblock \bibinfo{title}{Manage and {M}onitor {G}{P}{U}s in {C}luster
  {E}nvironments}.
\newblock \bibinfo{howpublished}{\url{https://developer.nvidia.com/dcgm}}.
\newblock
\newblock
\shownote{Accessed: 2022-10-08}.


\bibitem[Oyama et~al\mbox{.}(2016)]%
        {Oyama2016}
\bibfield{author}{\bibinfo{person}{Yosuke Oyama}, \bibinfo{person}{Akihiro
  Nomura}, \bibinfo{person}{Ikuro Sato}, \bibinfo{person}{Hiroki Nishimura},
  \bibinfo{person}{Yukimasa Tamatsu}, {and} \bibinfo{person}{Satoshi
  Matsuoka}.} \bibinfo{year}{2016}\natexlab{}.
\newblock \showarticletitle{Predicting {S}tatistics of {A}synchronous SGD
  {P}arameters for a {L}arge-scale {D}istributed {D}eep {L}earning {S}ystem on
  GPU {S}upercomputers}. In \bibinfo{booktitle}{\emph{2016 IEEE International
  Conference on Big Data (Big Data)}}.
\newblock


\bibitem[Paszke et~al\mbox{.}(2019)]%
        {pytorch}
\bibfield{author}{\bibinfo{person}{Adam Paszke}, \bibinfo{person}{Sam Gross},
  \bibinfo{person}{Francisco Massa}, \bibinfo{person}{Adam Lerer},
  \bibinfo{person}{James Bradbury}, \bibinfo{person}{Gregory Chanan},
  \bibinfo{person}{Trevor Killeen}, \bibinfo{person}{Zeming Lin},
  \bibinfo{person}{Natalia Gimelshein}, \bibinfo{person}{Luca Antiga},
  \bibinfo{person}{Alban Desmaison}, \bibinfo{person}{Andreas Kopf},
  \bibinfo{person}{Edward Yang}, \bibinfo{person}{Zachary DeVito},
  \bibinfo{person}{Martin Raison}, \bibinfo{person}{Alykhan Tejani},
  \bibinfo{person}{Sasank Chilamkurthy}, \bibinfo{person}{Benoit Steiner},
  \bibinfo{person}{Lu Fang}, \bibinfo{person}{Junjie Bai}, {and}
  \bibinfo{person}{Soumith Chintala}.} \bibinfo{year}{2019}\natexlab{}.
\newblock \showarticletitle{Py{T}orch: {A}n {I}mperative {S}tyle,
  {H}igh-{P}erformance {D}eep {L}earning {L}ibrary}. In
  \bibinfo{booktitle}{\emph{Advances in Neural Information Processing Systems}}
  \emph{(\bibinfo{series}{NIPS'19})}.
\newblock


\bibitem[Pei et~al\mbox{.}(2019)]%
        {Pei2019}
\bibfield{author}{\bibinfo{person}{Ziqian Pei}, \bibinfo{person}{Chensheng Li},
  \bibinfo{person}{Xiaowei Qin}, \bibinfo{person}{Xiaohui Chen}, {and}
  \bibinfo{person}{Guo Wei}.} \bibinfo{year}{2019}\natexlab{}.
\newblock \showarticletitle{Iteration Time Prediction for CNN in Multi-GPU
  Platform: Modeling and Analysis}.
\newblock \bibinfo{journal}{\emph{IEEE Access}} (\bibinfo{year}{2019}).
\newblock


\bibitem[Peng et~al\mbox{.}(2018)]%
        {optimus}
\bibfield{author}{\bibinfo{person}{Yanghua Peng}, \bibinfo{person}{Yixin Bao},
  \bibinfo{person}{Yangrui Chen}, \bibinfo{person}{Chuan Wu}, {and}
  \bibinfo{person}{Chuanxiong Guo}.} \bibinfo{year}{2018}\natexlab{}.
\newblock \showarticletitle{Optimus: An Efficient Dynamic Resource Scheduler
  for Deep Learning Clusters}. In \bibinfo{booktitle}{\emph{Proceedings of the
  Thirteenth EuroSys Conference}} (Porto, Portugal)
  \emph{(\bibinfo{series}{EuroSys '18})}. \bibinfo{publisher}{Association for
  Computing Machinery}, \bibinfo{address}{New York, NY, USA}, Article
  \bibinfo{articleno}{3}, \bibinfo{numpages}{14}~pages.
\newblock
\showISBNx{9781450355841}
\urldef\tempurl%
\url{https://doi.org/10.1145/3190508.3190517}
\showDOI{\tempurl}


\bibitem[Qi et~al\mbox{.}(2017)]%
        {paleo}
\bibfield{author}{\bibinfo{person}{Qi}, \bibinfo{person}{Evan~R. Sparks}, {and}
  \bibinfo{person}{Ameet~S. Talwalkar}.} \bibinfo{year}{2017}\natexlab{}.
\newblock \showarticletitle{Paleo: A Performance Model for Deep Neural
  Networks}. In \bibinfo{booktitle}{\emph{The International Conference on
  Learning Representations}} \emph{(\bibinfo{series}{ICLR'17})}.
\newblock


\bibitem[Radovanovic et~al\mbox{.}(2022)]%
        {Radovanovic2021CarbonAwareCF}
\bibfield{author}{\bibinfo{person}{Ana Radovanovic}, \bibinfo{person}{Ross
  Koningstein}, \bibinfo{person}{Ian Schneider}, \bibinfo{person}{Bokan Chen},
  \bibinfo{person}{Alexandre Duarte}, \bibinfo{person}{Binz Roy},
  \bibinfo{person}{Diyue Xiao}, \bibinfo{person}{Maya Haridasan},
  \bibinfo{person}{Patrick Hung}, \bibinfo{person}{Nick Care},
  \bibinfo{person}{Saurav Talukdar}, \bibinfo{person}{Eric Mullen},
  \bibinfo{person}{Kendal Smith}, \bibinfo{person}{Mariellen Cottman}, {and}
  \bibinfo{person}{Walfredo Cirne}.} \bibinfo{year}{2022}\natexlab{}.
\newblock \showarticletitle{Carbon-Aware Computing for Datacenters}.
\newblock \bibinfo{journal}{\emph{IEEE Transactions on Power Systems}}
  (\bibinfo{year}{2022}), \bibinfo{pages}{1--1}.
\newblock
\urldef\tempurl%
\url{https://doi.org/10.1109/TPWRS.2022.3173250}
\showDOI{\tempurl}


\bibitem[Reuther et~al\mbox{.}(2018)]%
        {REUTHER201876}
\bibfield{author}{\bibinfo{person}{Albert Reuther}, \bibinfo{person}{Chansup
  Byun}, \bibinfo{person}{William Arcand}, \bibinfo{person}{David Bestor},
  \bibinfo{person}{Bill Bergeron}, \bibinfo{person}{Matthew Hubbell},
  \bibinfo{person}{Michael Jones}, \bibinfo{person}{Peter Michaleas},
  \bibinfo{person}{Andrew Prout}, \bibinfo{person}{Antonio Rosa}, {and}
  \bibinfo{person}{Jeremy Kepner}.} \bibinfo{year}{2018}\natexlab{}.
\newblock \showarticletitle{{Scalable system scheduling for HPC and big data}}.
\newblock \bibinfo{journal}{\emph{J. Parallel and Distrib. Comput.}}
  \bibinfo{volume}{111} (\bibinfo{year}{2018}), \bibinfo{pages}{76--92}.
\newblock
\showISSN{0743-7315}
\urldef\tempurl%
\url{https://doi.org/10.1016/j.jpdc.2017.06.009}
\showDOI{\tempurl}


\bibitem[Sergeev and Balso(2018)]%
        {sergeev2018horovod}
\bibfield{author}{\bibinfo{person}{Alexander Sergeev} {and}
  \bibinfo{person}{Mike~Del Balso}.} \bibinfo{year}{2018}\natexlab{}.
\newblock \showarticletitle{Horovod: Fast and Easy Distributed Deep Learning in
  {TensorFlow}}.
\newblock \bibinfo{journal}{\emph{arXiv preprint arXiv:1802.05799}}
  (\bibinfo{year}{2018}).
\newblock


\bibitem[Shehabi et~al\mbox{.}(2016)]%
        {osti_1372902}
\bibfield{author}{\bibinfo{person}{Arman Shehabi}, \bibinfo{person}{Sarah
  Smith}, \bibinfo{person}{Dale Sartor}, \bibinfo{person}{Richard Brown},
  \bibinfo{person}{Magnus Herrlin}, \bibinfo{person}{Jonathan Koomey},
  \bibinfo{person}{Eric Masanet}, \bibinfo{person}{Nathaniel Horner},
  \bibinfo{person}{Inês Azevedo}, {and} \bibinfo{person}{William Lintner}.}
  \bibinfo{year}{2016}\natexlab{}.
\newblock \showarticletitle{{United States Data Center Energy Usage Report}}.
\newblock  (\bibinfo{date}{6} \bibinfo{year}{2016}).
\newblock
\urldef\tempurl%
\url{https://doi.org/10.2172/1372902}
\showDOI{\tempurl}


\bibitem[Shi et~al\mbox{.}(2018)]%
        {performance_modeling}
\bibfield{author}{\bibinfo{person}{Shaohuai Shi}, \bibinfo{person}{Qiang Wang},
  {and} \bibinfo{person}{Xiaowen Chu}.} \bibinfo{year}{2018}\natexlab{}.
\newblock \showarticletitle{Performance Modeling and Evaluation of Distributed
  Deep Learning Frameworks on GPUs}. In \bibinfo{booktitle}{\emph{2018 IEEE
  16th Intl Conf on Dependable, Autonomic and Secure Computing, 16th Intl Conf
  on Pervasive Intelligence and Computing, 4th Intl Conf on Big Data
  Intelligence and Computing and Cyber Science and Technology
  Congress(DASC/PiCom/DataCom/CyberSciTech)}}. \bibinfo{pages}{949--957}.
\newblock
\urldef\tempurl%
\url{https://doi.org/10.1109/DASC/PiCom/DataCom/CyberSciTec.2018.000-4}
\showDOI{\tempurl}


\bibitem[SIGs(2022)]%
        {metrics-server}
\bibfield{author}{\bibinfo{person}{Kubernetes SIGs}.}
  \bibinfo{year}{2022}\natexlab{}.
\newblock \bibinfo{booktitle}{\emph{{Kubernetes Metrics Server}}}.
\newblock Kubernetes SIGs.
\newblock
\urldef\tempurl%
\url{https://github.com/kubernetes-sigs/metrics-server}
\showURL{%
\tempurl}


\bibitem[Souza et~al\mbox{.}(2023)]%
        {ecovisor}
\bibfield{author}{\bibinfo{person}{Abel Souza}, \bibinfo{person}{Noman Bashir},
  \bibinfo{person}{Jorge Murillo}, \bibinfo{person}{Walid Hanafy},
  \bibinfo{person}{Qianlin Liang}, \bibinfo{person}{David Irwin}, {and}
  \bibinfo{person}{Prashant Shenoy}.} \bibinfo{year}{2023}\natexlab{}.
\newblock \showarticletitle{Ecovisor: A Virtual Energy System for
  Carbon-Efficient Applications}. In \bibinfo{booktitle}{\emph{Proceedings of
  the 28th ACM International Conference on Architectural Support for
  Programming Languages and Operating Systems, Volume 2}} (Vancouver, BC,
  Canada) \emph{(\bibinfo{series}{ASPLOS 2023})}.
  \bibinfo{publisher}{Association for Computing Machinery},
  \bibinfo{address}{New York, NY, USA}, \bibinfo{pages}{252–265}.
\newblock
\showISBNx{9781450399166}
\urldef\tempurl%
\url{https://doi.org/10.1145/3575693.3575709}
\showDOI{\tempurl}


\bibitem[Staples(2006)]%
        {2006torque}
\bibfield{author}{\bibinfo{person}{Garrick Staples}.}
  \bibinfo{year}{2006}\natexlab{}.
\newblock \showarticletitle{{TORQUE resource manager}}. In
  \bibinfo{booktitle}{\emph{Proceedings of the 2006 ACM/IEEE conference on
  Supercomputing}}. \bibinfo{publisher}{ACM}, \bibinfo{address}{New York, NY,
  USA}, \bibinfo{pages}{8}.
\newblock


\bibitem[Stewart(2023)]%
        {netflix-net-zero}
\bibfield{author}{\bibinfo{person}{Emma Stewart}.}
  \bibinfo{year}{2023}\natexlab{}.
\newblock \bibinfo{title}{Net {Z}ero + {N}ature: {O}ur {C}ommitment to the
  {E}nvironment}.
\newblock
  \bibinfo{howpublished}{\url{https://about.netflix.com/en/news/net-zero-nature-our-climate-commitment}}.
\newblock


\bibitem[Sukprasert et~al\mbox{.}(2023)]%
        {sukprasert2023quantifying}
\bibfield{author}{\bibinfo{person}{Thanathorn Sukprasert},
  \bibinfo{person}{Abel Souza}, \bibinfo{person}{Noman Bashir},
  \bibinfo{person}{David Irwin}, {and} \bibinfo{person}{Prashant Shenoy}.}
  \bibinfo{year}{2023}\natexlab{}.
\newblock \bibinfo{title}{{Quantifying the Benefits of Carbon-Aware Temporal
  and Spatial Workload Shifting in the Cloud}}.
\newblock
\newblock
\showeprint[arxiv]{2306.06502}~[cs.DC]


\bibitem[Tan and Le(2019)]%
        {efficient-net}
\bibfield{author}{\bibinfo{person}{Mingxing Tan} {and} \bibinfo{person}{Quoc
  Le}.} \bibinfo{year}{2019}\natexlab{}.
\newblock \showarticletitle{{E}fficient{N}et: Rethinking Model Scaling for
  Convolutional Neural Networks}. In \bibinfo{booktitle}{\emph{Proceedings of
  the 36th International Conference on Machine Learning}}
  \emph{(\bibinfo{series}{Proceedings of Machine Learning Research},
  Vol.~\bibinfo{volume}{97})}. \bibinfo{publisher}{PMLR},
  \bibinfo{pages}{6105--6114}.
\newblock
\urldef\tempurl%
\url{https://proceedings.mlr.press/v97/tan19a.html}
\showURL{%
\tempurl}


\bibitem[Tirmazi et~al\mbox{.}(2020)]%
        {Tirmazi2020BorgTN}
\bibfield{author}{\bibinfo{person}{Muhammad Tirmazi}, \bibinfo{person}{Adam
  Barker}, \bibinfo{person}{Nan Deng}, \bibinfo{person}{Md~E. Haque},
  \bibinfo{person}{Zhijing~Gene Qin}, \bibinfo{person}{Steven Hand},
  \bibinfo{person}{Mor Harchol-Balter}, {and} \bibinfo{person}{John Wilkes}.}
  \bibinfo{year}{2020}\natexlab{}.
\newblock \showarticletitle{Borg: The next Generation}. In
  \bibinfo{booktitle}{\emph{Proceedings of the Fifteenth European Conference on
  Computer Systems}} (Heraklion, Greece) \emph{(\bibinfo{series}{EuroSys
  '20})}. \bibinfo{publisher}{Association for Computing Machinery},
  \bibinfo{address}{New York, NY, USA}, Article \bibinfo{articleno}{30},
  \bibinfo{numpages}{14}~pages.
\newblock
\showISBNx{9781450368827}
\urldef\tempurl%
\url{https://doi.org/10.1145/3342195.3387517}
\showDOI{\tempurl}


\bibitem[WattTime(2022)]%
        {watttime}
\bibfield{author}{\bibinfo{person}{WattTime}.} \bibinfo{year}{2022}\natexlab{}.
\newblock \bibinfo{title}{Watt{T}ime}.
\newblock \bibinfo{howpublished}{\url{https://www.watttime.org/}}.
\newblock


\bibitem[Weng et~al\mbox{.}(2022)]%
        {weng2022mlaas}
\bibfield{author}{\bibinfo{person}{Qizhen Weng}, \bibinfo{person}{Wencong
  Xiao}, \bibinfo{person}{Yinghao Yu}, \bibinfo{person}{Wei Wang},
  \bibinfo{person}{Cheng Wang}, \bibinfo{person}{Jian He},
  \bibinfo{person}{Yong Li}, \bibinfo{person}{Liping Zhang},
  \bibinfo{person}{Wei Lin}, {and} \bibinfo{person}{Yu Ding}.}
  \bibinfo{year}{2022}\natexlab{}.
\newblock \showarticletitle{MLaaS in the wild: Workload analysis and scheduling
  in Large-Scale heterogeneous GPU clusters}. In \bibinfo{booktitle}{\emph{19th
  USENIX Symposium on Networked Systems Design and Implementation (NSDI 22)}}.
  USENIX Association, \bibinfo{pages}{945--960}.
\newblock


\bibitem[Wiesner et~al\mbox{.}(2021)]%
        {Wiesner2021LetsWA}
\bibfield{author}{\bibinfo{person}{Philipp Wiesner}, \bibinfo{person}{Ilja
  Behnke}, \bibinfo{person}{Dominik Scheinert}, \bibinfo{person}{Kordian
  Gontarska}, {and} \bibinfo{person}{Lauritz Thamsen}.}
  \bibinfo{year}{2021}\natexlab{}.
\newblock \showarticletitle{Let's Wait Awhile: How Temporal Workload Shifting
  Can Reduce Carbon Emissions in the Cloud}. In
  \bibinfo{booktitle}{\emph{Proceedings of the 22nd International Middleware
  Conference}} (Qu\'{e}bec city, Canada) \emph{(\bibinfo{series}{Middleware
  '21})}. \bibinfo{publisher}{Association for Computing Machinery},
  \bibinfo{address}{New York, NY, USA}, \bibinfo{pages}{260–272}.
\newblock
\showISBNx{9781450385343}
\urldef\tempurl%
\url{https://doi.org/10.1145/3464298.3493399}
\showDOI{\tempurl}


\bibitem[Yoo et~al\mbox{.}(2003)]%
        {2003slurm}
\bibfield{author}{\bibinfo{person}{Andy~B Yoo}, \bibinfo{person}{Morris~A
  Jette}, {and} \bibinfo{person}{Mark Grondona}.}
  \bibinfo{year}{2003}\natexlab{}.
\newblock \showarticletitle{{Slurm: {S}imple {L}inux {U}tility for {R}esource
  {M}anagement}}. In \bibinfo{booktitle}{\emph{Workshop on Job Scheduling
  Strategies for Parallel Processing}}. \bibinfo{publisher}{Springer},
  \bibinfo{address}{New York, NY, USA}, \bibinfo{pages}{44--60}.
\newblock


\bibitem[Zhang and Chien(2021)]%
        {variable_capacity_challanges}
\bibfield{author}{\bibinfo{person}{Chaojie Zhang} {and}
  \bibinfo{person}{Andrew~A. Chien}.} \bibinfo{year}{2021}\natexlab{}.
\newblock \showarticletitle{{Scheduling Challenges for Variable Capacity
  Resources}}. In \bibinfo{booktitle}{\emph{{Job Scheduling Strategies for
  Parallel Processing}}}, \bibfield{editor}{\bibinfo{person}{Dalibor
  Klus{\'a}{\v{c}}ek}, \bibinfo{person}{Walfredo Cirne}, {and}
  \bibinfo{person}{Gonzalo~P. Rodrigo}} (Eds.). \bibinfo{publisher}{Springer
  International Publishing}, \bibinfo{address}{Cham},
  \bibinfo{pages}{190--209}.
\newblock
\showISBNx{978-3-030-88224-2}


\bibitem[Zheng et~al\mbox{.}(2020)]%
        {Zheng2020MitigatingCA}
\bibfield{author}{\bibinfo{person}{Jiajia Zheng}, \bibinfo{person}{Andrew~A.
  Chien}, {and} \bibinfo{person}{Sangwon Suh}.}
  \bibinfo{year}{2020}\natexlab{}.
\newblock \showarticletitle{Mitigating Curtailment and Carbon Emissions through
  Load Migration between Data Centers}.
\newblock \bibinfo{journal}{\emph{Joule}} \bibinfo{volume}{4},
  \bibinfo{number}{10} (\bibinfo{year}{2020}), \bibinfo{pages}{2208--2222}.
\newblock
\showISSN{2542-4351}
\urldef\tempurl%
\url{https://doi.org/10.1016/j.joule.2020.08.001}
\showDOI{\tempurl}


\bibitem[Zhou et~al\mbox{.}(2013)]%
        {Zhou2013CarbonAwareLB}
\bibfield{author}{\bibinfo{person}{Zhi Zhou}, \bibinfo{person}{Fangming Liu},
  \bibinfo{person}{Yong Xu}, \bibinfo{person}{Ruolan Zou},
  \bibinfo{person}{Hong Xu}, \bibinfo{person}{John~C.S. Lui}, {and}
  \bibinfo{person}{Hai Jin}.} \bibinfo{year}{2013}\natexlab{}.
\newblock \showarticletitle{Carbon-{A}ware {L}oad {B}alancing for
  {G}eo-distributed {C}loud {S}ervices}. In
  \bibinfo{booktitle}{\emph{International Symposium on Modelling, Analysis and
  Simulation of Computer and Telecommunication Systems}}.
  \bibinfo{publisher}{IEEE}, \bibinfo{address}{New York, NY, USA},
  \bibinfo{pages}{232--241}.
\newblock
\urldef\tempurl%
\url{https://doi.org/10.1109/MASCOTS.2013.31}
\showDOI{\tempurl}


\end{thebibliography}

\appendix
\section{CarbonScaler Optimality}
\label{app:optimality}
The carbon scaling problem addressed by \systemName is a marginal resource allocation problem, where greedily selecting the local optimum (maximum marginal capacity per unit carbon), as in Algorithm~\ref{alg:proposed_algorithm}, yields the global optimum solution \cite{greedy_optimal}. Consequently, the optimality of the \texttt{Carbon Scaling Algorithm} follows from the theoretical results of \cite{greedy_optimal} and is shown below.

\begin{theorem}
Consider a distributed batch job with a known monotonically decreasing marginal capacity curve, s.t. $MC_m > MC_{m+1}>.. > MC_{M}$. The job needs to finish work $W$, within $n$ time slots with known carbon costs $c_1, c_2,..., c_n$, respectively. Greedily selecting the slot $i$ and scaling the job to $j$ servers with the highest marginal capacity per unit carbon $MC_j/c_i$\footnote{We assume that switching cost (scaling up or down) between time slots is negligible.}, in each step, results in the lowest (optimal) amount of carbon consumption.
\end{theorem}

\begin{proof}
We prove Theorem 1 by contradiction. 
Let $S$ be an optimal solution schedule that finishes work $W$ and has a carbon cost $C_S$. 
The schedule $S$ is constructed by allocating time slots and number of servers, until $W$ is completed. The tuple $(i,j)$ denotes the $i$-th time slot and the $j$-th server allocated to the job. $MC_j$ is the marginal work done when allocating the $j$-th server, and $c_i$ is the carbon cost used per server at time slot $i$, where we assume perfect knowledge of both.
The total carbon cost is $C_S=\sum_{i\in n}{c_i \times S[i]}$, where $S[i]$ is the used number of servers at time slot $i$.

The tuple $(k,l)$ denotes the $k$-th time slot and the $l$-th server, with marginal capacity per unit carbon of $MC_{l}/c_{k}$. 
Assume that there exists a time slot $i$ and a number of servers $j$, where $MC_{l}/c_{k}>MC_{j}/c_{i}$, s.t., $(i,j) \in S$ and $(k,l) \notin S$. 
We denote $S^{'}$ as a new schedule, where we only switch the $i$-th time slot and $j$-th server with $k$-th time slot and the $l$-th server, which has the higher marginal capacity per unit carbon.
To ensure that the schedule $S^{'}$ finishes work $W$, the amount of work  $MC_j$ must be incorporated into the new schedule. We denote $c_i$ and $\gamma$, as the old carbon and new carbon costs to perform work $MC_j$, respectively. $\gamma$ is computed based on the relationship between $l$ and $j$, where:
\begin{equation}
  \gamma=\begin{cases}
    c_k\cdot \frac{MC_j}{MC_l}, & \text{if $l\leq j$} . \\
    c_k + (\frac{MC_j - MC_l}{MC_j}) \cdot c_i, & \text{otherwise $(l>j)$}. \\
  \end{cases}
\end{equation}
In the first case $(l\leq j)$, $MC_l\geq MC_j$ and job will use part or all of the time slot $k$. In the second case $(l>j)$, $MC_l < MC_j$. We perform $MC_l$ work in time slot $k$, and run the overflow work, $MC_j - MC_l$, in time slot $i$, utilizing $\frac{MC_j - MC_l}{MC_j}$ of time slot $i$ and all of time slot $k$.

Next, to show that $c_i > \gamma$ and $(C_{S} > C_{S^{'}})$, we consider both cases. In the first case, since $MC_{l}/c_{k}>MC_{j}/c_{i}$, then:
\begin{align}
    c_i > c_k\cdot\frac{MC_{j}}{MC_l}\\
    c_i > \gamma
\end{align}
In the second case, since $\frac{MC_{l}}{c_{k}}>\frac{MC_j}{c_i}$, then:
\begin{align}
c_k < \frac{MC_l}{MC_j} \cdot c_i
\end{align}
By substituting $c_k$ in case 2:
\begin{align}
\frac{MC_l}{MC_j} \cdot c_i + (\frac{MC_j - MC_l}{MC_j}) \cdot c_i > \gamma \\
\frac{MC_l \cdot c_i + MC_j \cdot c_i - MC_l \cdot c_i}{MC_j} > \gamma \\
c_i > \gamma 
\end{align}
Therefore, carbon consumption of $S^{'}$, denoted as $C_{S^{'}} = C_{S} - c_i + \gamma$ is less than carbon cost of $S$ $(C_{S})$, since $c_i > \gamma$. Hence, $S$ is not optimal, a contradiction. 
\end{proof}



\received{February 2023}
\received[revised]{October 2023}
\received[accepted]{October 2023}
\end{document}